\documentclass[preprint,pre,aps,amsmath,amssymb,a4paper,superscriptaddress,nofootinbib,11pt,tightenlines]{revtex4-1}
\usepackage{graphicx}	
\usepackage{dcolumn}	
\usepackage{bm}			
\usepackage{xcolor}		
\usepackage{hyperref}	
\usepackage[capitalize]{cleveref}	
\usepackage{amsthm} 	
\usepackage{lineno} 	

\usepackage{url}							

\newcommand{\angles}[1]{\langle{#1}\rangle}

\newtheorem*{result}{\noindent Main Result}  

\hypersetup{				
	colorlinks   = true, 	
	urlcolor     = blue, 	
	linkcolor    = blue, 	
	citecolor    = blue     
}
\crefformat{equation}{Eq.~#2(#1)#3}				
\crefformat{figure}{Fig.~#2#1#3}				

\begin{document}
\title{Lorentzian geometry based analysis of airplane boarding policies highlights slow passengers first as better}
\author{Sveinung Erland}
\affiliation{Department of Maritime Studies,\\ {Western Norway University of Applied Sciences, N-5528 Haugesund, Norway}}
\email[Author to whom correspondence should be addressed: ]{sver@hvl.no}
\author{Jevgenijs Kaupu\ifmmode \check{z}\else \v{z}\fi{}s}
\affiliation{Institute of Mathematics and Computer Science, University of Latvia, LV-1459 Riga, Latvia}
\author{Vidar Frette}
\affiliation{Department of Fire Safety and HSE engineering, Western Norway University of Applied Sciences, N-5528 Haugesund, Norway}
\author{Rami Pugatch}
\affiliation{Department of Industrial Engineering and Management, Ben-Gurion University, Beer-Sheva 84105, Israel}
\author{Eitan Bachmat}
\affiliation{Department of Computer Science, Ben-Gurion University, Beer-Sheva 84105, Israel}
\date{\today}

\begin{abstract}
We study airplane boarding in the limit of large number of passengers using geometric optics in a Lorentzian metric.  The airplane boarding problem is naturally embedded in a 1+1 dimensional space-time with a flat Lorentzian metric. The duration of the boarding process can be calculated based on a representation of the one-dimensional queue of passengers attempting to reach their seats, into a  two-dimensional space-time diagram. The ability of a passenger to delay other passengers depends on their queue positions and row designations. This is equivalent to the causal relationship between two events in space-time, whereas two passengers are time-like separated if one is blocking the other, and space-like if both can be seated simultaneously. Geodesics in this geometry can be utilized to compute the asymptotic boarding time, since space-time geometry is the many-particle (passengers) limit of airplane boarding.
	
Our approach naturally leads to a novel definition of an effective refractive index. The introduction of an effective refractive index enables, for the first time, an analytical calculation of the average boarding time for groups of passengers with different aisle-clearing time distribution. In the past, airline companies attempted to shorten the boarding times by trying boarding policies that either allow slow or fast passengers to board first. 
Our analytical calculations, backed by discrete-event simulations, support the counter-intuitive result that the total boarding time is shorter with the slow passengers boarding before the fast passengers. This is a universal result, valid for any combination of the parameters that characterize the problem --- percentage of slow passengers, ratio between aisle-clearing times between the fast and the slow groups, and the density of passengers along the aisle. We find an improvement of up to 28\% compared with the Fast-First boarding policy. Our approach opens up the possibility to unify numerous simulations-based case studies under one framework.
\end{abstract}


\maketitle
\clearpage
\section{Introduction}\label{sec:introduction}
A main theme in statistical physics is the connection between the microscopic dynamics of an ensemble of interacting particles or units --- and the macroscopic observables. In this paper we consider the problem of airplane boarding, where the microscopic level is the structure of the passenger queue --- and the main macroscopic observable is the required time for all $N$ passengers to get settled in their assigned seats. 
The connection between the microscopic level and the macroscopic observable is presented by a novel and simple two-dimensional diagram, with a geometric interpretation that is directly linked to special relativity.

The average boarding time can be calculated analytically in the large-$N$ limit. This average boarding time has been found to be a \emph{square-root law} in the number of passengers \citep{Bachmat/Berend/Sapir/Skiena/Stolyarov:2006,Frette/Hemmer:2012,Bernstein:2012,Bachmat/Khachaturov/Kuperman:2013,Baek/Ha/Jeong:2013,Martins/Kaupuzs/Mahnke:2013,Bachmat:2014,Mahnke/Kaupuzs/Brics:2015}. Our analytical result further enables a direct approach for optimization over the three main parameters of the boarding process, to be elucidated in what follows. {Interestingly, our approach presents a very simple and straight-forward interpretation to the causal set program of quantum gravity \citep{Bombelli/Lee/Meyer/Sorkin:1987,myrheim:1978,tHooft:1979,Brightwell/Gregory:1991}. In terms of the causal set program, the main novelty of our recent work is that we consider a scenario in which there is more than one type of space-time event, with different types of events having different proper time contributions.}

During airplane boarding, the passengers have reserved seats, but arrive at the gate in arbitrary order, and a queue of passengers is formed in the jet bridge. Recently it has been shown that any delay in the boarding process immediately adds to the overall airplane turnaround time, especially for intra-continental flights, i.e., delays in the boarding process will delay the flight departure time \cite{Jaehn/Neumann:2015,neumann:2019}. 
Yet, attempts to optimize the boarding process are not so successful, as many travelers still experience slow queue advancement, with few busy passengers arranging carry-on luggage and taking their seat at any given moment during boarding.

In fact, it can be easy to arrange a boarding queue optimally {(an example is given in Appendix~\ref{app:figures})}. Still, optimal queue arrangements can hardly be used, for two reasons: Firstly, even though specific queue position for each passenger may be implemented, such instructions tend to reduce passenger satisfaction, for example through splitting groups of passengers traveling together. Secondly, the optimal queue arrangements are usually not robust to deviations in terms of passengers that do not take their designated position in the queue. So, unlike many other problems, even though optimal solutions could be easily found, the optimal solutions for airplane boarding are of little interest for the airlines due to their impracticality.

Weaker constraints, for example with passengers assigned to groups according to seat or row number, are employed more often. We will refer to an imposed arrangement of the queue prior to boarding as a \emph{boarding policy}. The most common policy is the unorganized Random Boarding policy, where the passengers enter the queue in random order. Surprisingly, random boarding is relatively efficient. Another familiar policy, is the Back-to-Front policy, where the passengers are divided into two or more groups, and those who have designated seats in the back of the airplane are instructed to enter first. This is a widely used policy, but both simulations and analytical computations show that it is usually detrimental compared to Random Boarding \citep{Bachmat/Khachaturov/Kuperman:2013}.

In this paper we investigate two simple, group-based policies that can bring us closer to a near-optimal solution, which is nevertheless practical, namely passengers that tend to use more time to complete the seating are separated from the others and can be requested to enter the airplane either before or after the remaining passengers. One such type of passengers are those with overhead bin luggage. 

Moreover, the size of carry-on luggage is also an easy and practical criteria for separating the passengers into groups of what we denote slow and fast passengers, respectively. Fast-First boarding policies where passengers without overhead bin luggage are allowed to board before other passengers enter the plane, have been implemented in the past \citep{Reed:2013}. The opposite Slow-First policy, has been implemented in the way that certain small groups of slow passengers usually are allowed to board before other passengers. Such groups typically consist of, e.g., small children and those who need special assistance.

That the Slow-First policy is superior to the Fast-First policy is surprising at first. We apply the analytical tools of Lorentzian geometry to prove that this is indeed the case. While Lorentzian geometry has been used previously to analyze the boarding process \citep{Bachmat:2014,Bachmat/Berend/Sapir/Skiena/Stolyarov:2006}, the two group scenario brings a new aspect. The different group speeds
can be analyzed in terms of the concept of a refraction index, i.e., they scale the metric by different amounts in different regions of space, bringing an optics perspective into the picture.

To our knowledge there are no other techniques in the airplane boarding literature that are able to analyze such boarding policies in a general setting (see, e.g., \cite{Jaehn/Neumann:2015}). Simulation algorithms may indicate which policy is superior to others, but fail to provide generality, proof or insight.

To be more specific, the simulation-based result in \cite{Audenaert/Verbeeck/Berghe:2009} is here verified for far more general model settings. Policies that organize the fast and slow passengers in more optimal ways according to their designated row positions have been constructed \citep{Milne/Kelly:2014,Qiang/Jia/Xie/Gao:2014,Notomistaetal:2016,Bachmat:2019}. However, as mentioned before, policies that require detailed control on each of the passengers cause great discomfort and noncompliance. Thus, while these boarding policies are of theoretical interest, they are of little use and are not expected to be implemented. 

The calculations to be discussed in detail below, demonstrate that it is possible to obtain analytical results for complex situations using visual representations (diagrams) and geometry. 

The structure of the paper is as follows. The boarding process is described in Sec.~\ref{sec:boardingprocess}, and the main results are summarized in Sec.~\ref{sec:mainresults}. Relevant parameters of the boarding process, its representation in diagrams, and the connection to spacetime geometry are presented in Sec.~\ref{sec:model}. The asymptotic boarding time is presented in Sec.~\ref{sec:analysis} for general, group-based policies in the many passenger limit ($N\rightarrow \infty$). In Sec.~\ref{sec:kgt0} we compute analytically the asymptotic boarding time for the Slow-First (SF) and the Fast-First (FF) policies. We also show by simulations that the large-$N$ limit results hold widely for realistic number of passengers.

\section{The Boarding Process}\label{sec:boardingprocess}
We consider the boarding process from the time when the passengers have queued up in the jet bridge outside the airplane entrance, until the last passenger is seated. Most passengers wait most of the time during boarding since they are blocked by the other passengers from reaching their designated seat. We assume that the queue order is maintained throughout the process, i.e., passengers cannot pass other passengers that are in front of them in the aisle. 

The boarding is modeled as an iterative, two-step process: First all passengers move until they either reach their designated row or until they are blocked on the way to their seat by another passenger. This is assumed to take an insignificant amount of time compared to the next step where those passengers who stand next to their designated row use a certain \emph{aisle-clearing time} in order to organize luggage and take a seat.

\begin{figure*}[htb]
	\begin{center}
		\includegraphics[width=7.0cm]{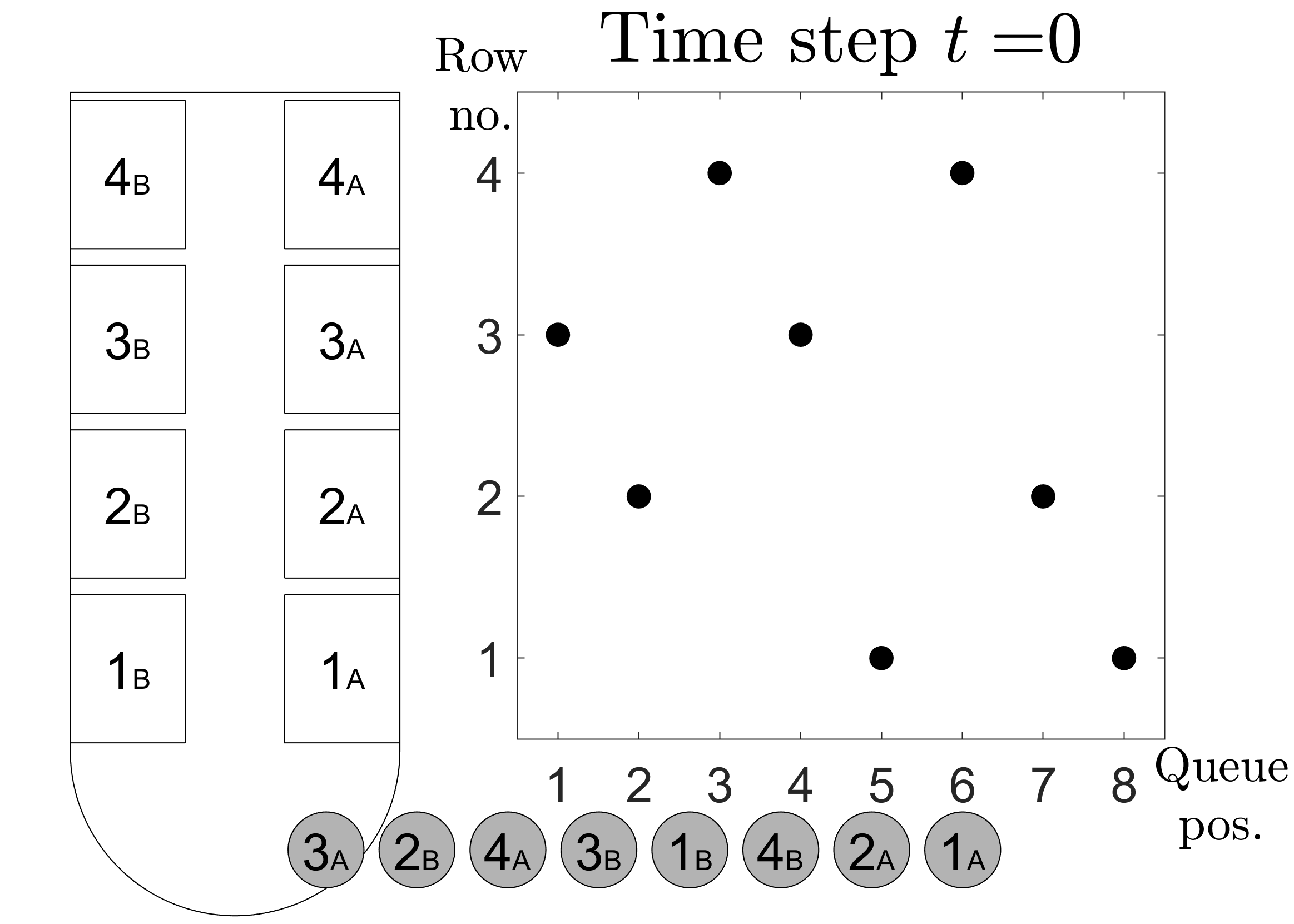}
		\parbox[b][3.6cm][c]{1cm}{$\quad$}
		\includegraphics[width=7.0cm]{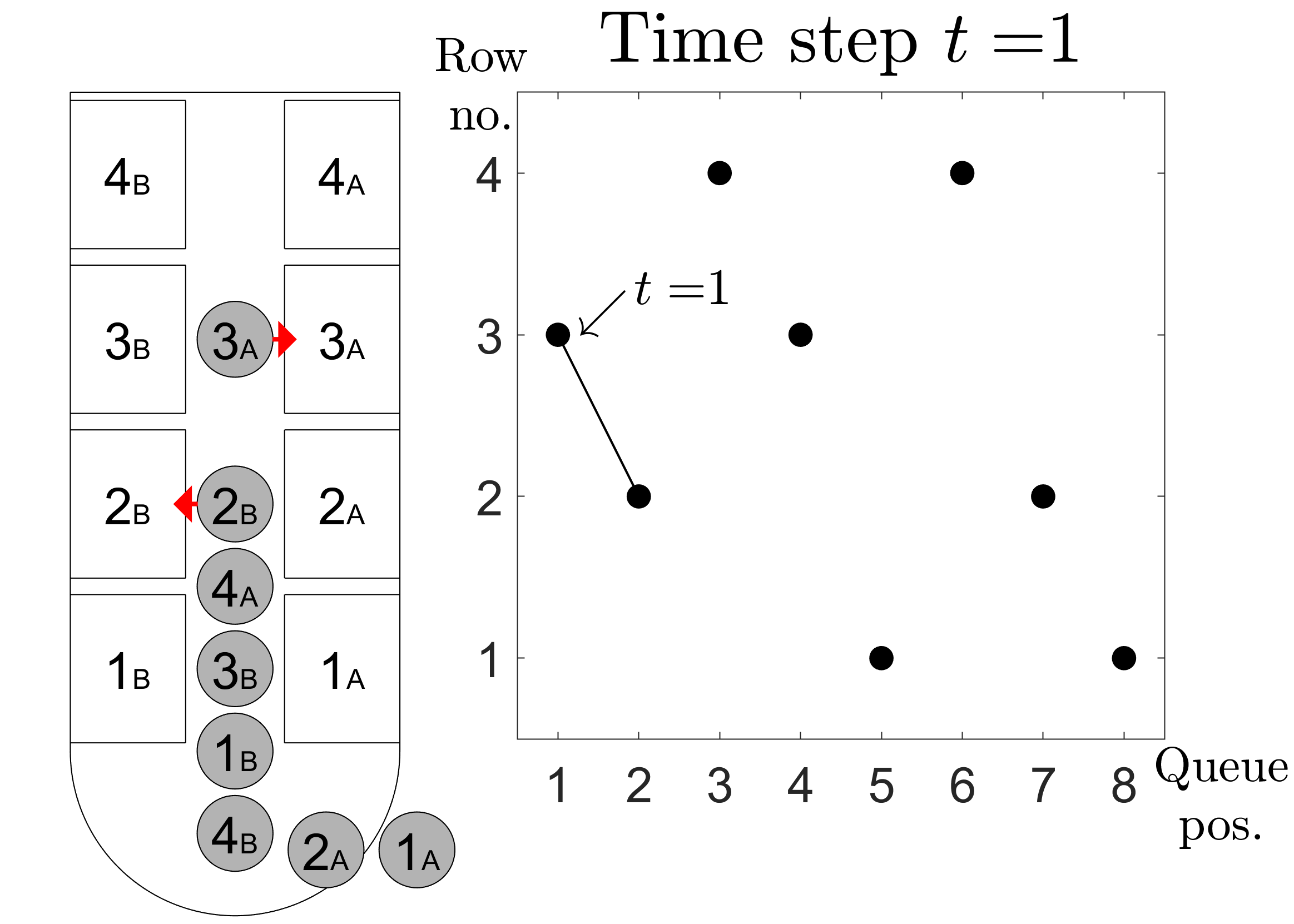}\\[5ex]
		\includegraphics[width=7.0cm]{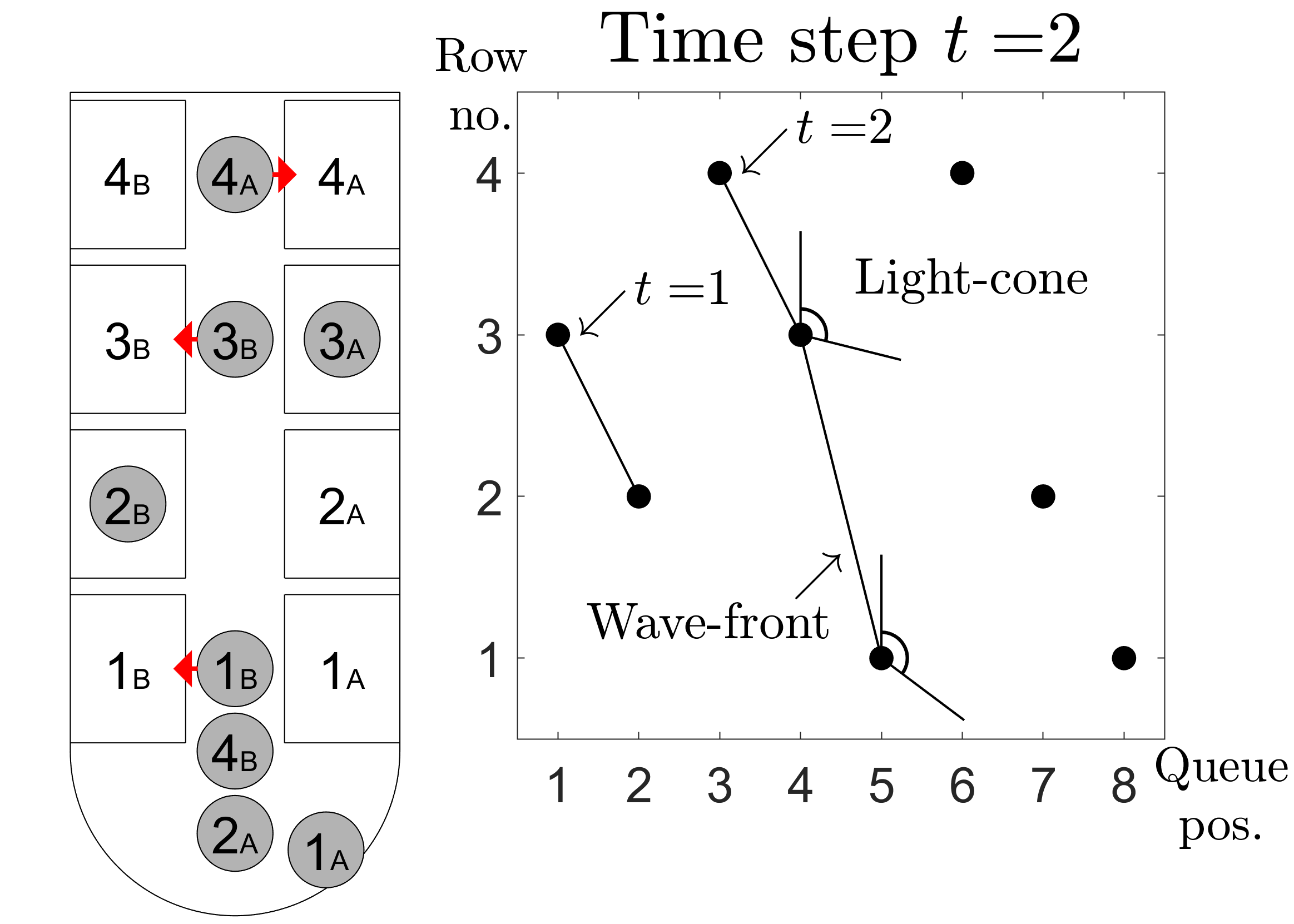}
		\parbox[b][3.6cm][c]{1cm}{$\quad$}
		\includegraphics[width=7.0cm]{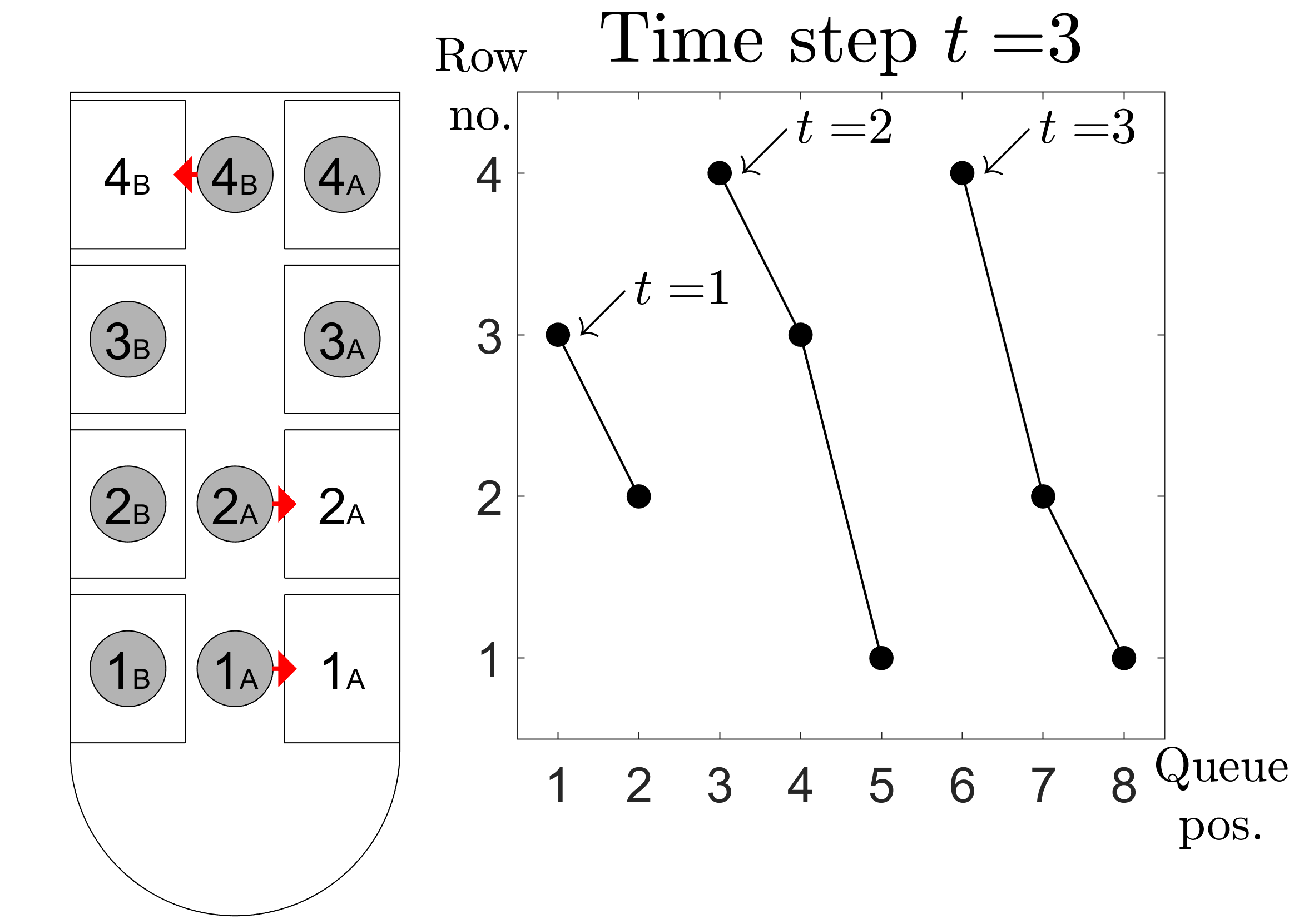}\\
	\end{center}	
	\caption{\label{fig:boarding_illustration} Illustration of the stepwise advance of the queue during the boarding process, with $N=8$ passengers, $4$ rows and $2$ seats per row. There is space for two passenger per row along the aisle. When $t=0$, the passengers are lined up in a queue outside the airplane at the left-hand side of each inset. Each passenger is marked as a circle with designated row number. At each time step, the queue moves forward, and passengers that have reached their designated rows sit down simultaneously. Red arrows indicate passengers that take their seat in that time step. At the right-hand side of each figure part, each passenger is marked as a point in a qr-diagram: the initial queue position of the passenger is on the horizontal axis and the designated row number is on the vertical axis. The points of passengers that sit down simultaneously are joined by line segments (\emph{wave-fronts}). 
	Fellow passengers that are within another passenger's \emph{light-cone} must wait for that passenger to settle in his seat before they can sit down themselves.
	}
\end{figure*} 

A simple example with only $N=8$ passengers, all with the same aisle-clearing time, is presented in Fig.~\ref{fig:boarding_illustration}. At each time step, a group of passengers is able to sit down simultaneously. A passenger can be delayed by the passenger in front of her in two ways. Firstly, she could have a higher row number than the passenger in front of her (in time step $t=1$, the third passenger which is heading for row 4 must wait for the passenger taking a seat at row 2). Secondly, she could be displaced by passengers who are waiting for other passengers to take a seat (in time step $t=1$, the fifth passenger is heading for row 1, but must wait until passengers in front of her have been let through). This displacement effect is less significant if passengers stand closer to each other and thus occupy less space in the aisle. 
The time until the last passenger is seated we denote the boarding time $T$. In Fig.~\ref{fig:boarding_illustration}, $T=3$ time steps.

By reorganizing the queue in Fig.~\ref{fig:boarding_illustration}, it is possible to obtain a minimal boarding time, as shown 
in Appendix \ref{app:figures}. For such optimal solutions, it is necessary to impose a specific position in the queue for each passenger. Optimization at the level of individual passengers will not be pursued in this article.

For the sake of visualizing and analyzing the boarding process, we present the \emph{qr-diagram}. In the qr-diagram a point $(q,r)$ represents a particular passenger's initial queue position $q$ and designated row number $r$ in the airplane (see diagrams in Fig.~\ref{fig:boarding_illustration}). 
Passengers that take seats simultaneously are linked by lines. We call each such group of passengers a \emph{wave-front} in analogy to wave-fronts in physics, as they represent all the events that share the same phase, i.e., all the passengers that are seated simultaneously. The boarding time can be found by counting the number of equidistant wave-fronts, multiplying by the time-difference. 

We use qr-diagrams as a tool to analyze the boarding process. Such diagrams convey the entire hierarchy of blocking between passengers for a given queue configuration.

\section{Main Results}\label{sec:mainresults}
\begin{figure*}
	\begin{center}
		\includegraphics[width=7.5cm,clip]{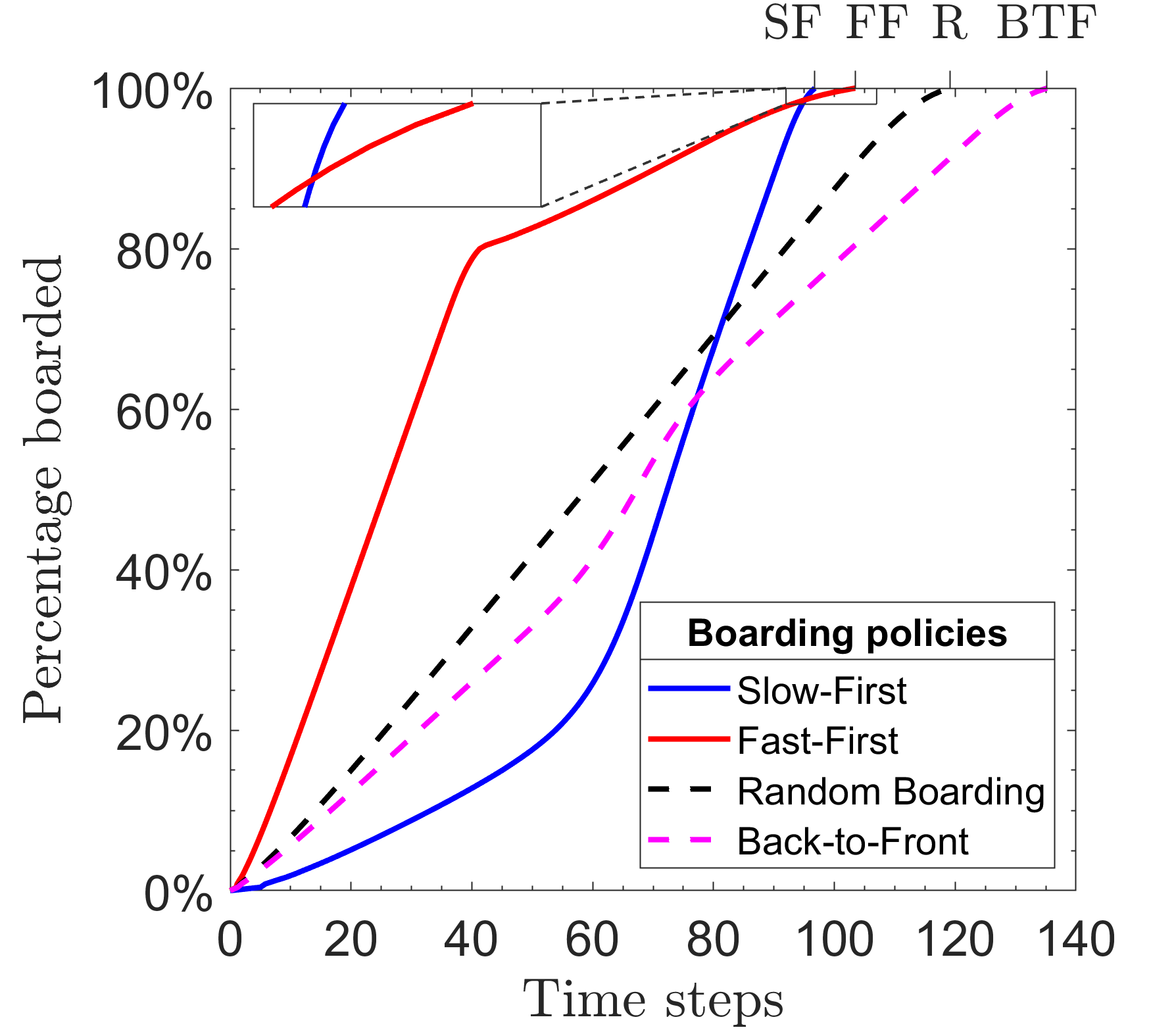}  
	\end{center}
	\caption{\label{fig:mainresultsa} 
		Comparison of 4 different boarding policies. We used a realistic congestion parameter $k=4$, $20\%$ slow passengers ($p=0.2$) and assumed that the slow passengers clear the aisle on average 5 times longer than the fast passengers ($C=5$). We assumed there is a single aisle and 6 seats per row, and a total of $N = 240$ passengers. The percentage of seated passengers is plotted as a function of time. 
		Remarkably, on average the Fast-First policy is leading all the way to around $\sim 98\%$. However, the Slow-First policy eventually seats all passengers in a shorter time --- relative to all the other policies.
		The Fast-First policy is second (FF; $+7\%$), Random Boarding comes third (R; $+23\%$) and the Back-to-Front policy turns out to be the worst (BTF; $+40\%$).
		That the Slow-First policy is superior can be intuitively explained by that it is the most parallel among all the policies --- it better exploits the possibility to seat passengers simultaneously while the other policies are more serial in structure. 
		The graph is an average of 10,000 discrete-event simulations.
	}
\end{figure*}

The main results of this paper are shown in Fig.~\ref{fig:mainresultsa}, for a particular realization of the three governing parameters to be defined below. 
In the subsequent sections we will rigorously prove that the main features in Fig.~\ref{fig:mainresultsa} are universal and apply for any set of parameters.

The airplane boarding problem we consider here is characterized by three key parameters. The first parameter is the \emph{congestion} --- $k$, which is the ratio between the queue length before boarding to the aisle length. 
Values of $k$ are typically in the range of $3$ to $5$.
The second parameter is the fraction $p$ of the passengers that are considered slow, i.e., passengers with long aisle-clearing time. The remaining fraction, $(1-p)$, are considered fast passengers. 
The third parameter $C$ is the ratio of the  aisle-clearing time of the fast passengers to the aisle-clearing time of the slow passengers.
This plays the role of a refraction index, as we discuss below.

In Fig.~\ref{fig:mainresultsa} we take the congestion parameter to be  $k=4$, the fraction of slow passengers to be $p=20\%$ and the ratio of the aisle-clearing time of the fast passengers to that of the slow passengers to be $C=0.2$, i.e., the aisle-clearing time of the slow passengers is 5 times longer than of the fast passengers.

The graphs in Fig.~\ref{fig:mainresultsa} show the percentage of seated passengers as a function of time for the four different boarding policies described in Sec.~\ref{sec:introduction}, namely Slow passengers first (SF), Fast passengers first (FF), Back to front boarding (BTF) and Random Boarding (R). 
The boarding is completed when the fraction of seated passengers equals $100\%$. The boarding time $T$ is equal to the first time when all the passengers are seated. In the figure, the following  ranking of the policies can be observed. The best boarding policy is Slow-First, with a boarding time of $97$ time steps. The second fastest boarding policy is Fast-First, with boarding time of $103$. The third best policy is Random Boarding with boarding time of $119$, and the worst policy is Back-to-Front with boarding time of $135$. 
In Ref. \cite{Bachmat/Khachaturov/Kuperman:2013} it is shown that Random Boarding is typically superior to Back-to-Front. Random Boarding and Back-to-Front are included in this figure for reference only, since they have been studied in the literature and are often implemented by airlines. Our main focus remains the comparison between Slow-First and Fast-First.

The graphs  for Slow-First (SF) and Fast-First (FF) in Fig.~\ref{fig:mainresultsa} both consist of two curve segments with different slopes. For Fast-First, the steep segment comes first, followed by the less steep, with opposite order for Slow-First. The steep segments correspond to boarding dominated by fast passengers, while the less steep segments are dominated by slow passengers. As boarding starts, the queue of passengers is four times as long as the aisle ($k=4$). 
\begin{figure*}
	\begin{center}
		\includegraphics[width=7.5cm,clip]{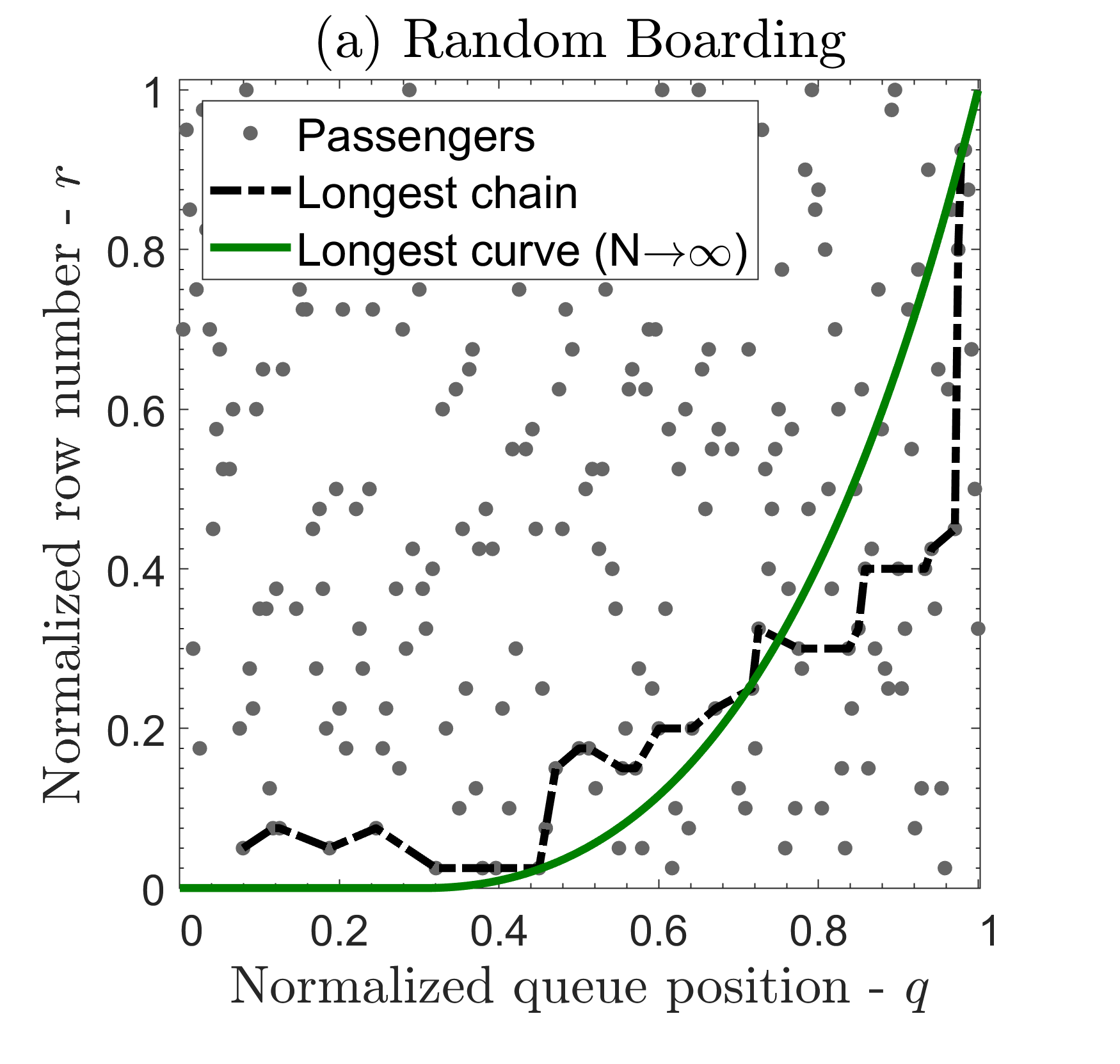}
		\includegraphics[width=7.5cm,clip]{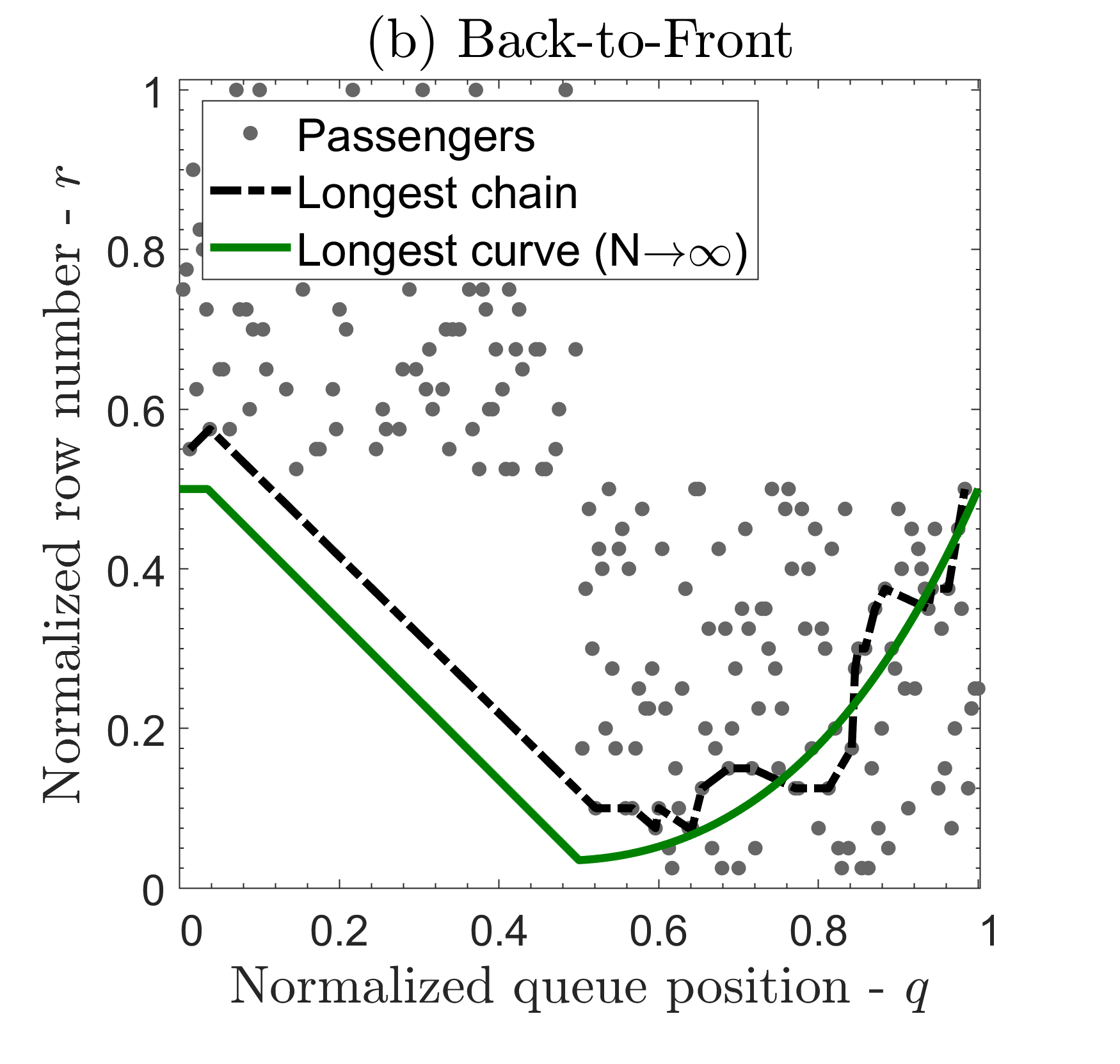}\\
		\includegraphics[width=7.5cm,clip]{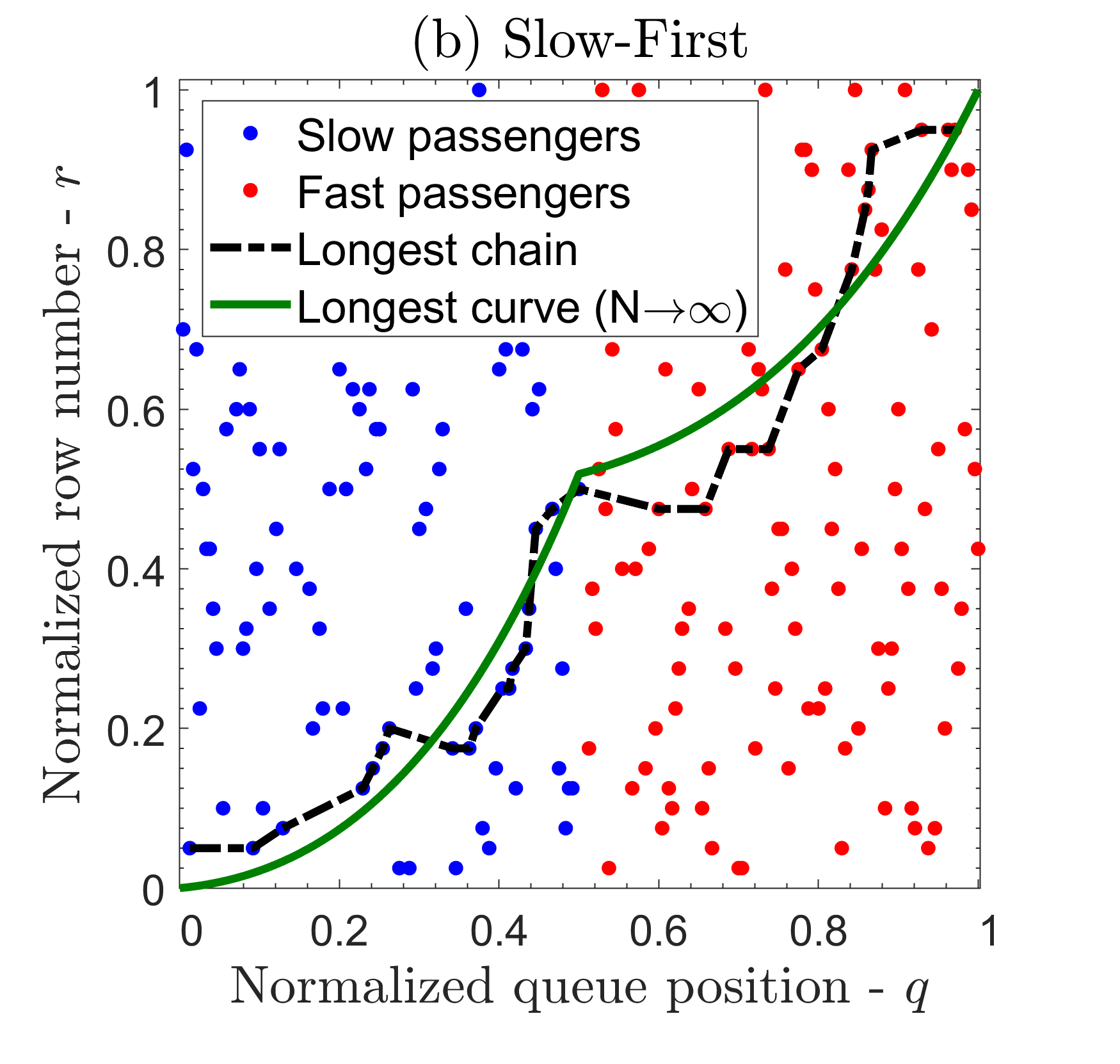}
		\includegraphics[width=7.5cm,clip]{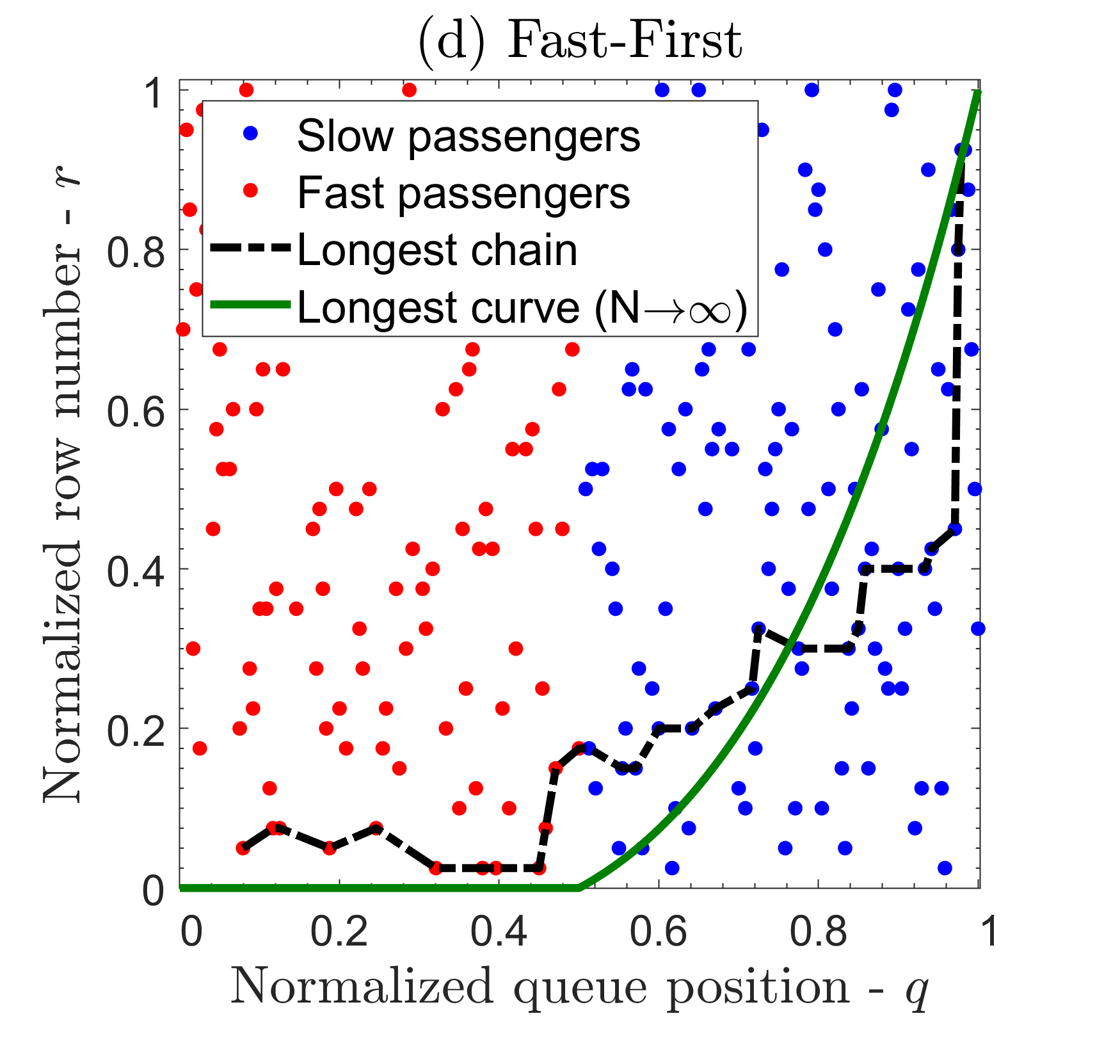}
	\end{center}
	\caption{\label{fig:policies} qr-diagrams for some boarding policies, with each of the $N=240$ passengers marked as a point, $h=6$ seats per row, and congestion $k=1$. 
		(a) \emph{Random Boarding policy}: the passengers are uniformly distributed over the diagram; 
		(b) \emph{Back-to-Front policy} with two equal-sized groups: the first part of the queue is heading for the rows in the back of the airplane; 
		(c) \emph{Slow-First policy} with two equal-sized groups: the slow passengers are in the first part of the queue (blue points); 
		(d) \emph{Fast-First policy} with two equal-sized groups: the fast passengers are in the first part of the queue (red points).
		For all policies the boarding time is the sum of the aisle-clearing times for passengers that belong to the longest chain (dashed lines). 
		The preceding passenger in a chain must take his seat before the next in the chain can sit down. 
		In all four diagrams, the longest chain that determines the boarding time follows the asymptotic limit (the geodesic), up to statistical fluctuations that are diminishing as the number of passengers increases (see Sec.~\ref{sec:kgt0} for further details).
	}
\end{figure*}
Thus, in the Fast-First case, the first slow passenger arrive in the aisle late during the boarding process. Similarly, for Slow-First, only after a significant portion of the slow passengers are seated, the first fast passengers enter the airplane.

Comparing the graphs of Fast-First and Slow-First in Fig.~\ref{fig:mainresultsa} more closely, it is clear that while the curves are quite similar during the fast and the slow regimes, the transitions between the regimes are different.  
The trajectory of the Fast-First policy has a distinct change in slope around time step 40. 
The corresponding transition in the Slow-First policy is much smoother. This indicates that a significant proportion of the fast passengers are able to take their seat simultaneously with the last slow passengers. This is not the case with the Fast-First policy, since only few slow passengers are able to take their seat during the relatively short time period that it takes the last few fast passengers to sit. This asymmetry explains why the Slow-First policy is superior.

For all four policies in Fig.~\ref{fig:mainresultsa}, the boarding time is determined by a \emph{longest chain}, similar to the ones shown in the qr-diagrams in Fig.~\ref{fig:policies}. The preceding passenger in a chain must take his seat before the next in the chain can sit down. The boarding time is the sum of the aisle-clearing times for passengers that belong to the longest chain. The longest chain follows the asymptotic longest curve (the geodesic), up to statistical fluctuations that are diminishing as the number of passengers increases ($N\rightarrow\infty$).
For Slow-First and Fast-First in Fig.~\ref{fig:policies}, the aisle-clearing time is twice as long for passengers in the slow group than for those in the fast group ($C=0.5$). Notice that the aisle-clearing time acts as a refraction index as the longest curve breaks on the border between the two groups.

\begin{figure*}[htb]
	\begin{center}
		\mbox{\makebox[10.5cm][c]{\bf Boarding time relative difference --- Fast-First vs. Slow-First}}\\[0ex]
		\includegraphics[width=7.85cm,clip]{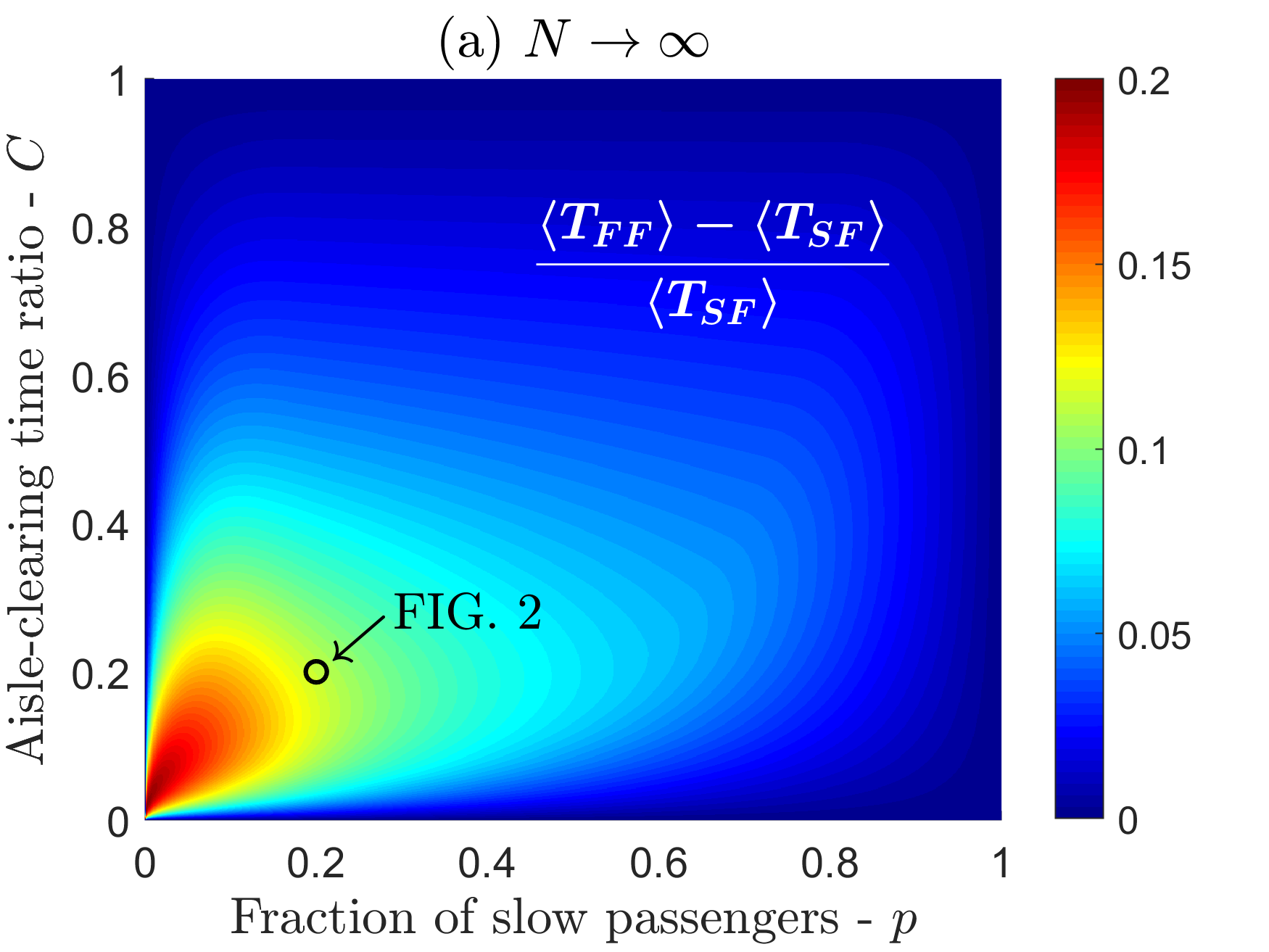}
		\includegraphics[width=8cm,clip]{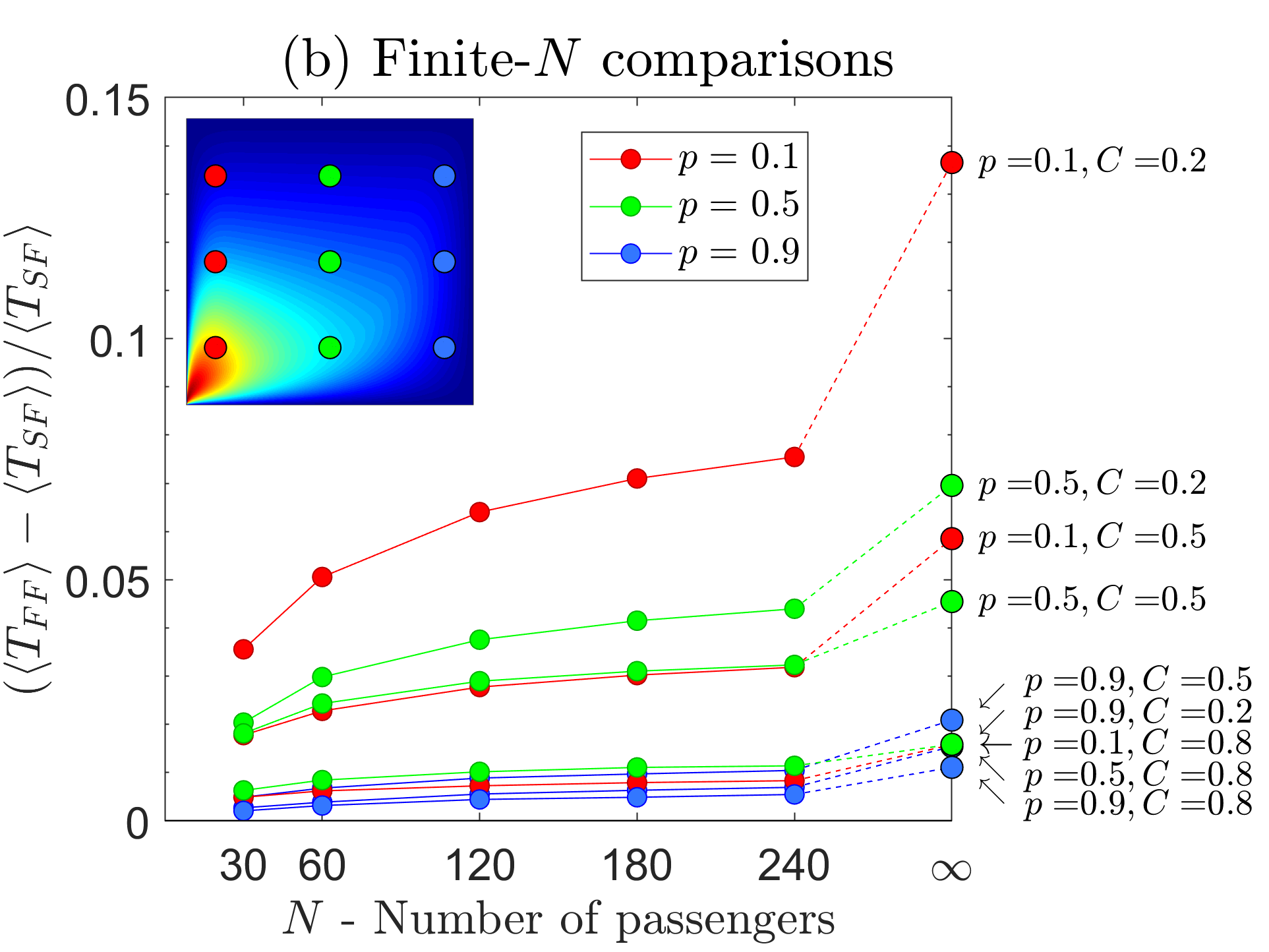}\\
	\end{center}
	\caption{\label{fig:mainresultsb} (a) The relative difference in average boarding time $D= (\angles{T_{FF}}-\angles{T_{SF}})/\angles{ T_{SF}}$ between the Fast-First and the Slow-First policies when $k=4$ and the number of passengers $N\rightarrow \infty$. The Slow-First policy is superior for all $(p,C)-$combinations, and the maximum relative difference is $20\%$ for small $p$ and $C$. For the parameter choice in Fig.~\ref{fig:mainresultsa}, the relative difference is $11\%$ (black circle).
	(b) Simulation results for finite number of passenger, $N$, confirm that Slow-First is superior to Fast-First, i.e.,  $D=(\langle T_{FF} \rangle -\langle T_{SF} \rangle)/{\langle T_{SF} \rangle}>0$ for increasing number of passengers for all combinations of parameter values $p\in\{0.1,0.5,0.9\}$ and $C\in\{0.2,0.5,0.8\}$. There are 6 seats per row, $k=4$, and the accuracy is $\pm 0.0002$ (as a result of $10^6$ scenarios  for each finite-$N$ data point). The rightmost points are asymptotic values taken from the indicated positions in the inset contour plot from (a).
}
\end{figure*}

Fig.~\ref{fig:mainresultsa} reports boarding times for one specific choice of the parameters $k,p,C$. Under variations in these parameters, the comparison can conveniently be made through the relative difference 
\begin{equation*}
D(k,p,C,N)=\frac{\angles{T_{FF}}-\angles{T_{SF}}}{\angles{T_{SF}}}
\end{equation*}    
between the average boarding times of the Fast-First and Slow-First policies.
The contour plot in \crefformat{figure}{Fig.~#2#1{(a)}#3}\cref{fig:mainresultsb} 
shows $D$ in the $(p,C)$-unit square for $k=4$ in the asymptotic case when $N\rightarrow\infty$. It is obtained by the spacetime geometry approach. The average boarding time is larger for Fast-First than for Slow-First for all values of $p$ and $C$. For $(k,p,C)=(4,0.2,0.2)$, the relative difference is $D=11\%$ for $N\rightarrow\infty$, compared to $D=7\%$ for $N=240$ in Fig.~\ref{fig:mainresultsa}. The maximum of $D=20\%$ (for $k=4$) is obtained for $(p,C)$ very small. 

Our main result can be stated as follows.
\begin{result}
	The expected boarding time $\angles{T}$ is shorter for the Slow-First policy than for the Fast-First policy for all values of $k>0$ and $p,C \in (0,1)$, in the asymptotic regime when $N\rightarrow\infty$. The maximum relative difference between the policies when $N\rightarrow\infty$, is at least $D=28.4\%$.
\end{result}
This result is proved analytically in Appendix \ref{app:proofs}. The maximum relative difference is $D=28.4\%$ for $k= 1.594$, $C= 0.513\sqrt{p}$ and $p$ small.
Even when the fraction of slow passengers is fixed to the more realistic value $p=0.1$, the maximum relative difference is $D=24.4\%$ (for $k=1.54$, $C=0.16$).

In \crefformat{figure}{Fig.~#2#1{(b)}#3} \cref{fig:mainresultsb} the relative difference in boarding times for $k=4$ and finite $N$ are compared with the asymptotic results when $N\rightarrow \infty$. All simulation results show $D\geq 0$. 
The relative difference is larger in the $N\rightarrow\infty$ limit compared with finite-$N$ cases, but still the relative ranking between the results for different parameter settings is to a large degree preserved for smaller values of $N$.

\section{The Boarding Process and spacetime geometry}\label{sec:model}

In this section we explain further the analogy between airplane boarding and spacetime geometry. The reader is referred to Ref. \cite{Bachmat:2014} for a more rigorous mathematical description. 
	
During boarding, a given passenger may be blocked from reaching his/her designated row by another passenger, which in turn may be blocked by others. This blocking hierarchy can be visualized in the qr-diagrams through what is known as blocking chains. Importantly, \emph{the longest blocking chain determines the boarding time}. 
	
A condition for blocking to occur is formulated in \cref{eq:blocking_def} (Sec.~\ref{ssec:blockingchains}). 
The condition is extended to the continuous case when $N\rightarrow\infty$, and then the passengers correspond to events in spacetime geometry. In this setting, the blocking condition also determines --- up to a proportionality constant --- the appropriate Lorentzian metric that should be used to calculate the distance (proper time) along a trajectory between two events (Sec.~\ref{ssec:curvelength}). Finally, the boarding time can be found by computing the longest blocking chain (Sec.~\ref{ssec:lorentzianmetric}) which, in the limit $N\rightarrow\infty$ tends to the length of the geodesic line.

\subsection{Main parameters}\label{ssec:mainparameters}
The boarding process is governed by the following parameters; 

(i) \emph{$N=$ The total number of passengers}. For simplicity, we assume that there are no empty seats in the airplane, i.e., the airplane is full. Hence, the total number of passengers equals the number of seats in the airplane. In Fig.~\ref{fig:boarding_illustration}, $N=8$.

(ii) \emph{$k=$ Congestion:} the length of the queue before boarding $(t=0)$ relative to the length of the aisle. 
Let $h$ be the number of seats per row, $w$ the distance between passengers needed for each to stand comfortably, one after the other along the aisle, and $d$ the distance between consecutive rows. Then $k=hw/d$. The parameter $k$ reflects the interior design of the airplane and the maximum density of passengers queuing along the aisle.\footnote{The parameter $k$ can also be modified to include the number of aisles and the relative occupancy of the airplane \cite{Bachmat/Berend/Sapir/Skiena/Stolyarov:2009}.} In Fig.~\ref{fig:boarding_illustration}, $k=h\cdot w/d=2\cdot 1/2=1$.

(iii) \emph{$p=$ Fraction of slow passengers.} In Fig.~\ref{fig:boarding_illustration}, all passengers have equal aisle-clearing time, so $p=0$. 

(iv) \emph{$\tau=$ Aisle-clearing time:} the time needed for a passenger to organize bin luggage and take a seat. In Fig.~\ref{fig:boarding_illustration} all passengers have an aisle-clearing time of $\tau=1$ time steps.

(v) \emph{$q=$ Queue position of a passenger} normalized by the total number of passengers $N$. In Fig.~\ref{fig:boarding_illustration}, the fourth passenger in the queue outside the airplane (aiming for row 3) has $q=4/8=0.5$.

(vi) \emph{$r=$ Designated row number for a passenger} normalized by the total number of rows. The fourth passenger in the queue in Fig.~\ref{fig:boarding_illustration} has $r=3/4=0.75$. 

In Fig.~\ref{fig:boarding_illustration} the actual queue and row numbers are given on the axes, while in the remaining part of the paper the normalized $(q,r)$-values will be used.

\subsection{Boarding policies --- visualized}\label{ssec:boardingpolicies}
A boarding policy is the way the queue of passengers is organized. The most common policy is the unorganized Random Boarding policy, where the passengers enter the queue in random order. A typical scenario with the Random Boarding policy is illustrated in the qr-diagram in \crefformat{figure}{Fig.~#2#1{(a)}#3}\cref{fig:policies}. The points representing each of the $N=240$ passengers are uniformly distributed over the unit square.\footnote{This uniformity applies in general for a coarse-grained description when $N$ is sufficiently large. On the microscopic level, however, the point cloud has a structure since the $q$- and $r$-directions are not equivalent: for a given $q$-value one has only one point in the diagram, while for a given $r$-value there are as many as there are seats in a row.} 

A scenario with the Back-to-Front policy is shown in \crefformat{figure}{Fig.~#2#1{(b)}#3}\cref{fig:policies}. The passengers are divided into two groups, where those who have designated seats in the back of the airplane constitutes the first part of the queue.  Within each group, the passengers are randomly distributed in the queue. 
In \crefformat{figure}{Fig.~#2#1{(c)}#3}\cref{fig:policies} a scenario with the Slow-First policy is shown. The diagram resembles the one of the Random Boarding policy, but passengers assumed to use long time to take a seat, are placed in a separate group in the first part of the queue. The designated row numbers are randomly distributed within both groups as in Random Boarding. The Fast-First policy in \crefformat{figure}{Fig.~#2#1{(d)}#3}\cref{fig:policies} has a diagram similar to that of the Slow-First policy, except that the red and the blue regions are exchanged.

\subsection{Blocking chains and blocking relation}\label{ssec:blockingchains}
\begin{figure*}[htb]
	\begin{center}
		\includegraphics[width=7.5cm,clip]{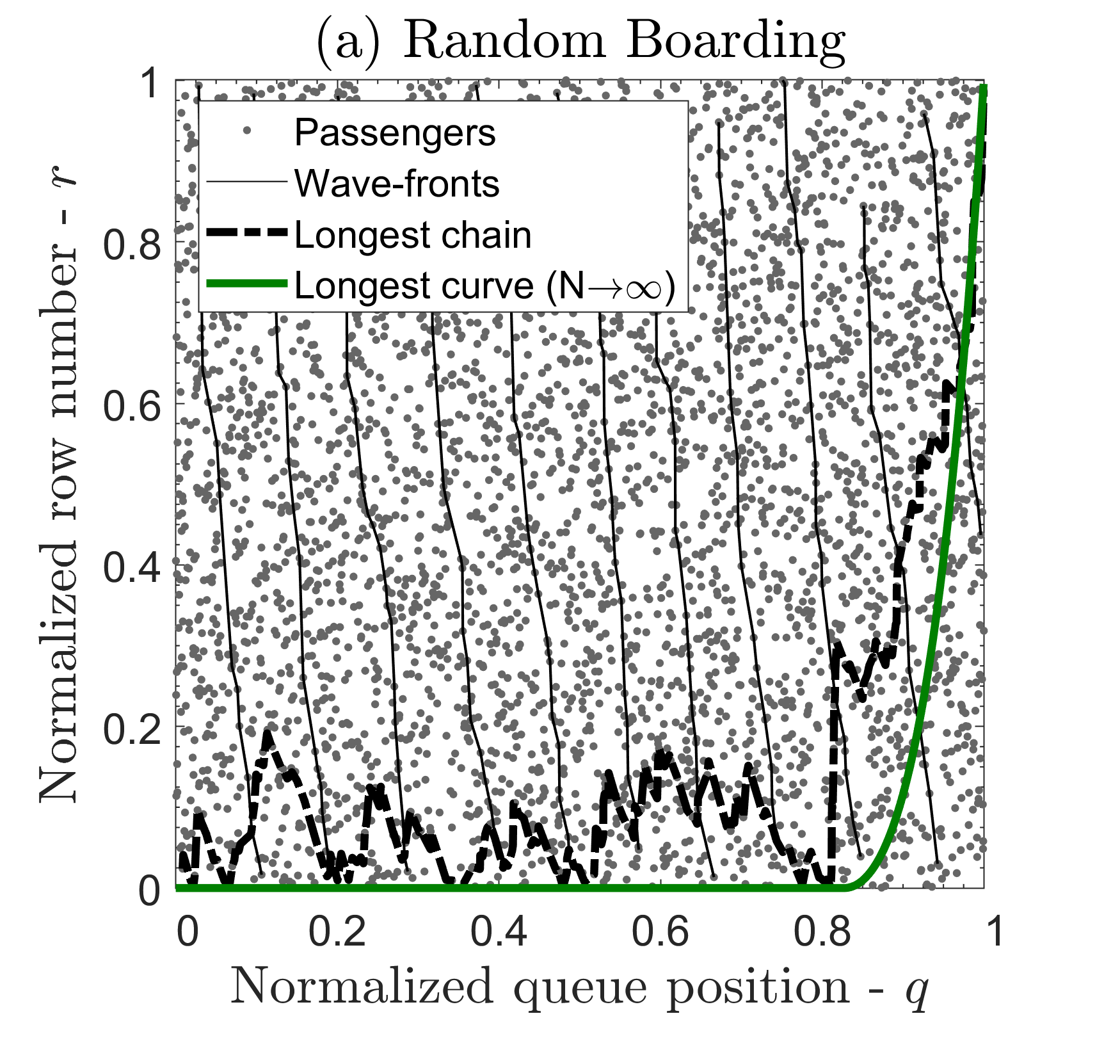}
		\includegraphics[width=7.5cm,clip]{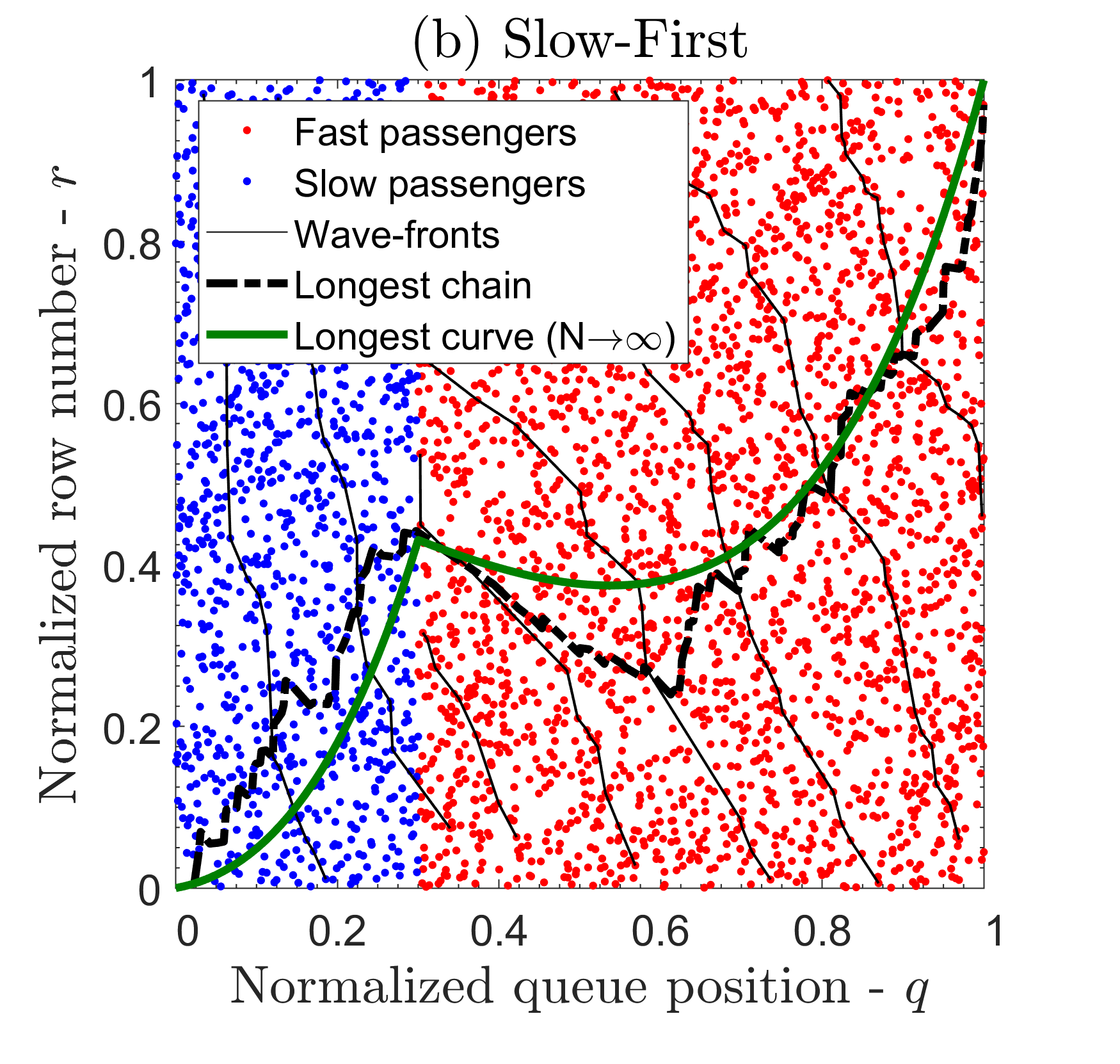}\\
	\end{center}
	\caption{\label{fig:maximal_chains} With $N=4000$ passengers, there is closer correspondence between the longest chain and the longest curve than in Fig.~\ref{fig:policies} (where $N=240$). Passengers (points) seat simultaneously in consecutive wave-fronts (black thin lines). For clarity, only every 20th wave-front is shown. 
	(a) \emph{Random Boarding policy:} $k=4$, with same aisle-clearing time for all passengers.  
	(b) \emph{Slow-First policy:} $(k,p,C)=(1.5,0.3,0.33)$. The longest curve breaks when the aisle-clearing time changes value on the border between the slow and the fast passengers. The aisle-clearing time plays the role of a refractive index.} 
\end{figure*}

As shown in Fig.~\ref{fig:boarding_illustration}, the boarding process can be thought of as wave-fronts of passengers that take their seats simultaneously. When all passengers that stand next to their designated row have taken their seat, the remaining passengers in the aisle move rapidly forward, and a new wave-front of passengers sits down. Hence, the boarding time is the product of the aisle-clearing time times the number of wave-fronts needed to seat all passengers. 
Wave-fronts are shown in Fig.~\ref{fig:boarding_illustration}, with $N=8$ passengers and congestion parameter $k=1$. 
When $k=4$, as in the Random Boarding case in \crefformat{figure}{Fig.~#2#1{(a)}#3}\cref{fig:maximal_chains}, 
the wave-fronts are steeper, and a single wave-front spans less of the $q$-axis since there is no room for more than a quarter of the initial queue in the aisle.

The direct approach of finding all the wave-fronts and counting them to determine the boarding time is impractical when the number of passengers $N$ is large. Furthermore, \emph{average} boarding time --- obtained when averaging over all possible queue configurations --- is even harder, and closed-form analytical results for finite-$N$ are not known. In the following we will describe an indirect way to calculate the number of wave-fronts (the boarding time) by the introduction of blocking chains, which later will be shown to correspond to causal chains in spacetime geometry. These will be essential in order to establish the asymptotic boarding time as $N\rightarrow\infty$.

We say that passenger $A$ \emph{blocks} passenger $B$ if $A$ must be seated before $B$ can sit down. Two passengers in the same wave-front cannot block each other, and $A$ cannot block $B$ if $B$ is in front of $A$ in the queue. A \emph{blocking chain} consists of passengers that consecutively block each other. 
The \emph{length} of the chain is the sum of aisle-clearing times for passengers that belong to the chain, and the length of the \emph{longest chain} of all blocking chains equals the boarding time.
Given a queue, we can construct a longest chain by starting with one of the passengers in the last wave-front. Several passengers in the preceding wave-front may be blocking this passenger. The one that is closest in the queue is chosen as the next passenger in the chain. The longest chain is obtained by proceeding like this, until reaching a passenger in the first wave. Examples of longest chains are shown in Figs.~\ref{fig:policies} and \ref{fig:maximal_chains}. Notice that the longest chains are approaching an asymptotic longest curve when $N$ increases (see details in the following sections).

\begin{figure*}[htb]
	\begin{center}
		\includegraphics[height=5.3cm,clip]{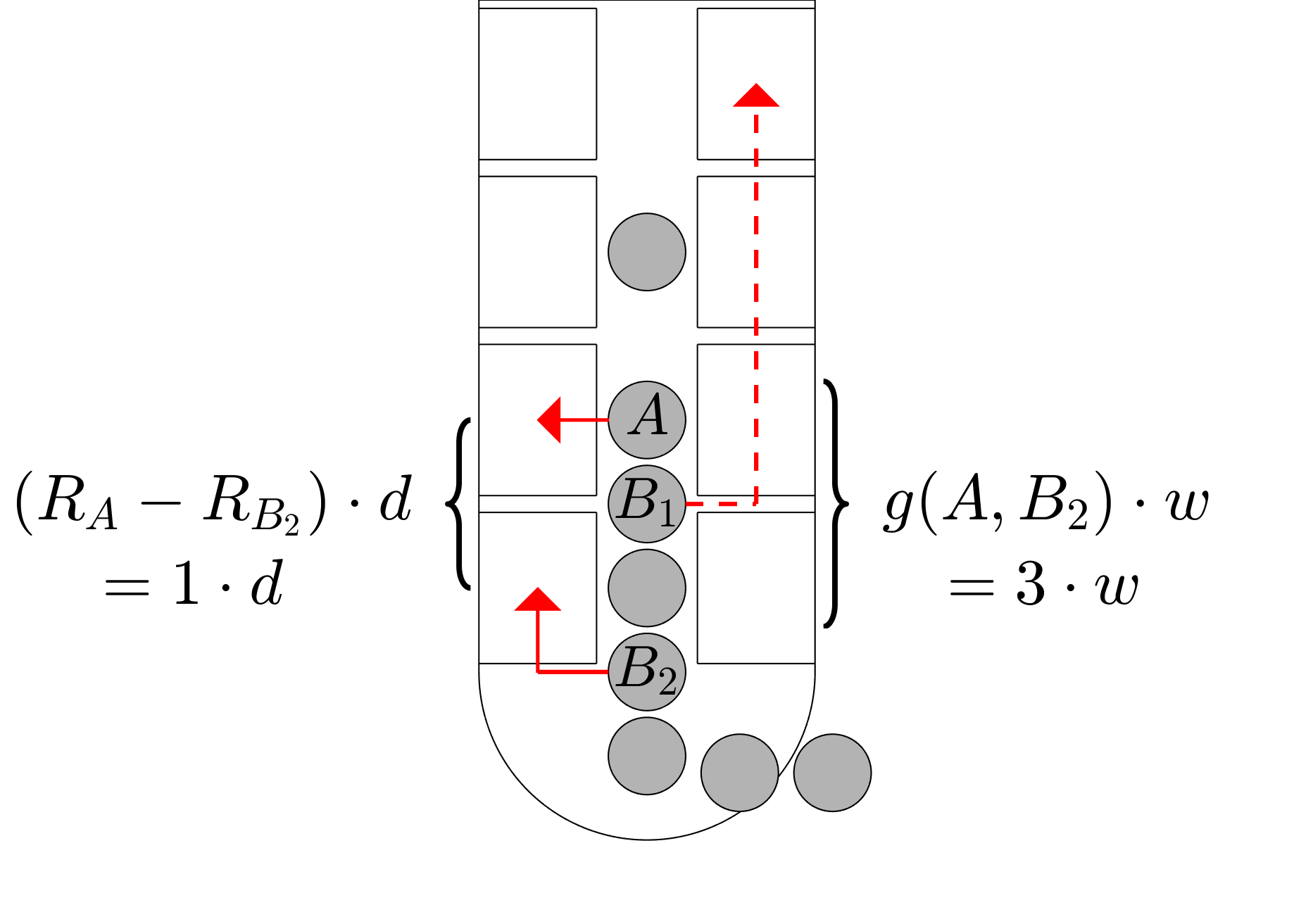}
		\includegraphics[width=6.0cm,clip]{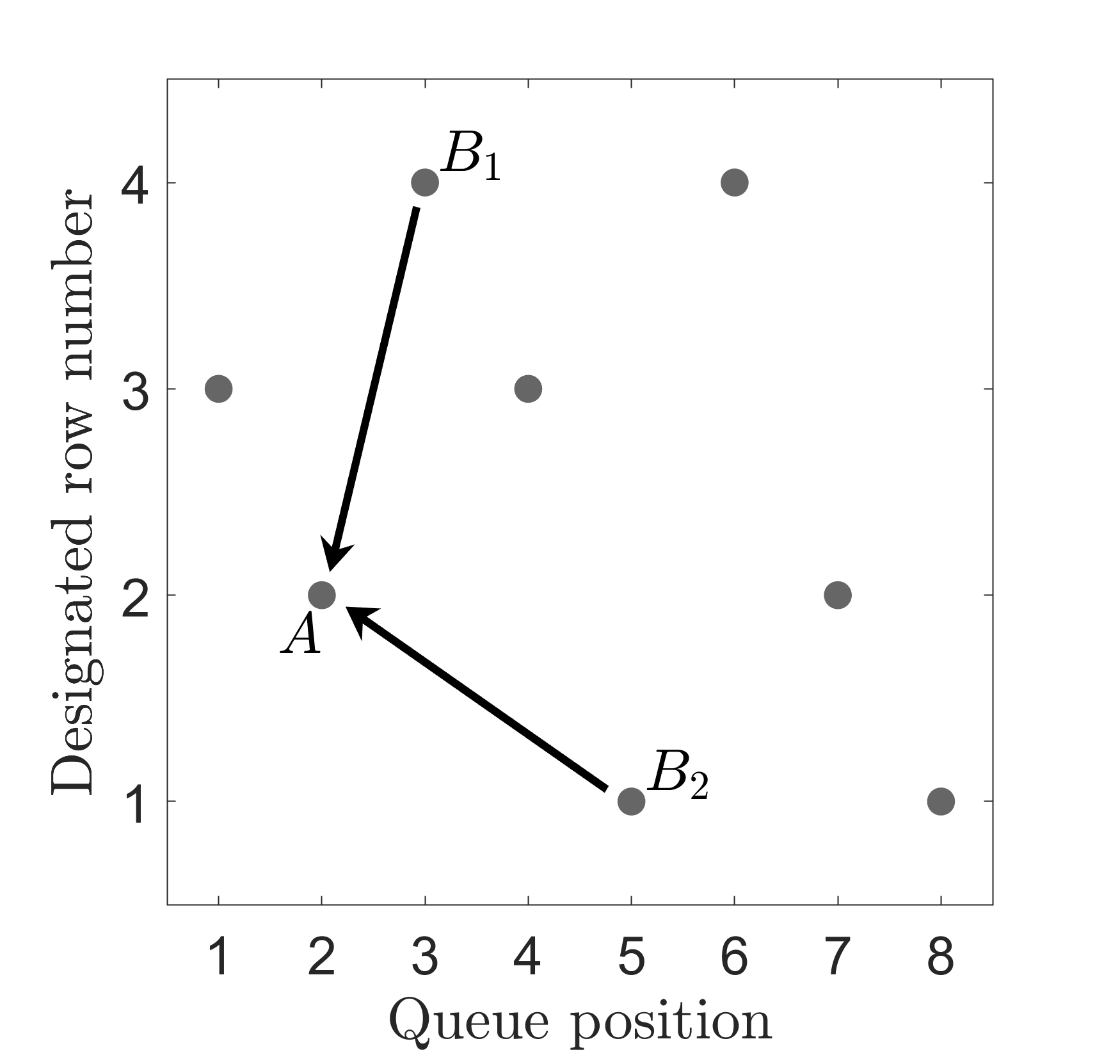}\\
		\mbox{\makebox[7cm][c]{(a)}\makebox[6cm][c]{(b)}}
	\end{center}
	\caption{\label{fig:blocking_relation} Example of blocking relations for the same case as in Fig.~\ref{fig:boarding_illustration}. The distance $d$ between consecutive rows is twice the distance $w$ occupied by each passenger in the aisle. Passenger $A$ blocks both passengers $B_1$ and $B_2$. (a) Passenger $B_1$ is blocked by $A$ since she is heading for row $4$, which is beyond row 2 where $A$ is taking a seat. Passenger $B_2$, is blocked by $A$ by displacement as in \cref{eq:blocking_def}: the space $3w$ occupied in the aisle by passenger $A$ and the $2$ passengers between $A$ and $B_2$ is larger than the distance $d=2w$ between the designated rows of $A$ and $B_2$. (b) The blocking relations are indicated by arrows. There are several other blocking relations that are not shown in the figure. 
	}
\end{figure*}

The blocking chain can be defined in terms of a \emph{blocking relation}. Let passenger $A$ be in front of passengers $B_1$ and $B_2$ in the queue, as shown in Fig.~\ref{fig:blocking_relation}. Passenger $B_1$ is heading for a row further back in the airplane, so $A$ is obviously blocking her. Passenger $B_2$ is heading for row 1, which is in front of $A$'s row. However, due to the displacement caused by the two passengers in between $A$ and $B_2$, $A$ blocks $B_2$ from reaching his seat.\footnote{For $k=0$, blocking through displacement never occurs.}

More generally, let $g(A,B)$ be the number of passengers (including passenger $A$) standing in between passengers $A$ and $B$, just before $A$ sits down. Passengers $A$ is in front of $B$, and they are heading for rows $R_A$ and $R_B$, respectively. We say passenger $A$ \emph{blocks} passenger $B$ if the distance between the designated rows of $A$ and $B$ is less than the space (displacement) in the aisle occupied by the passengers between $A$ and $B$:
\begin{equation}\label{eq:blocking_def}
(R_A-R_B) \cdot d < g(A,B) \cdot w.
\end{equation} 
Here, $d$ is the distance between each row, and $w$ is the space (length) occupied by each passenger along the aisle. 

In Fig.~\ref{fig:blocking_relation}, where $d=2w$, both $B_1$ and $B_2$ are blocked by $A$ according to this definition. For $B_1$ the row distance is $(R_A-R_{B_1})\cdot d=-2\cdot d=-4w$ and the displacement is $g(A,B_1)\cdot w = w$. For $B_2$ the respective values are $(R_A-R_{B_2})\cdot d=1\cdot d=2w$ and $g(A,B_2)\cdot w=3w$. This means that $A$ blocks both $B_1$ and $B_2$, since the inequality in \cref{eq:blocking_def} is satisfied for both. In fact, most of the other passengers are blocked by $A$ in this example.

We call the inequality in \cref{eq:blocking_def} a blocking relation, and it turns out that the relation also determines how the number of passengers in a blocking chain can be computed when $N\rightarrow\infty$.

\subsection{Blocking relation and curve length in spacetime geometry ($N\rightarrow\infty$)}\label{ssec:curvelength}
In relativity theory, events are given by space and time coordinates, and the mathematical description is continuous. Events can be classified through the past-future (or causality) relation. Event $A$ is in the past of an event $B$ if it is possible to move from $A$ to $B$ under or at the speed of light. In other words, a future event $B$ can only be affected by an event $A$ if $B$ is within the future light-cone of $A$. 

Proper time is the time passing on a clock attached to a particle passing through a (continuous) set of events.
In the \emph{causal set} approach to gravity, spacetime is discrete and composed of a finite number of events, each contributing 1 time unit to proper time.  In this approach, continuous spacetime emerges as the limit of discrete spacetime as the number of events increases. 

For airplane boarding, the passengers play the role of the events. Passengers have a natural causal (past/future) structure defined by the blocking relation in \cref{eq:blocking_def}. We consider a passenger $A$ as being in the past of passenger $B$ if passenger $B$ is blocked by $A$. 
In this setting, we may denote the blocking chain a \emph{causal chain}, where each passenger in the chain contributes its aisle-clearing time to the proper time of the chain. Thus, the proper time of the longest chain equals the boarding time.

The continuous analogue of the notion of a causal chain is a \emph{causal curve}, i.e., the possible trajectory of an object traveling below the speed of light.
In spacetime geometry the proper time of a particle's path (causal curve) is given by its length as found by integration using the spacetime interval $ds$. In its simplest form with one spatial dimension, it is given by $ds^2 = dt^2 - dx^2$.
The past-future relation, which ensures that future events stays within the future light cone of past events, is given by $ds^2 \geqslant 0$. 

\begin{figure*}[htb]
	\begin{center}
		\includegraphics[width=6.0cm,clip]{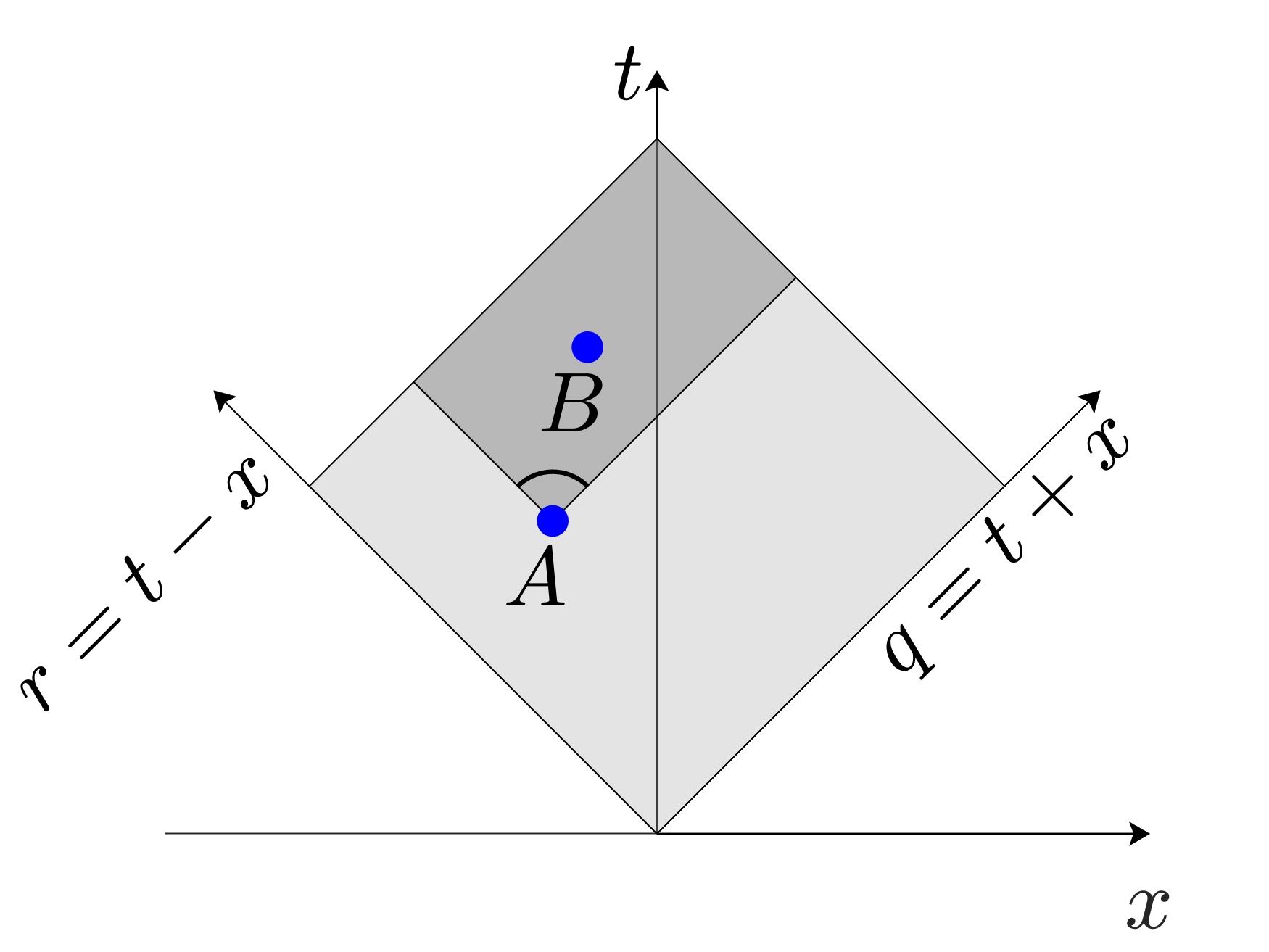}
		\parbox[b][3.6cm][c]{0.2cm}{$\quad$}
		\includegraphics[width=4.5cm,clip]{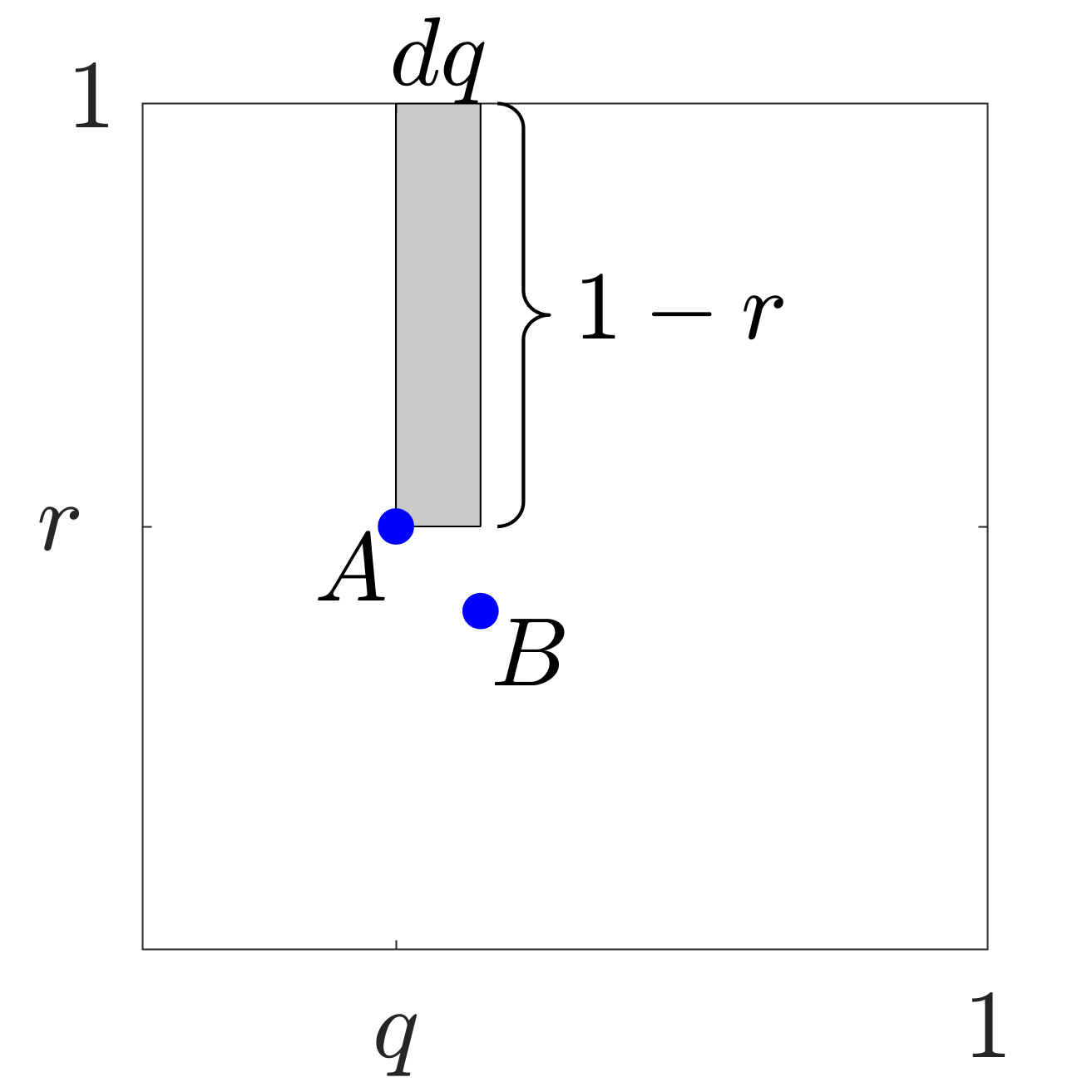}
		\parbox[b][3.6cm][c]{0.2cm}{$\quad$}
		\includegraphics[width=4.5cm,clip]{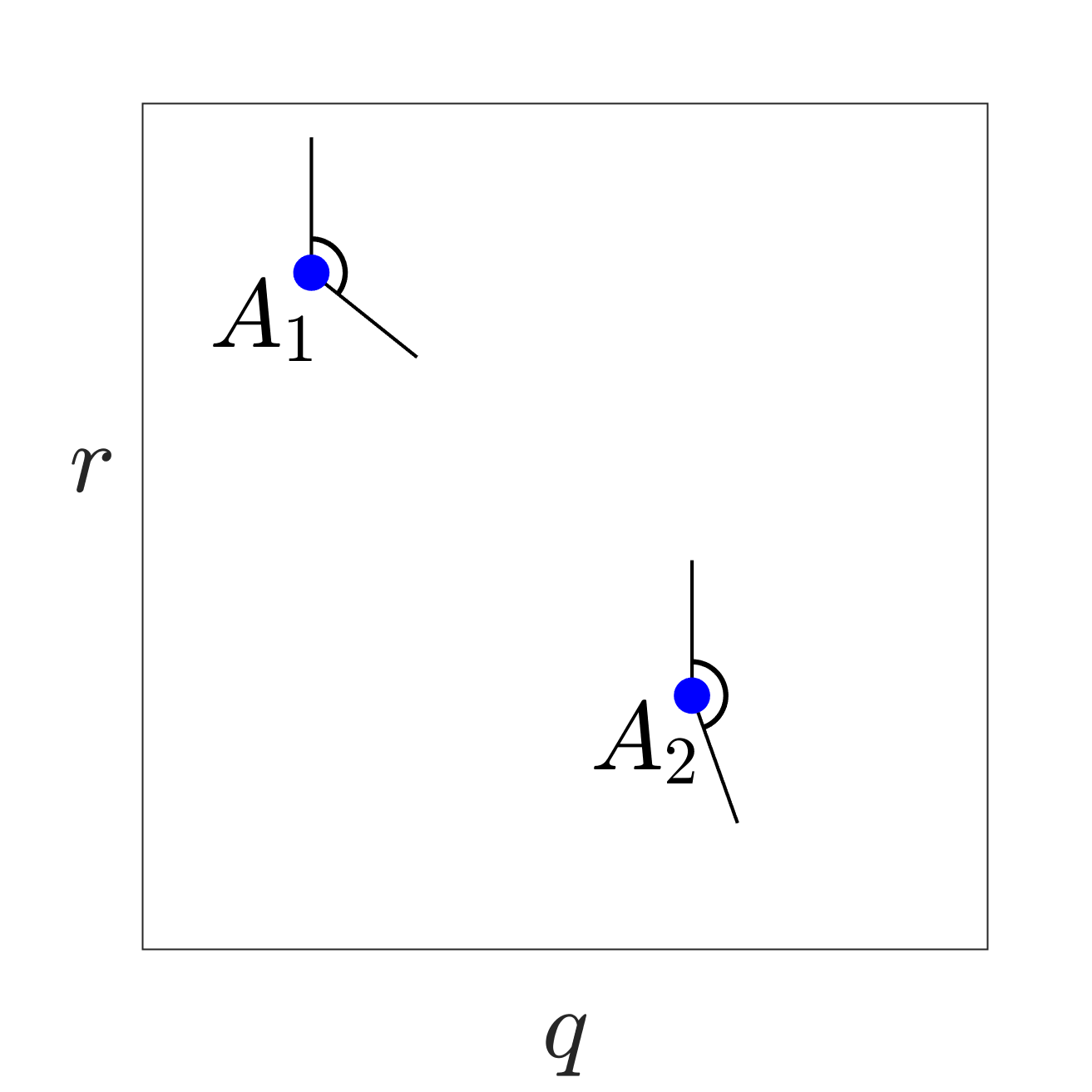}
		\mbox{\makebox[6.0cm][c]{(a)}\makebox[5.1cm][c]{(b)}\makebox[4.7cm][c]{(c)}}
	\end{center}
	\caption{\label{fig:geometric_spacetime} (a) Coordinate transformation from spacetime to queue-row when $k=0$. The qr-diagram is in the future light cone of the origin. Passenger $B$ can only be blocked by $A$ if she is within the "light-cone" of $A$. (b) Blocking by displacement for $k>0$ according to \cref{eq:blocking_def}. Since passenger $B$'s designated row is in front of $A$'s designated row, $A$ can only block $B$ by displacement. The value of $k$ and the number of passengers in between $A$ and $B$ in the queue determine the extent of the displacement. Most of the passengers in between are heading for the rows behind $A$'s designated row (shaded area). (c) Passengers nearby $A_1$ and $A_2$ must be within their respective future light-cones in order to be blocked. The cones of each blocking passenger have a wider angle when the designated row is near the front of the airplane, reflecting a larger potential for blocking fellow passengers (here $k=4$). Passengers can never block passengers who are in front of them in the queue, and therefore the line emanating upwards is always part of the light-cone boundary.}
\end{figure*}

Using the coordinate transformation $q=t+x$ and $r=t-x$, the queue-row diagram emerges from the future light cone of a spacetime diagram as shown in \crefformat{figure}{Fig.~#2#1{(a)}#3}\cref{fig:geometric_spacetime}. 
The spacetime interval now reads
\begin{equation*}
ds^2 = dt^2 - dx^2 = (dt+dx)(dt-dx) = dqdr.
\end{equation*} 

If passengers had no width, ($w=0$), the congestion parameter would be $k=hw/d=0$, and the blocking relation in \cref{eq:blocking_def} reduces to $R_A < R_B$, when passenger $A$ is in front of $B$ in the queue. This means that passenger $A$ in \crefformat{figure}{Fig.~#2#1{(a)}#3}\cref{fig:geometric_spacetime} can only block passengers in the shaded, upper right-hand rectangle of point $A$. Notice the past-future relation $ds^2=dqdr \geqslant 0$ (i.e., $dr/dq\geqslant 0$) is equivalent to the blocking relation: line segments between events (passengers) must be non-decreasing in $q$.

Real passengers are not infinitely thin, and the congestion parameter $k=hw/d>0$. Assume that $N$ is large. Let passengers $A$ and $B$ be close in the queue, $A$ in front of $B$, separated by $dq>0$ (see \crefformat{figure}{Fig.~#2#1{(b)}#3}\cref{fig:geometric_spacetime}). The normalized difference in assigned row position is $dr=(R_B-R_A)/(N/h)$, where $h$ is the number of seats per row. Just before passenger $A$ sits down, the number of passengers between $A$ and $B$ in the queue is essentially those who are heading for the rows behind $R_A$. Since the passengers are uniformly distributed on the qr-diagram, this number is given by $g(A,B)\approx dq(1-r)N$, shown as the shaded area in \crefformat{figure}{Fig.~#2#1{(b)}#3}\cref{fig:geometric_spacetime}. The blocking relation in \cref{eq:blocking_def} can now be written
\begin{equation}\label{eq:blocking_def_infinitesimal}
\frac{h}{Nd}\left[(R_B-R_A)d + g(A,B)w\right]
\;\approx\; dr + k(1-r)dq \;>\; 0.
\end{equation}
Thus, a causal curve must satisfy $r'(q) > -k(1-r)$. The sectors of passengers blocked by $A_1$ and $A_2$ in \crefformat{figure}{Fig.~#2#1{(c)}#3}\cref{fig:geometric_spacetime} indicate that the sector angle increases for smaller $r$ --- the potential for blocking other passengers is larger when the designated row is in the front of the airplane. 

The proper time of a causal chain when each event in the chain contributes 1 time unit to the proper time and $N\rightarrow\infty$, corresponds to the proper time of a causal curve. There exists a Lorentzian metric $ds$, that can be used to compute the length of a causal curve (its proper time), and this metric is defined uniquely --- up to a constant scaling factor --- by the blocking relation. For airplane boarding and the blocking relation in \cref{eq:blocking_def_infinitesimal}, the metric\footnote{This is a simplified metric. The more general metric in \cite{Bachmat:2014}, which includes the density distribution of passengers, has, e.g., been used to analyze the Back-to-Front boarding policy.} is \cite{Bachmat:2014}: 
\begin{equation}\label{eq:Lorentzian_metric}
ds^2=dq(dr + k(1-r)dq). 
\end{equation}
The length of a curve $r(q)$ between two points $q_0$ and $q_1$ is then given by:
\begin{equation}\label{eq:length}
L(r) = \int_{q_0}^{q_1} \sqrt{r'(q) + k(1-r(q))} \, dq.
\end{equation}
This definition of length, together with an appropriate scaling factor, will in the following sections be used to calculate the number of passengers in a longest chain in airplane boarding in the large-$N$ limit.

\subsection{Longest curves under the Lorentzian metric}\label{ssec:lorentzianmetric}
According to general relativity theory, among all possible paths between two events, a free falling
particle (a particle only under the influence of gravity) will follow a trajectory that maximizes proper time (locally, between any two nearby points on the trajectory).
Such free fall trajectories are known as \emph{geodesics}. Longest chains, which determines the boarding time, correspond to \emph{longest curves} in the continuous version. The longest curves are geodesics when not constrained by boundary conditions.

\begin{figure*}[htb]
	\begin{center}
		\includegraphics[width=6.5cm,clip]{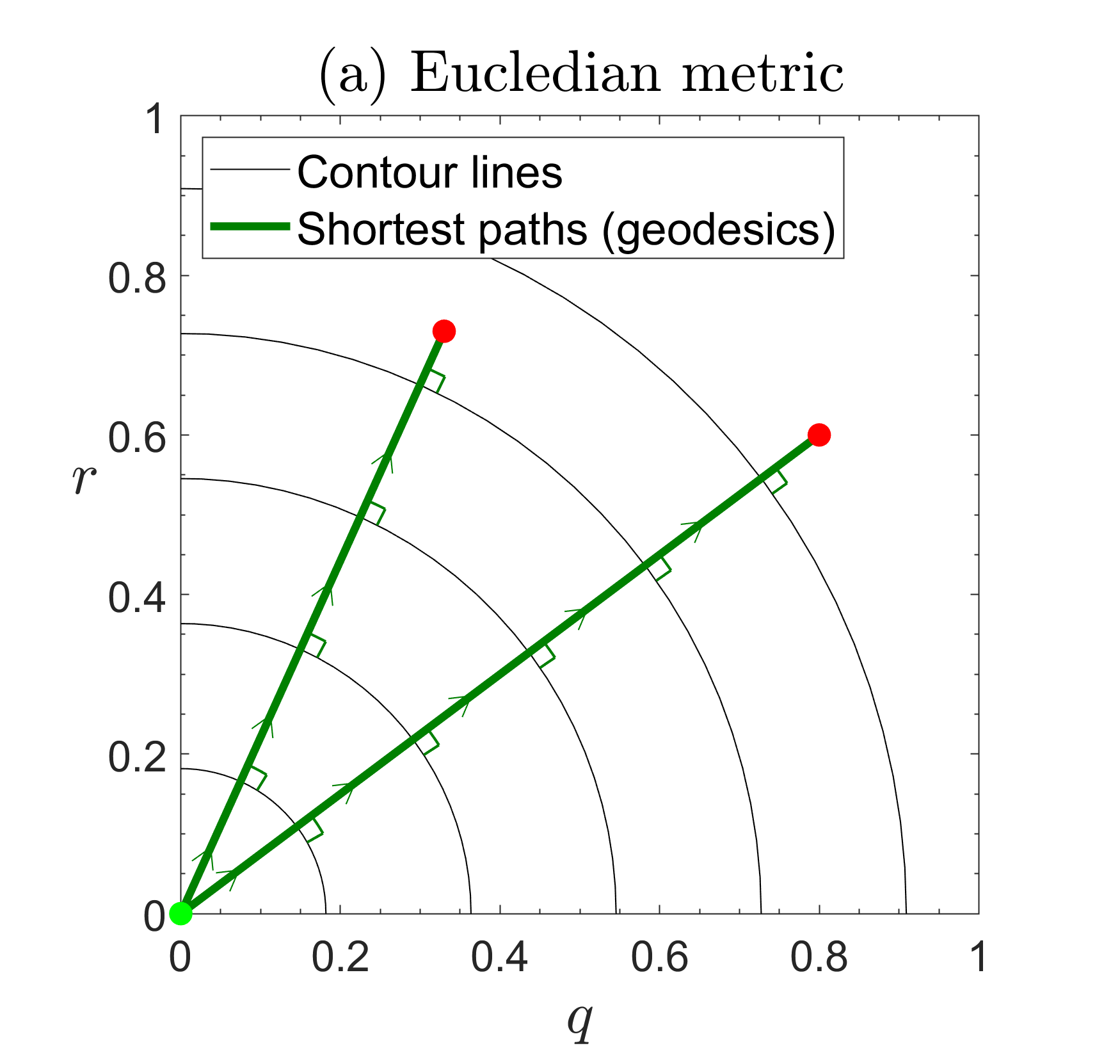}
		\includegraphics[width=6.5cm,clip]{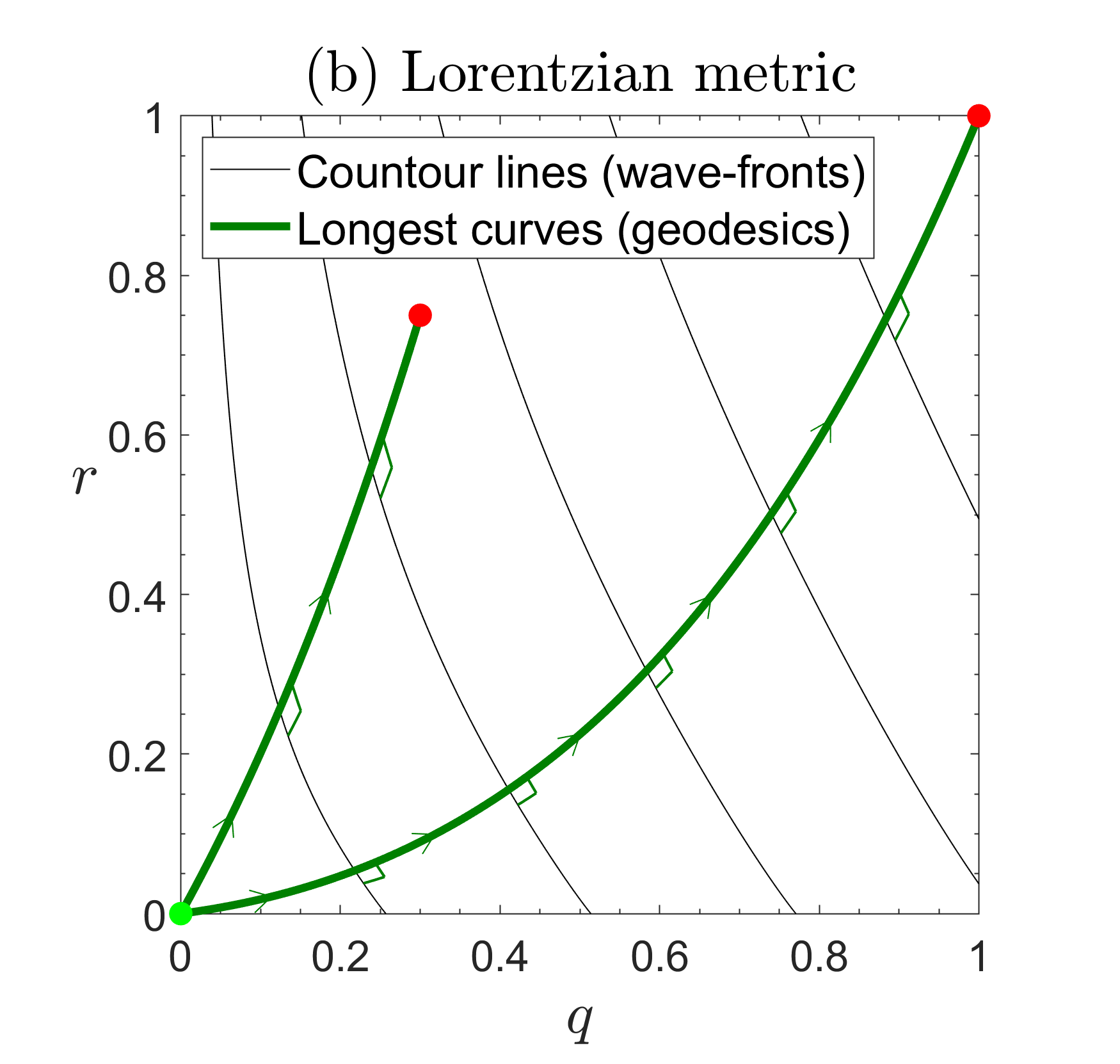}\\
	\end{center}
	\caption{\label{fig:steepestascent} The optimal path (green thick curves) depends on the metric and is orthogonal to the contour lines (black thin curves) which defines the equidistant points from the starting point (green bullet).
		(a) Under the Euclidean metric, the contour lines are circle-shaped. The shortest paths from the starting point to any other point (red bullets) are straight lines.  
		(b) In airplane boarding we use an appropriate Lorentzian metric, and the starting point is $(0,0)$ in the qr-diagram (here $k=0.6$). The contour lines coincide with passenger wave-fronts when $N\rightarrow \infty$. The longest curves from the starting point to any other point are everywhere orthogonal to the wave-fronts.
	}
\end{figure*}

We first look at the Euclidean metric, where the length of a curve is $\int \sqrt{r'(q)^2 + 1}\, dq$. The equidistant points relative to a starting point in $(0,0)$ are circles, as shown in \crefformat{figure}{Fig.~#2#1{(a)}#3}\cref{fig:steepestascent}. 
The shortest curve from the starting point to any other point are straight lines that are orthogonal to the circles. This can be compared to a ball rolling down a hill where the contour lines indicate the height. The shortest path is always in the steepest direction, orthogonal to the contour lines, and the ball will take the same time to reach any point on a chosen contour line.

Under the Lorentzian metric in \cref{eq:length}, the contour lines in \crefformat{figure}{Fig.~#2#1{(b)}#3}\cref{fig:steepestascent} where all points are equidistant to the starting point in $(0,0)$, are not circle-shaped. The contour lines are asymptotically equal to the wave-fronts in airplane boarding when $N\rightarrow\infty$. The path which is orthogonal to the contour lines is the \emph{longest} curve (geodesic) under the Lorentzian metric, and it will take the same amount of time to reach any point on a contour line, i.e., all passengers on this line (wave) sit down simultaneously.

The start point of the longest chain correspond to the first passenger to sit during the boarding process. Likewise, the end point of the longest chain corresponds to the last passenger seated. 
If the first passenger in the queue is seated at the first row  ($q=0, r=0$), he can block all other passengers and will be in the first wavefront. And if the last passenger in the queue is seated at the last row ($q=1, r=1$), he is blocked by all the other passengers and must be in the last wavefront. For large $N$ and uniform distribution there will be passengers with $(q,r)$-coordinates arbitrarily close to those points. Hence, the curve that approximates a longest chain when $N$ is large, should be the longest continuous path under the Lorentzian metric between $(0,0)$ and $(1,1)$ within the $(q,r)$-unit square. Examples of such longest paths in airplane boarding are shown in Figs.~\ref{fig:policies} and \ref{fig:maximal_chains}, and their expressions are given in Appendix \ref{app:computations}.

\section{Asymptotic boarding time}\label{sec:analysis}

\subsection{Asymptotic boarding time with one group}
A cornerstone result in the causal set approach is a limiting result by Myrheim \cite{myrheim:1978} which links the number of elements in a longest chain with the length of the longest curve, up to a scaling factor depending only on the dimension of the domain. The scaling factor was found independently by Vershik and Kerov \cite{Vershik/Kerov:1977} and Logan and Shepp \cite{Logan/Shepp:1977} for dimension two --- the same dimension as in airplane boarding. The $\sqrt{N}$-law follows from simple sub-additivity arguments.
	
Let $\tau$ be the common aisle-clearing time for each of the passengers, $N$ the number of passengers, and $\max_{r}L(r)$ the length in \cref{eq:length} of the longest causal curve $r(q)$ between $(0,0)$ and $(1,1)$ within the $(q,r)$-unit square. For large $N$, {the longest chain follows closely the trajectory of the longest causal curve (see Fig.~\ref{fig:maximal_chains})}, and both satisfy the blocking relation in \cref{eq:blocking_def_infinitesimal}. A generalization of a result of Deuschel and Zeitouni \cite{Deuschel/Zeitouni:1995} states that the boarding time converges to a multiple of the length of the longest causal curve \cite{Bachmat:2014},
\begin{equation*}
	\frac{T}{\sqrt{N}} \;\overset{\textrm{a.s.}}{\to}\; 2\tau\cdot\max_{r}L(r).
\end{equation*}

From this, the asymptotic average boarding time is given by:
\begin{equation}\label{eq:Tapprox_length}
\angles{T} \;\sim\; 2\tau\sqrt{N}\cdot\max_{r}L(r)  \;\equiv\; \hat{T}.
\end{equation}
The asymptotic boarding time $\hat{T}$ is a leading term, and has been shown to over-estimate the finite-$N$ average boarding time $\langle T\rangle$ by a relative error of order $o(N^{-\frac{1}{4}})$ \cite{Bachmat/Khachaturov/Kuperman:2013}. Still, the relative ranking of boarding policies has shown to be maintained for small $N$, as demonstrated by discrete-event simulations in \crefformat{figure}{Fig.~#2#1{(b)}#3} \cref{fig:mainresultsb} and Ref. \cite{Bachmat/Berend/Sapir/Skiena/Stolyarov:2006}.

The procedure for computing the length of the longest curves is presented in Sec.~\ref{sec:kgt0}. For the Random Boarding examples in Figs.~\ref{fig:boarding_illustration} and \crefformat{figure}{#2#1{(a)}#3}\cref{fig:maximal_chains}, we set $\tau=1$. 
In \crefformat{figure}{Fig.~#2#1{(a)}#3}\cref{fig:maximal_chains} $k=4$, $N=4000$ and $r(q)\equiv 0$ for $q<0.83$ and $r(q)=4(e^{-2k(1-q)}-e^{-k(1-q)}) + 1$ for $q\geqslant 0.83$. The length of this curve is $2.153$, and from \cref{eq:Tapprox_length}, $\hat{T}=272$. 
In Fig.~\ref{fig:boarding_illustration}, $k=1$ and $N=8$, and a corresponding curve gives $\hat{T}=7.4$. The actual boarding times are $232$ and $3$ for the two examples, respectively. 
This illustrates that the asymptotic estimate $\hat{T}$ in \cref{eq:Tapprox_length} can be inaccurate for small $N$, but improves as $N$ increases.

\subsection{Asymptotic boarding time for two groups with different aisle-clearing times}\label{ssec:boardingtime_2groups}
In this paper we consider policies where all the slow (or fast) passengers are placed in the first part of the queue. Hence, the aisle-clearing time is different for the two different groups of passengers. Let $p$ be the fraction of slow passengers. $\tau_S,\tau_F$ are the aisle-clearing times for slow and fast groups, respectively.

The asymptotic average boarding time in \cref{eq:Tapprox_length} must be modified to reflect the fact that the queue now consists of two separate groups with different aisle-clearing times.  
The aisle-clearing time can be parameterized according to the queue position,  $\tau=\tau(q)$. E.g. for the Slow-First policy, $\tau(q)=\tau_S$ for $q\leqslant p$ and $\tau(q)=\tau_F$ for $q>p$.

The aisle-clearing time can be thought of as the proper time (Lorentzian metric length) between two successive passengers (events) in a chain. The boarding time is no longer given by the maximal length of a causal chain, but rather by the causal chain with maximal weight. The definition of length in \cref{eq:length} must be scaled to reflect that the aisle-clearing time of each passenger depends on the queue position. Following \cite{Bachmat:2014}, the \emph{curve weight} (proper time) of a causal curve is defined by
\begin{equation}\label{eq:weight}
W(r) = \int_{q_0}^{q_1} \tau(q) \sqrt{r'(q) + k(1-r(q))}dq,
\end{equation}
where $\tau(q)$ also can be considered a weight function applied to the curve in \cref{eq:length}.

When there are two groups, as in, e.g., the Slow-First policy, the curve weight on the interval $q\in (0,1)$ is given by
\begin{align}
W_{SF}(r) &= \int_{0}^{p} \tau_S \sqrt{r'(q) + k(1-r(q))}dq + \int_{p}^{1} \tau_F \sqrt{r'(q) + k(1-r(q))}dq \nonumber\\[3ex]
&= \tau_SL_S(r) + \tau_FL_F(r) \label{eq:weight_SF}
\end{align}
where $L_S,L_F$ are curve lengths as defined in \cref{eq:length}. 

The boarding time is (for $N \gg 1$) proportional to the {longest} (i.e., heaviest) curve $r(q)$ from $(0,0)$ to $(1,1)$ within the unit square:
\begin{equation}\label{eq:Tapprox}
\angles{T} \;\sim\; 2\sqrt{N}\cdot\max_{r}W(r)  \;\equiv\; \tilde{T}.
\end{equation}
An additional constraint on the curve is that $r'(q)$ must be continuous whenever $\tau(q)$ is continuous. 
The result in \cref{eq:Tapprox} is used in Sec.~\ref{sec:kgt0} to derive analytical expressions for the expected boarding time for both the Slow-First  and the Fast-First boarding policies (again, for $N \gg 1$).

\subsection{Airplane boarding and geometric optics}
The result above for airplane boarding states that the boarding time can be derived from the curve (geodesic) that maximizes the proper time (curve weight) in spacetime under an appropriate Lorentzian metric. In geometric optics Fermat's principle states that light will travel between two points along a path that minimizes the amount of travel time, which is a function of the local index of refraction. 

In \cref{eq:weight,eq:weight_SF} the aisle-clearing time $\tau$ plays the same role in the Lorentzian space for airplane boarding as the refractive index does in the Euclidean space for light. This can be seen in the qr-diagrams in \crefformat{figure}{Figs.~#2#1{(c-d)}#3}\cref{fig:policies} and \crefformat{figure}{#2#1{(b)}#3}\cref{fig:maximal_chains}, where the curve breaks at the boundary between different groups of passengers with different aisle-clearing time.  
While light moves in straight lines in homogeneous media, the longest curve under the Lorentzian metric has a curved shape in the spacetime domains filled with passengers with equal aisle-clearing time. Interestingly, when $k=0$, those lines become straight also under the Lorentzian metric, and break on the border between passenger groups according to a principle similar to Snell's law.

\section{Boarding time for Slow-First and Fast-First}\label{sec:kgt0}
We now turn to  compute the asymptotic average boarding time in \cref{eq:Tapprox} for $k>0$, and $p$, $C\equiv \tau_F/\tau_S$ both in the range $(0,1)$. We show that the Slow-First policy is more efficient than the Fast-First policy in the entire $(k,p,C)$-parameter space in the large-$N$ limit ($N\rightarrow \infty$). Comparisons to simulation results for smaller $N$ are also made.

\subsection{Analysis of the Random Boarding policy}\label{ssec:Tcompute_RA_kgt0}
To better explain our analytical approach we first illustrate the computations of the average boarding time of the Random Boarding policy.

With one group, $\tau(q)\equiv \tau_{}$ when all passengers have the same aisle-clearing time. The curve weight in \cref{eq:weight} becomes $W(r)=\tau\cdot L(r)$. The curve length $L(r)$ in \cref{eq:length} can be maximized straightforwardly using the variational method. This leads to general solutions of the form 
$r^*(q)=ae^{2kq}+be^{kq}+1$ when $k>0$.

\begin{figure*}[htb]
	\begin{center}
		\includegraphics[width=4.0cm,clip]{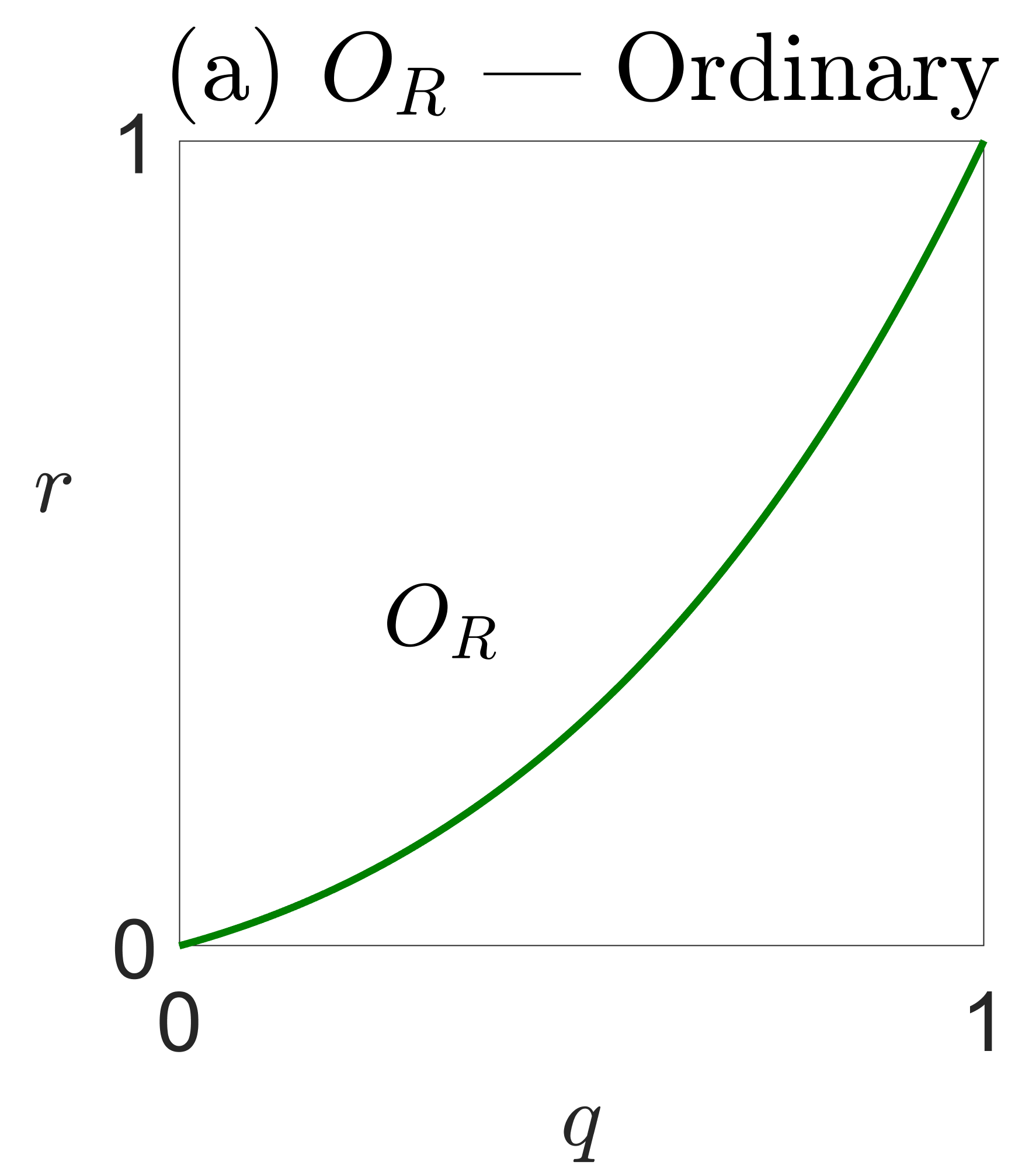}
		\includegraphics[width=4.0cm,clip]{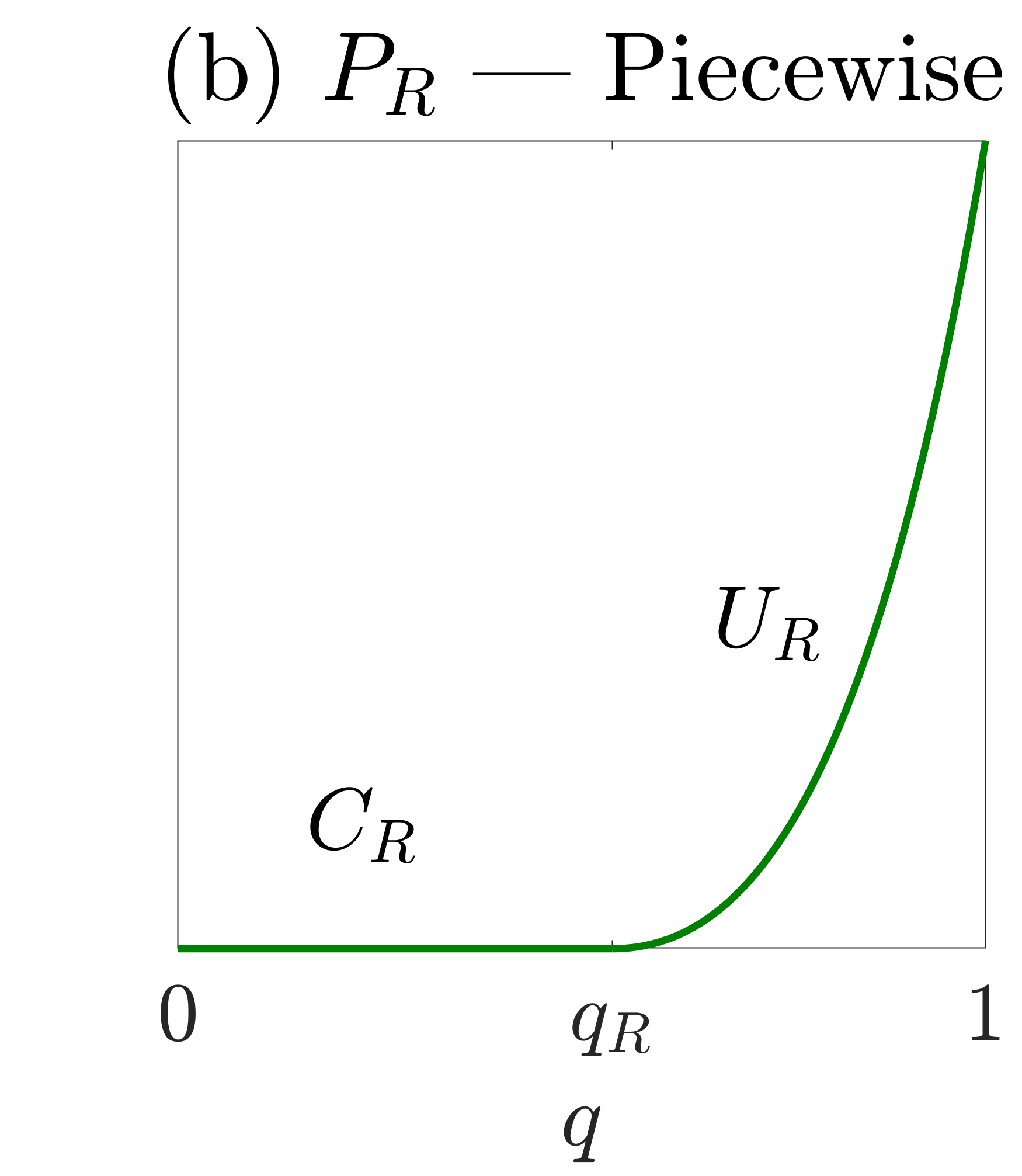}
	\end{center}
	\caption{\label{fig:rcurve_kgt0_g1} The shape of the longest curve for Random Boarding can be either ordinary or piecewise. 
		(a) Ordinary type curve ($O_R$) when $0<k\leqslant\ln(2)$. 
		(b) Piecewise curve ($P_R$) consisting of a constant function ($C_R$) and an upward-going ordinary-type curve ($U_R$) when $k>\ln(2)$.}
\end{figure*}

The constants $a,b$ are determined using the values at the start and end points: $r^*(0)=0$ and $r^*(1)=1$. A typical shape is shown in \crefformat{figure}{Fig.~#2#1{(a)}#3}\cref{fig:rcurve_kgt0_g1} for $k\leqslant \ln(2)$. 
The resulting maximal length of what we denote an \textit{ordinary-type curve} is by \cref{eq:length}
\begin{equation}\label{eq:O_R}
L(r^*)=\sqrt{(e^k-1)/k}\equiv O_R.
\end{equation}

However, when $k> \ln(2)$, an ordinary-type curve $r^*$ will extend below the  $(q,r)$-unit square. Since the curve should be within the unit square,\footnote{All passengers in a longest chain are within the unit square} the first part of the curve should be horizontal at value zero along the $q$-axis ($C_R$ in \crefformat{figure}{Fig.~#2#1{(b)}#3}\cref{fig:rcurve_kgt0_g1}). The remaining part is an upward-going ordinary-type curve ($U_R$).
Continuity of $r$ and $r'$ in the transition point $q=q_R$ between the $C_R$ and $U_R$ curves, and the end point $r(1)=1$, determine the values of $a,b$ in the $U_R$ curve (see Appendix \ref{app:computations} for details). This gives $q_R=1-\ln(2)/k$, and the total length of the resulting \textit{piecewise curve}, $r^*_P$, is (for $k>\ln(2)$):
\begin{equation}\label{eq:P_R}
L(r^*_P)=\frac{1}{\sqrt{k}}\left(k-\ln(2)+1 \right)\equiv P_R.
\end{equation}
Hence, the expected boarding time with Random Boarding and equal aisle-clearing time $\tau$ is by leading order given by \cref{eq:Tapprox_length}, 
which gives
\begin{equation}\label{eq:T_R}
\tilde{T}_R =
\begin{cases}
2 \tau\sqrt{\frac{ N}{k}}\sqrt{e^k-1}
\qquad &  0<k\leqslant \ln(2) \\[0ex]
2 \tau\sqrt{\frac{ N}{k}}\left(k-\ln(2)+1\right)
\qquad & \ln(2) < k.
\end{cases}
\end{equation}

\subsection{Analysis of the Slow-First policy}
The curve weight for the Slow-First policy is given by \cref{eq:weight_SF}. 
The {longest} curve must be continuous, but does not have to be smooth in the crossing point $(p,r(p))$ between the regions of the slow and the fast passengers in the qr-diagram.
If we fix the crossing height $r(p)=\delta$, the longest curves in each part of the qr-diagram must be either ordinary-type or piecewise, as for the single group policy in Sec.~\ref{ssec:Tcompute_RA_kgt0}.

\begin{figure}[htb]
	\begin{center}
		\includegraphics[width=4.0cm,clip]{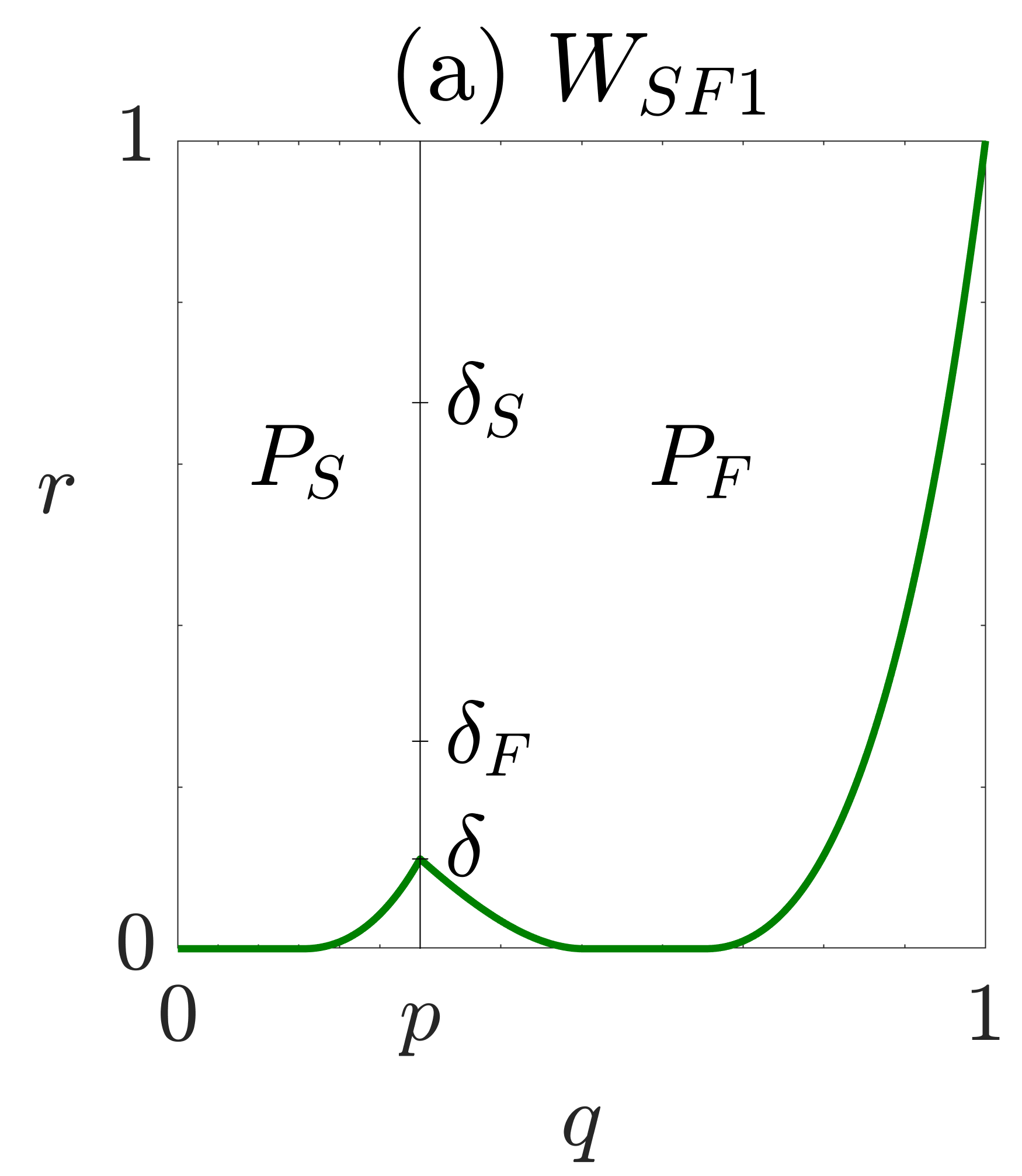}
		\includegraphics[width=4.0cm,clip]{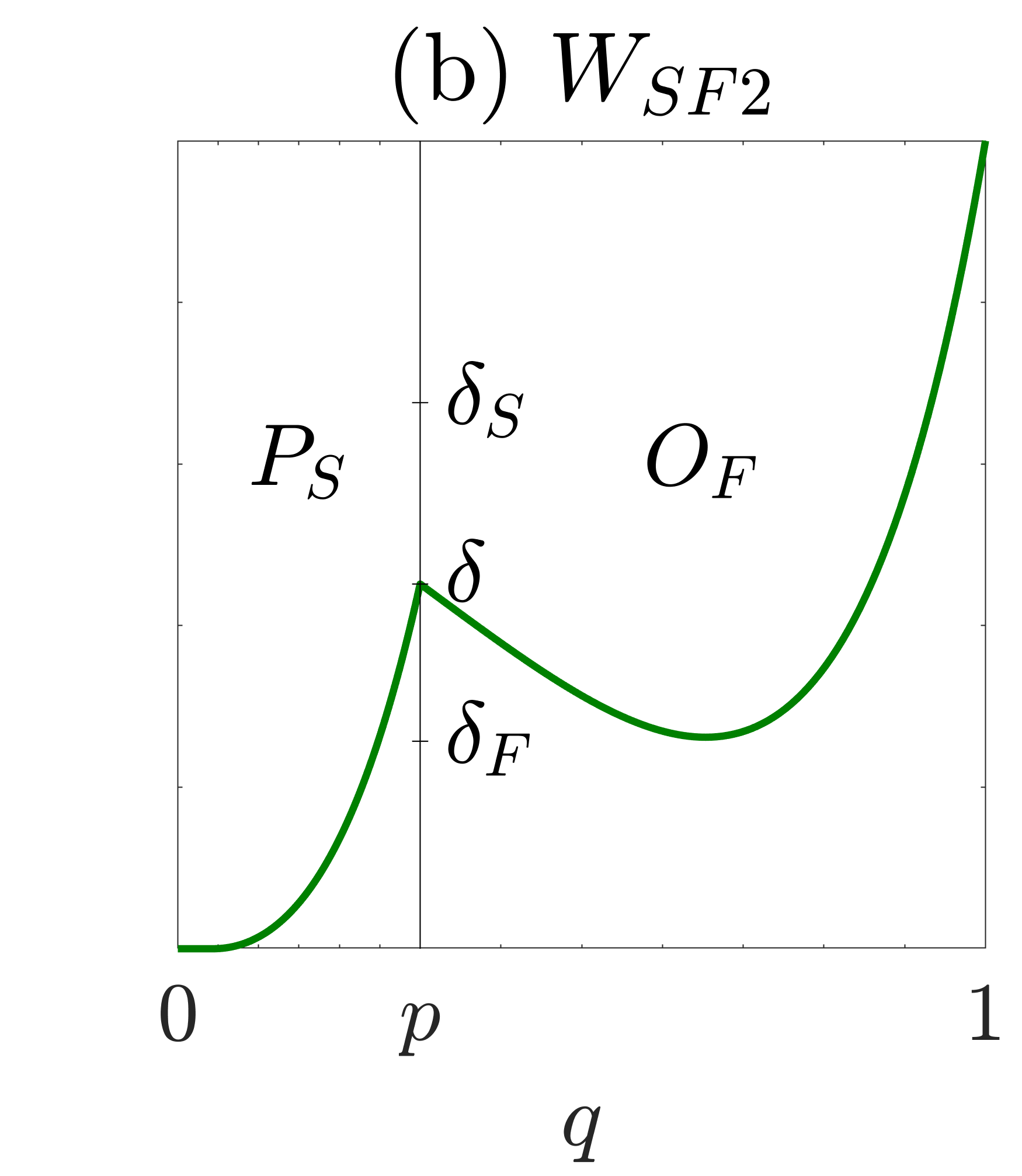}
		\includegraphics[width=4.0cm,clip]{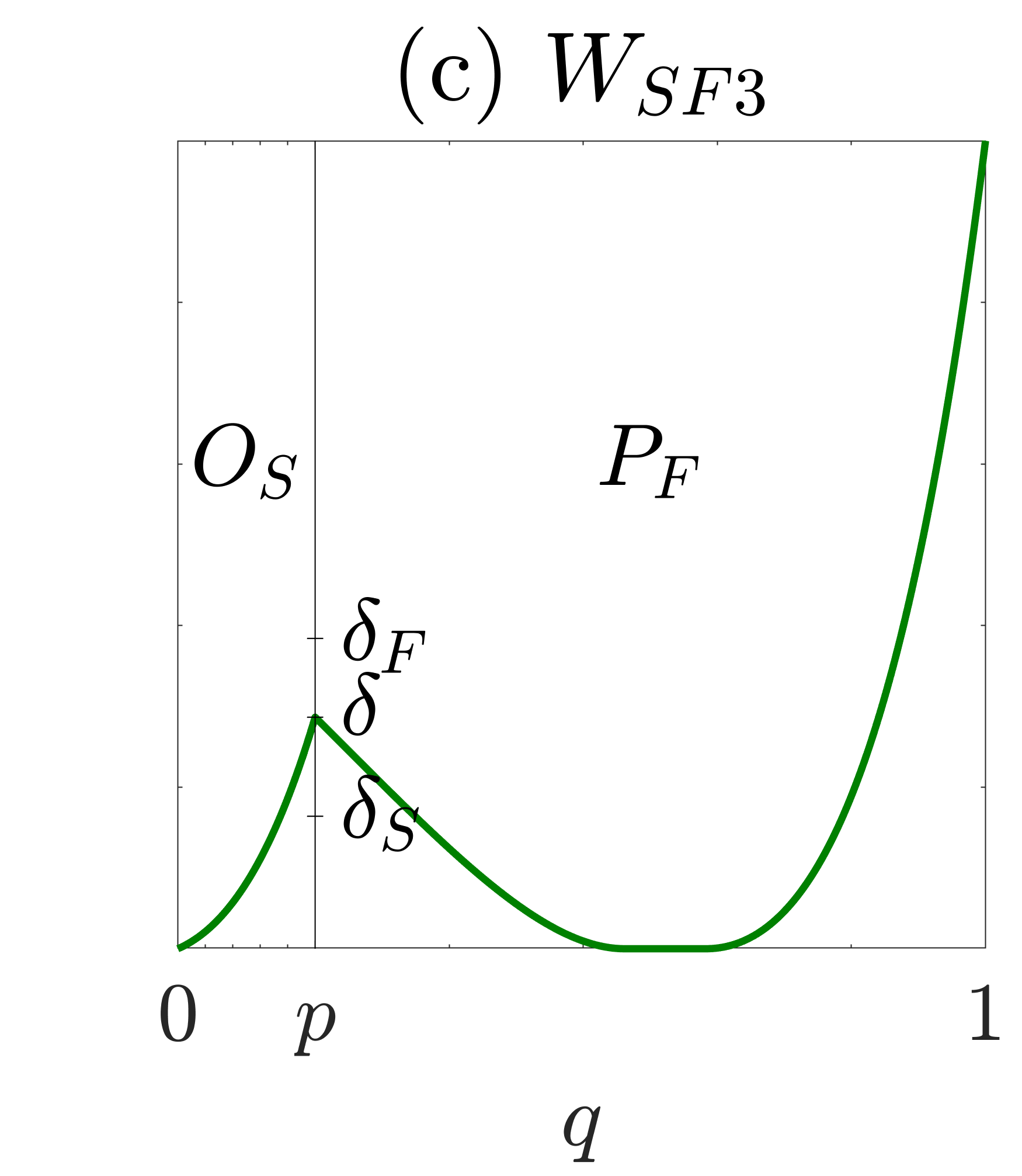}
		\includegraphics[width=4.0cm,clip]{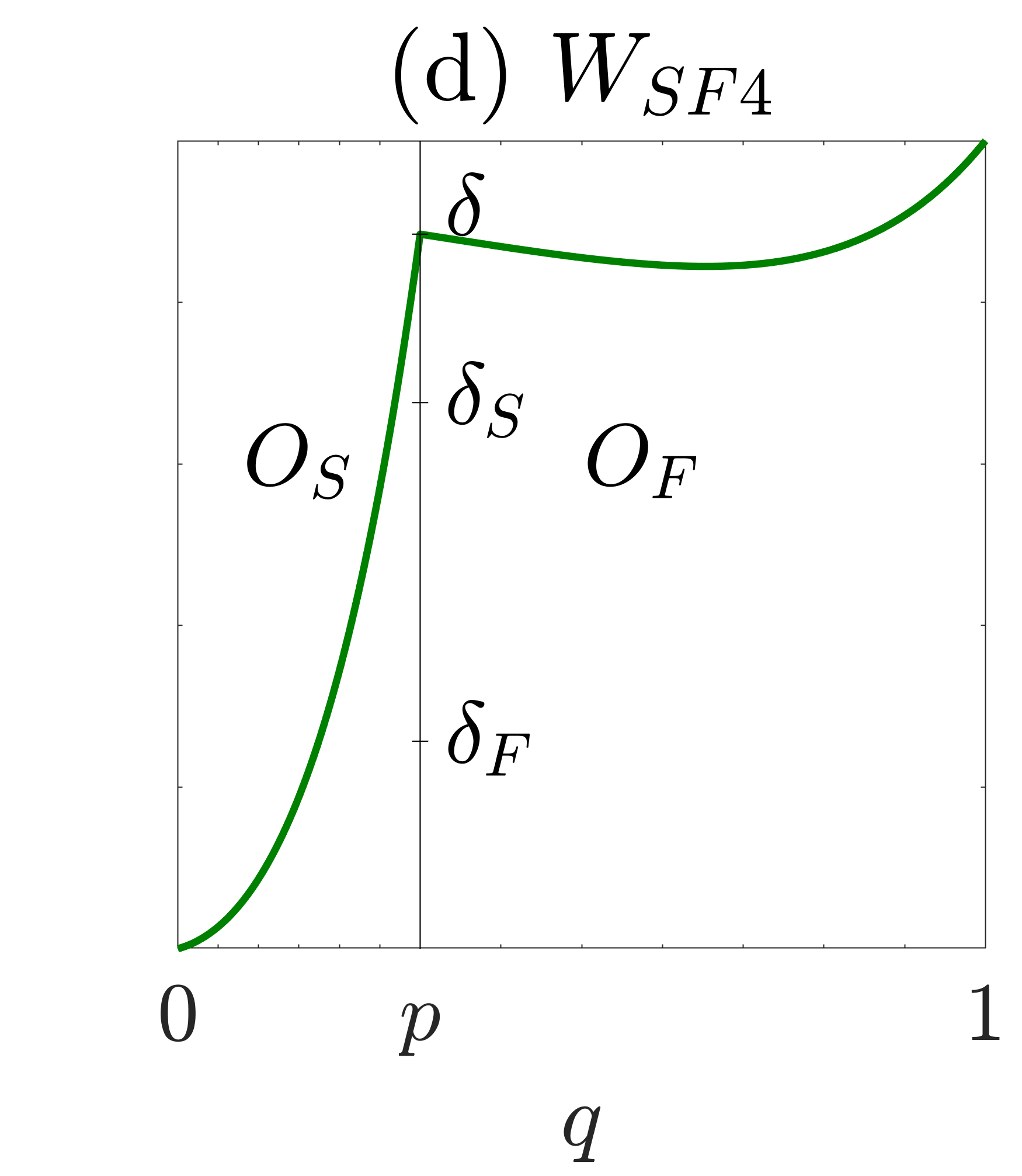}
	\end{center}
	\caption{\label{fig:rcurve_kgt0_g2} For fixed crossing height $\delta$, the shape of the longest curve for Slow-First can be either piecewise ($P_S,P_F$) or ordinary ($O_S,O_F$) in both the slow and fast region, respectively. Hence, the total curve can take four different shape-types. The sub-function that defines the weight $W_{SF}(\delta)$ in \cref{eq:W_SF}, depends on the values of $\delta_S(k,p)$ and $\delta_F(k,p)$ (expressions are given in the main text). (a) $W_{SF1}$: $\delta < \min\{\delta_S,\delta_F\}$, (b) $W_{SF2}$: $\delta_F < \delta <\delta_S$, (c) $W_{SF3}$: $\delta_S < \delta <\delta_F$, (d) $W_{SF4}$: $\max\{\delta_S,\delta_F\} < \delta$. The $\delta$ which maximizes $W_{SF}(\delta)$ also depends on the relative aisle-clearing time, $C=\tau_F/\tau_S$.}
\end{figure}

The length of the longest curve $L_S(\delta)$ in the first part of the diagrams in Fig.~\ref{fig:rcurve_kgt0_g2} is a piecewise curve ($L_S=P_S$) if  $\delta<\delta_S\equiv (e^{kp}-1)^2$ and an ordinary-type curve ($L_S=O_S$) if $\delta\geqslant\delta_S$. A similar parameter $\delta_F=\max\{0,1-2e^{-k(1-p)}\}^2$ determines the type of longest curve in the second part of the qr-diagrams. Explicit expressions for the curves ($O_S, P_S,$ etc.) are given in Appendix \ref{app:computations}.\footnote{The same curves are used for the Fast-First policy, only by exchanging $p \leftrightarrow 1-p$ and  $\tau_S \leftrightarrow \tau_F$.}

Depending on the value of the fixed crossing height $\delta$ (relative to the values of $\delta_S (k,p)$ and $\delta_F (k,p)$), the resulting total curves can be one of four different combinations of each of these curves.
Let $W_{SF}(\delta)$ be the weight of the {longest} piecewise curve for a fixed crossing height $\delta$
(the dependence on $k,p, \tau_S, \tau_F$ is suppressed in the following):
\begin{equation}\label{eq:W_SF}
W_{SF}(\delta;k,p,\tau_S,\tau_F) =
\begin{cases}
W_{SF1}(\delta) = \tau_S P_S(\delta) + \tau_F P_F(\delta)
&\qquad   \delta<\min\{\delta_S,\delta_F\} \\[0ex]
W_{SF2}(\delta) = \tau_S P_S(\delta) + \tau_F O_F(\delta)
&\qquad  \delta_F < \delta < \delta_S \\[0ex]
W_{SF3}(\delta) = \tau_S O_S(\delta) + \tau_F P_F(\delta)
&\qquad  \delta_S < \delta < \delta_F \\[0ex]
W_{SF4}(\delta) = \tau_S O_S(\delta) + \tau_F O_F(\delta)
&\qquad  \max\{\delta_S,\delta_F\} <\delta \\[0ex]
\end{cases}
\end{equation}

To find the {longest} curve, we must compute the $\delta=\delta^*$ that maximizes $W_{SF}(\delta)$. The function $W_{SF}(\delta)$ is differentiable and has negative curvature when $\delta \in (0,1)$. This means that there is never more than one local maximum on the domain. 

Moreover, each of the four sub-functions in \cref{eq:W_SF} has a maximum point $\delta_i^*\in [0,1], i\in\{1,2,3,4\}$ with maximum value $W_{SFi}^*$. Each of these are global maximum points for $W_{SF}(\delta)$ if and only if $\delta_i^*$ lies within the respective subdomain in \cref{eq:W_SF}. Hence, the weight of the {longest} curve for the Slow-First policy is, e.g., given by $W_{SF}^*=W_{SF1}^*$ when $\delta_1^*<\min\{\delta_S,\delta_F\}$. 

The maximum points $\delta=\delta_i^*$, for each of the subfunctions in \cref{eq:W_SF} yield the following global maxima $W_{SF}^*$ of $W_{SF}$:
\begin{align}\label{eq:W*_SF}
\begin{array}{ll}
W_{SF1}^* = \frac{\tau_S}{\sqrt{k}} \left[kp(1-C) + kC +1 + C\ln\left(\frac{C}{1+C}\right)  - \ln\left(\frac{2}{1+C}\right) \right]
&\max\left\{C_2,C_1\right\} \leqslant C \\[0.5ex]
W_{SF2}^* = \frac{\tau_S}{\sqrt{k}} \left[kp +1  - \ln\left(\frac{2}{1+C^2(e^{k(1-p)}-1)}\right) \right]
&C_3^2 \leqslant C^2 \leqslant C_1^2\\[0ex]
W_{SF3}^* = \frac{\tau_S}{\sqrt{k}}\left[\sqrt{(1-e^{-kp})(e^{kp}-1+\delta_3^*)}\right. \quad +&  \\[0.5ex]
\qquad\qquad\qquad
\left. C\left(k(1-p)+1+\sqrt{\delta_3^*}+\ln((1-\sqrt{\delta_3^*})/2)\right) \right]
&C_4^2 \leqslant C^2 \leqslant C_2^2 \\[0.5ex]
W_{SF4}^* = \frac{\tau_S}{\sqrt{k}} \sqrt{(e^{kp}-1) + C^2(e^k - e^{kp})}
&C^2 \leqslant \min\{C_3^2,C_4^2 \}, 
\end{array}
\end{align}
Here $C\equiv \tau_F/\tau_S \in(0,1)$. Since $\delta_i^*(k,p,C),\delta_S(k,p),\delta_F(k,p)$ are functions of $(k,p,C)$, the conditions on $\delta=\delta_i^*$  in \cref{eq:W_SF} have been rewritten as conditions on $C$, where
$C_1\equiv (e^{k(1-p)}-1)^{-1}$, $C_2\equiv 2e^{-kp}-1$, $C_3^2 \equiv (2-e^{kp})/(e^k-e^{kp})$, and 
$C_4^2\equiv 4(e^{kp}-1)/(e^{2k} - 4(e^k-e^{kp}))$.\footnote{The expression for the maximum point, $\delta_3^*$, for $W_{SF3}(\delta)$ is given in Appendix \ref{app:computations}.} 

\begin{figure*}[htb]
	\begin{center}
		\includegraphics[width=17.0cm,clip]{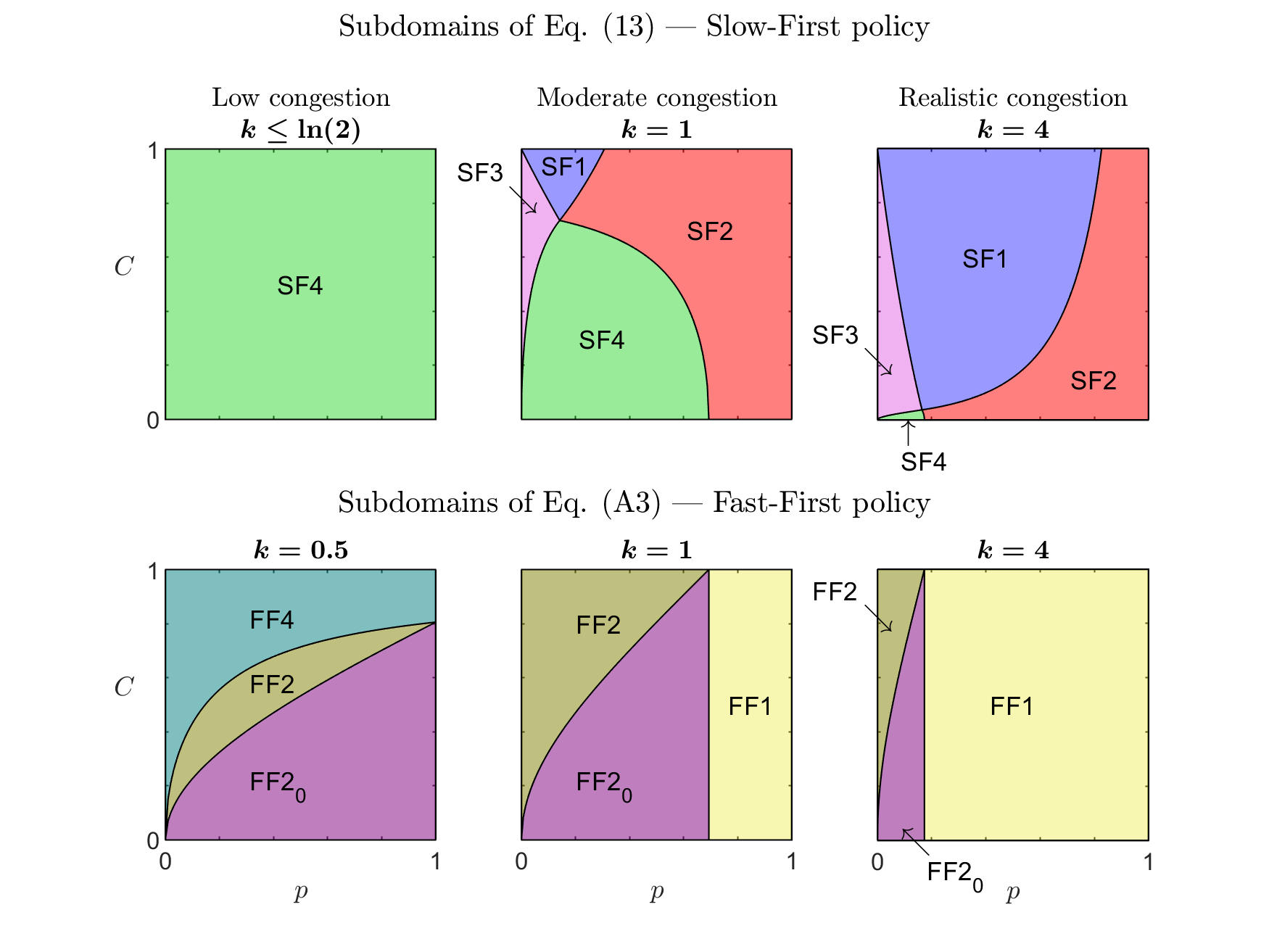}
	\end{center}
	\caption{\label{fig:SF_FF_activesubfunctions} \emph{Upper row:} The subdomains of the $(p,C)$-unit square where the Slow-First boarding time is represented by the different sub-functions in \cref{eq:W*_SF}. \emph{Lower row:} Corresponding subdomains for the Fast-First policy in \cref{eq:W*_FF}. 
	}
\end{figure*}

The subdomains where the conditions in \cref{eq:W*_SF} are satisfied are shown in the upper row of Fig.~\ref{fig:SF_FF_activesubfunctions} for $k\in \left\{0.5, 1,4\right\}$. E.g. for $p>\ln(2)/k$, the conditions are simplified such that $W_{SF1}^*$ is the maximum when $C\geqslant C_1$ and $W_{SF2}^*$ when $C\leqslant C_1$. 

\begin{figure*}[htb]
	\begin{center}
		\includegraphics[width=8cm,clip]{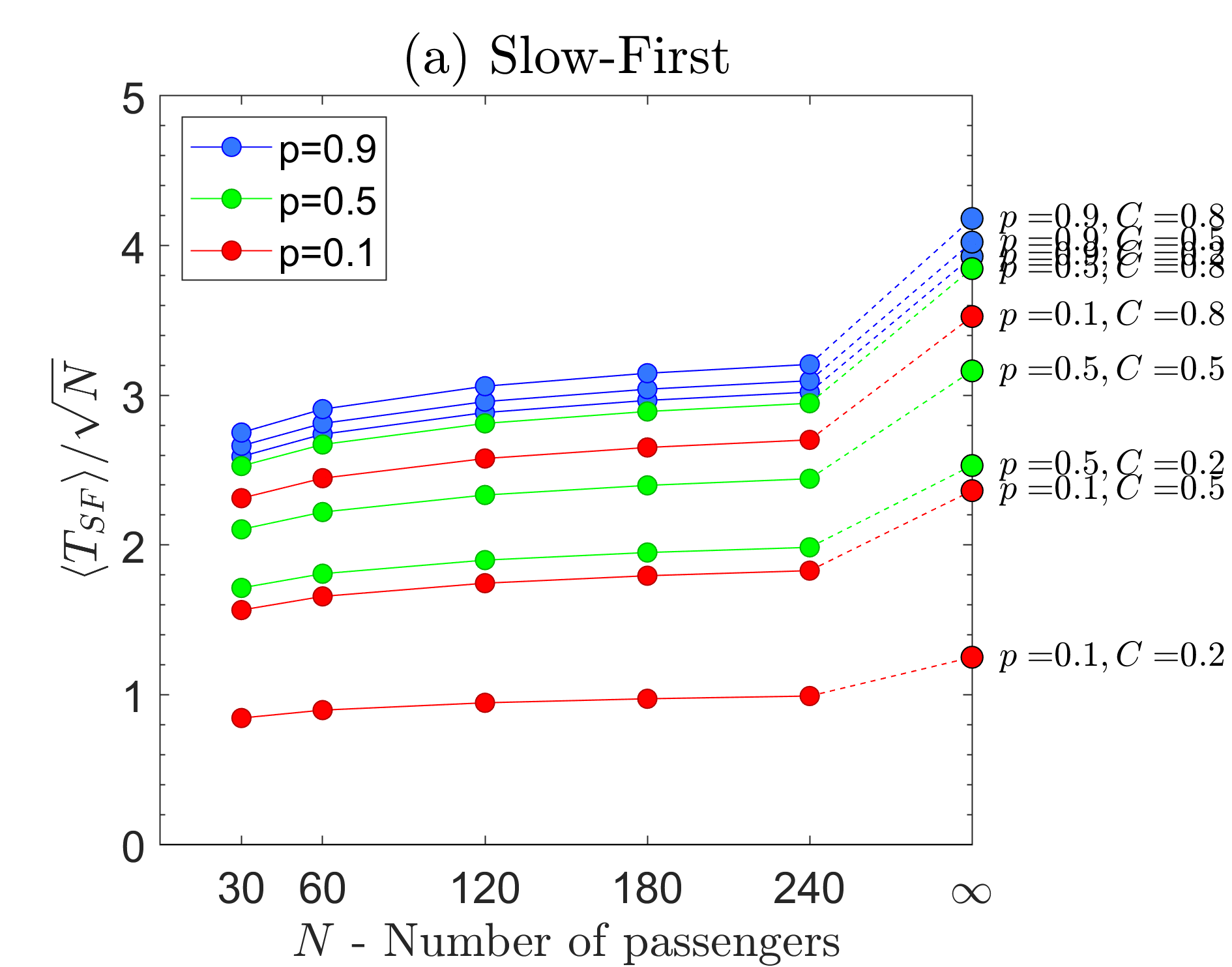}
		\includegraphics[width=8cm,clip]{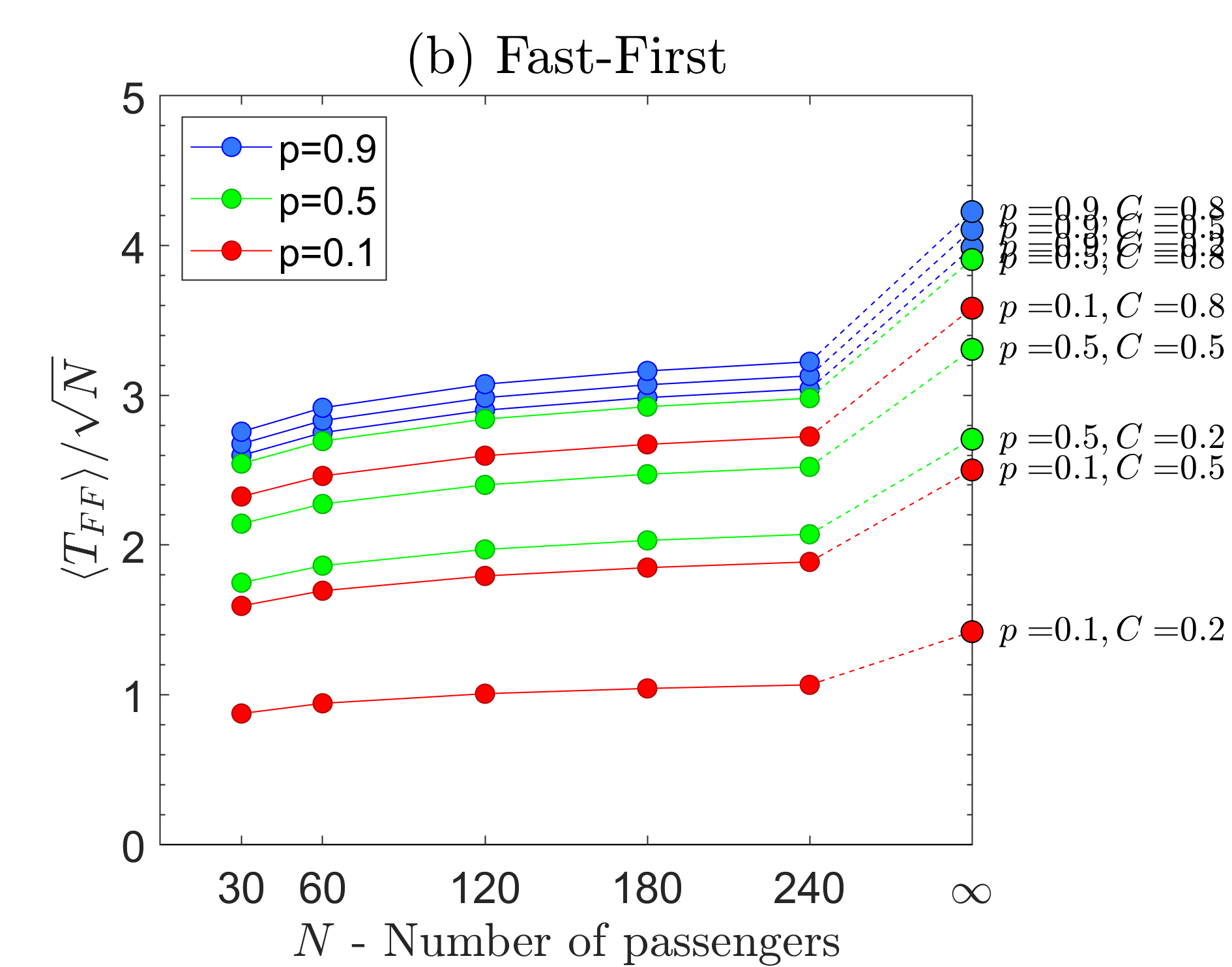}
	\end{center}
	\caption{\label{fig:SFandFF_boardingtimes} Average boarding time estimates for the Slow-First and Fast-First policies for different $(k,p,C)$-parameter settings. Simulation results for increasing number of passengers are compared to the asymptotic results for all combinations of parameter values $p\in\{0.1,0.5,0.9\}$ and $C\in\{0.2,0.5,0.8\}$. $k=4$, and the accuracy is $\pm 0.002$ (as a result of $10^6$ scenarios  for each finite-$N$ data point). The rightmost points are asymptotic values. 
	}
\end{figure*}

The maximal weight $W_{SF}^*$ in \cref{eq:W*_SF} is used to calculate the corresponding asymptotic boarding time $\tilde{T}_{SF}$ in \cref{eq:Tapprox}. In \crefformat{figure}{Fig.~#2#1{(a)}#3}\cref{fig:SFandFF_boardingtimes} comparisons of the asymptotic boarding time for the Slow-First policy with simulation results for $N\leqslant 240$ show that the asymptotic result in \cref{eq:Tapprox} tend to overestimate the boarding time, but the relative ranking between different parameter settings is maintained.

\subsection{Analysis of the Fast-First policy}\label{ss:FF_kgt0}
The same procedure as in the section above can be repeated for the Fast-First policy by exchanging $p\leftrightarrow 1-p$, $\tau_S \leftrightarrow \tau_F$, and $C\rightarrow 1/C$. 

For $W_{FF1}(\delta)$ (which corresponds to $W_{SF1}$), $W_{FF1}'(\delta)<0, \forall \delta\in(0,1)$. Consequently, the weight of the {longest} curve is given for $\delta=0$, and 
\begin{equation}\label{eq:W*_FF1}
W^*_{FF1}=W_{FF1}(0)=\frac{\tau_S}{\sqrt{k}} \left(kp(1-C) + kC +1 - \ln(2)\right).
\end{equation}
$W_{FF}(\delta)$ has negative curvature, and $\delta=0$ is therefore the global maximum point for $W_{FF}$ as long as $\delta=0$ satisfies the conditions corresponding to the subdomains for FF1 in \cref{eq:W_SF}. For Fast-First, $\delta_S(k,1-p)= (1-e^{k(1-p)})^2 >0$. Moreover, $\delta_F(k,1-p)=\max\{0,1-2e^{-kp}\}^2 $ is also positive when $kp>\ln(2)$. This means that when $kp>\ln(2)$, the condition $\delta= 0<\min\{\delta_S(k,1-p),\delta_F(k,1-p)\}$ is satisfied, and hence $W^*_{FF}=W^*_{FF1}$. The corresponding subdomain is indicated in the lower row of Fig.~\ref{fig:SF_FF_activesubfunctions}.

For $kp\leqslant \ln(2)$, results for the Fast-First policy corresponding to those in \cref{eq:W*_SF} are given in Appendix \ref{app:computations}. 
In \crefformat{figure}{Fig.~#2#1{(b)}#3}\cref{fig:SFandFF_boardingtimes} comparisons of the asymptotic boarding time for the Fast-First policy with simulation results for $N\leqslant 240$ are similar to those for the Slow-First policy.

\subsection{Comparing Slow-First and Fast-First policies}\label{ss:compareSFvsFF_kgt0}
The Slow-First policy outperforms the Fast-First policy for all values of $k>0$ and $p,C \in (0,1)$. 
Here, explicit results are only shown for $kp>\ln(2)$. The derivation of the other results are left for Appendix \ref{app:computations} and \ref{app:proofs}, and the results are summarized in \cref{sec:mainresults} with comparisons to finite-$N$ discrete-event simulations. 

Based on the results in \cref{eq:W*_SF,eq:W*_FF1}, $W^*_{FF}-W^*_{SF}$ is, for $kp>\ln(2)$, given by:
\begin{align}
W^*_{FF1}-W_{SF1}^* 
&= \frac{\tau_S}{\sqrt{k}} \left(\ln\left(\frac{1}{1+C}\right)  - C\ln\left(\frac{C}{1+C}\right) \right)
&&C_1 \leqslant C \label{eq:FF1-SF1}\\[0ex]
W^*_{FF1}-W_{SF2}^* 
&= \frac{\tau_S}{\sqrt{k}} \left(Ck(1-p)  - \ln\left(1+C^2(e^{k(1-p)}-1)\right) \right)\qquad
&&C \leqslant C_1,\label{eq:FF1-SF2}
\end{align}
where $C_1\equiv (e^{k(1-p)}-1)^{-1}$. In Appendix \ref{app:proofs}, both of these are shown to be positive, meaning that the Fast-First policy yields a longer average boarding time than the Slow-First policy. 

The difference approaches 0 for both $C\rightarrow 0$ and $C\rightarrow 1$.
{That $C\rightarrow 0$ means that the fast passengers are much faster than the slow ones. Then it does not matter who enters first since the fast fraction of the queue will sit down immediately anyway. When $C\rightarrow 1$, both groups have the same speed, and the two policies should not differ.}

The difference also vanishes when $k\rightarrow\infty$. {This means that each passenger takes up so much room that the first group will be seated before the next group enters the airplane. Consequently, the sequencing of the groups does not matter.} 

When $k(1-p)\rightarrow 0$ the difference also approaches 0. {This means that when the fraction $(1-p)$ of fast passengers vanishes, there is no difference between Slow-First and Fast-First. This is expected since the queue then reduces to slow passengers only.}

\section{Discussion - outlook}
In this paper, airplane boarding has been analyzed using Lorentzian geometry, which is exact in the limit of a large number of passengers.
We have showed that the boarding time with the Slow-First policy on average is faster than with the Fast-First policy in the large $N$-limit. The reason for Slow-First superiority is because this policy better utilizes parallelism, extending the time window during which fast passengers can sit down simultaneously with the last slow passengers. With the Fast-First policy on the other hand, the last fast passengers are quick to take a seating, with less time remaining for simultaneous seating to happen for the slow passengers.
	
Simulations we performed confirm that our asymptotic result still applies for a lower number of passengers $N$ of the order of hundreds, as in medium-sized airplanes. There are also some basic similarities with Back-To-Front boarding (BTF), which is a beneficial policy when the congestion factor $k$ is small, since then the two groups can sit down in parallel. In general, Slow-First policy has better parallelism compared to Fast-First, but of a different kind which is much less congestion dependent than the Back-To-Front policy. 
	
Future extension of this work include increasing the variability to more than just two groups of passengers' aisle-clearing time. For an inhomogeneous group of passengers with different aisle-clearing times, one can attach an \emph{effective} aisle-clearing time which measures by which factor the group as a total slows the boarding process \cite{Bachmat:2019}. This parameter plays the role of an \emph{effective refractive index},
where each group could consist of both slow and fast passengers.
Initial results indicate that the Slow-First policy is always better than the Random Boarding policy --- in which the fast and slow passengers are all mixed in one group. Assuming this is the case, one can argue that both the slow and the fast group should be divided over again according to speed. In the end, the slowest passenger would be in front of the queue and the fastest passenger in the back of the queue. Whether this would reduce the boarding time is still an open question.
	
For airlines it is not only the average boarding time that matters. The frequency of departure delays due to unexpected long boarding time could also be of relevance. Then the percentage of boardings exceeding a certain duration would be an appropriate measure. The same types of results as we present here must then be obtained in terms of percentiles. Such percentiles can be estimated using the Tracy-Widom distribution \cite{Bachmat:2014}, but requires extensive calculations.
	
The methods used to compute the asymptotic boarding time can also be used when assumptions and policies deviate from those used in this paper. The modifications needed to estimate boarding time when there is, e.g., half-empty airplane or varying passenger widths, $w$, are described in the supplementary information of Ref. \cite{Bachmat/Berend/Sapir/Skiena/Stolyarov:2009}. Moreover, in this paper we do not take into account that a window seat passenger takes longer time to settle down if the middle seat or aisle seat passenger in the same row is already seated. How this affects the boarding time can also be taken into account within our framework, but the computational details are more complex. However, this would let us compare the Random Boarding policy with the group-based Window-Middle-Aisle boarding policy which is applied by some airlines. 
	
Will the fast passengers accept the Slow-First policy? Some passengers prefer to spend as little time as possible in confined spaces and are very happy to be the last passenger to enter the airplane. However, other might dislike waiting or the idea that the "troublesome" passengers with much carry-on luggage are occupying the overhead lockers, leaving little room for the coats and smaller items of the light-traveling passengers. This could potentially lead to an unintended consequence where passenger  would have an incentive  to bring more carry-on luggage on board, leading to a kind of "tragedy of the commons" scenario where everybody will wait longer. This suggests that a strategic, i.e., game-theoretic point of view of boarding policies is of great interest for a future study, as it can pinpoint which policies would be less sensitive to adaptive airline boarders, who simply wish to reach their destination while maximizing their own comfort.

\begin{acknowledgments}
	The work of Eitan Bachmat was supported by the German Science Foundation (DFG) through
	the grant “Airplane Boarding” (JA 2311/3-1).	
\end{acknowledgments}

\appendix
\section{Calculation details}\label{app:computations}

\subsection{One group: Random Boarding}
In this section we show more details of the computations behind the maximized curve lengths of the Random Boarding policy in \cref{eq:O_R} and \cref{eq:P_R}. The problem is broken down by first noting that the general solution of maximized curve length \cref{eq:length} between any two points is of the form $r^*(q)=ae^{2kq}+be^{kq}+1$. There is one such function $O_R$ which goes between $(0,0)$ and $(1,1)$. When $k>\ln(2)$ this curve would dip below the $q$-axis and only the last part $U_R$ of a piecewise smooth function would be of the mentioned form. The parameters $a,b$ for these curves (shown in Fig.~\ref{fig:rcurve_kgt0_g1}) are given by
\begin{align*}
	\text{$O_R$-curve:}\quad
	&a=(e^{k}-1)^{-1},&    
	&b=-ae^k,&
	&q\in(0,1),&
	&k\leqslant \ln(2)\\[0ex]
	\text{$U_R$-curve:}\quad
	&a=4e^{-2k},&    
	&b=-4e^{-k},&
	&q \in (1-\ln(2)/k,\; 1),&
	&k > \ln(2)\;.
\end{align*}
In order to compute the length of each of the curves in Fig.~\ref{fig:rcurve_kgt0_g1}, we use that the length $L(r)$ in \cref{eq:length} of a function $r(q)$ between two points $q_1$ and $q_2$ is given by:
\begin{align}\label{eq:L_Ordinary_Constant}
	\begin{array}{ll}
		\text{Ordinary function: $\quad r(q)=ae^{2kq}+be^{kq}+1$,} \qquad\qquad
		& L(r)=\sqrt{\frac{a}{k}}\left(e^{kq_2}-e^{kq_1}\right)  \\[0ex]
		\text{Constant function: $\quad r(q)\equiv 0$,} 
		& L(r)=\sqrt{{k}}\left(q_2-q_1\right)\;.
	\end{array}
\end{align}
This gives the length of both $O_R$ and $P_R$ in \cref{eq:O_R} and \cref{eq:P_R}, respectively.

\subsection{Two groups: Slow-First policy}
In this section we show more details of the computations behind the maximized curve weight of the Slow-First policy in \cref{eq:weight_SF} (and the corresponding for the Fast-First policy). The problem is broken down by first noting that the general solution of maximized curve length \cref{eq:length} between any two points is of the form $r^*(q)=ae^{2kq}+be^{kq}+1$. There is one such function in each of the two regions shown in Fig.~\ref{fig:rcurve_kgt0_g2}(d) when the crossing height $r^*(p)=\delta$ is fixed. The $O_S$-curve goes between the points $(0,0)$ and $(p,\delta)$, while the $O_F$-curve continues to $(1,1)$. The parameters $a,b$ are given by
\begin{align*}
	&\text{$O_S$-curve:}&
	&a=\frac{e^{kp}-1+\delta}{e^{2kp}-e^{kp}},&
	&b=-(a+1),&	
	&q\in(0,p),&
	&\delta>\delta_S	\\[0ex]	
	&\text{$O_F$-curve:}&
	&a=\frac{1-\delta}{e^{kp}(e^k-e^{kp})},&
	&b=-ae^k,&
	&q\in (p,1),&
	& \delta>\delta_F\;.
\end{align*}
The restriction on $\delta$ is in order to avoid that the curve dips below the $q$-axis.
E.g. when $\delta<\delta_S$, $O_S$ in the first region in Fig.~\ref{fig:rcurve_kgt0_g2}(c-d) would have dipped below the $q$-axis. To maintain a positive continuous smooth curve as in Fig.~\ref{fig:rcurve_kgt0_g2}(a-b), a part of the curve must be constant along the $q$-axis. The remaining parts of these piecewise curves are ordinary type curves that are connected smoothly to the constant part of the curve. They have parameters:
\begin{align*}
	&\text{Last part of $P_S$-curve:}&	
	a&={e^{-2kp}}\left({1+\sqrt{\delta}}\right)^2,&	
	b&=-2e^{-kq_S},&	
	&q \in (q_S,\; p) \\[0ex]
	&\text{First part of $P_F$-curve:}&	
	a&={e^{-2kp}}\left({1-\sqrt{\delta}}\right)^2,&	
	b&=-2e^{-kq_F},&	
	&q \in (p,\; q_F)\\[0ex]
	&\text{Last part of $P_F$-curve:}&	
	a&=4e^{-2k},&	
	b&=-4e^{-k},&	
	&q \in (1-\ln(2)/k,\; 1) \;,
\end{align*}
where $q_S=p-\ln(1+\sqrt{\delta})/k$ and $q_F=p-\ln(1-\sqrt{\delta})/k$.

The length of each of the curves in Fig.~\ref{fig:rcurve_kgt0_g2} is found by \cref{eq:L_Ordinary_Constant}. This gives:
\begin{align*}
	O_S(\delta) &= \frac{1}{\sqrt{k}}
	\sqrt{(e^{-kp}-1)(e^{kp}-1+\delta)}\\[0ex]
	O_F(\delta) &= \frac{1}{\sqrt{k}}
	\sqrt{(1-\delta)(e^{k(1-p)}-1)}\\[0ex]
	P_S(\delta) &= \frac{1}{\sqrt{k}} 
	\left[kp + \sqrt{\delta} - \ln (1+\sqrt{\delta})\right]\\[0ex]
	P_F(\delta) &= \frac{1}{\sqrt{k}}
	\left[k(1-p) + (1+\sqrt{\delta}) + \ln(1-\sqrt{\delta}) - \ln(2) \right]\;.
\end{align*}

For Slow-First, these are combined to $W_{SF}(\delta)$ according to \cref{eq:W_SF}. Since $W_{SF}(\delta)$ is differentiable and has negative curvature when $\delta \in (0,1)$ there is never more than one local maximum on the domain. 
 
To show the differentiability, we need continuity of  $W_{SF}'(\delta)$ in the transition points when $\delta=\delta_S, \delta_F$. For $\delta_S<\delta_F$ we need that $W_{SF1}'(\delta_S) = W_{SF3}'(\delta_S )$ and $W_{SF3}'(\delta_F )=W_{SF4}'(\delta_F )$. For $\delta_F<\delta_S$, we must require that $W_{SF1}'(\delta_F )=W_{SF2}'(\delta_F )$ and $W_{SF2}'(\delta_S )=W_{SF4}'(\delta_S )$. This reduces to showing that $P_S' (\delta_S )=O_S'(\delta_S )$, and $P_F' (\delta_F )=O_F' (\delta_F )$, which is straightforward.
The negative curvature of $W_{SF}$ follows from $W_{SFi}''(\delta)<0,\forall\delta\in(0,1)$, for $i=1,2,3,4$.

The $\delta$'s that are maximizing each of the subfunctions in 
\cref{eq:W_SF} are given by:
\begin{align}
	\sqrt{\delta^*_1} &= \frac{1-C}{1+C} \nonumber\\[0ex]
	\sqrt{\delta^*_2} &= \frac{1-C^2(e^{k(1-p)}-1)}{1+C^2(e^{k(1-p)}-1)}\nonumber\\[0ex]
	\sqrt{\delta^*_3} &= \frac{\sqrt{(e^{kp}-1)(1-C^2)}  - C(e^{kp}-1)}{\sqrt{(e^{kp}-1)(1-C^2)}  + C} \label{eq:delta_SF3}\\[0ex]
	\delta^*_4 &= \frac{1-C^2 \left(e^k-e^{kp} \right)}{1 + C^2 \left(e^k-e^{kp} \right)/\left(e^{kp}-1 \right)}\nonumber
\end{align}
 If $\delta_i=\arg\max_{\delta\in(0,1)}  W_{SFi} (\delta)$ lies in the subdomain where $W_{SF}=W_{SFi}$, then $\delta_i$ is the global maximum of $W_{SF}$ due to the negative curvature of $W_{SF}(\delta)$. Inserting these into \cref{eq:W_SF} gives $W^*_{SF}$ in \cref{eq:W*_SF}.

When $k > \ln 2$, the boundaries shown in the upper row of Fig.~\ref{fig:SF_FF_activesubfunctions} meet at the vortex point $(p^*,C^*)$, where $C^*= 2e^{-k}$ and $p^*= \frac{1}{k} \ln \left(\frac{2}{1+C^*} \right)$. The boundaries
are given by
\begin{align*}
	\mbox{SF1 -- SF2} \quad : \qquad C_1 &= \frac{1}{e^{k(1-p)}-1} \;,&&
	p \in [ p^*,1- (\ln2/k)] \\
	\mbox{SF1 -- SF3} \quad : \qquad C_2 &= 2 e^{-kp} - 1 \;,&&
	p \in [0,p^*] \\
	\mbox{SF2 -- SF4} \quad : \qquad C_3 &= \sqrt{\frac{2-e^{kp}}{e^k-e^{kp}}} \;,&&
	p \in [ p^*, \ln2/k]\\
	\mbox{SF3 -- SF4} \quad : \qquad C_4 &= 2 \, \sqrt{\frac{e^{kp}-1}{e^{2k} -4 \left(e^k-e^{kp} \right)}} \;,&&
	p \in [0,p^*]\;.
\end{align*}

\subsection{Two groups: Fast-First policy}
The same procedure as in the section above can be repeated for the Fast-First policy by exchanging $p\leftrightarrow 1-p$, $\tau_S \leftrightarrow \tau_F$, and $C\rightarrow 1/C$. The $\delta$'s maximizing each of the subfunctions of $W_{FF}(\delta)\equiv W_{SF}(\delta;k,1-p,\tau_F,\tau_S)$ in 
\cref{eq:W_SF} are given by:
\begin{align*}
	\sqrt{\tilde{\delta}^*_1} &= 0\\[0ex]
	\sqrt{\tilde{\delta}^*_{2}}&=
	\begin{cases} 
		0,										& C<\tilde{C}_1\\[0ex]
		\frac{C^2-(e^{kp}-1)}{C^2+(e^{kp}-1)},	& C>\tilde{C}_1 
	\end{cases}\\[0ex]
	\tilde{\delta}^*_4 &= \frac{C^2-\left(e^k-e^{k(1-p)} \right)}
	{C^2 + \left(e^k-e^{k(1-p)} \right)/\left(e^{k(1-p)}-1 \right)}	\;,
\end{align*}
where $\tilde{C}_1=\sqrt{e^{kp}-1}$. $\tilde{\delta}^*_{3}$ is not computed since this will never be a global maximum for $W_{FF}(\delta)$. Inserting $\tilde{\delta}^*_1$ into $W_{FF}(\delta)$ gives $W^*_{FF}=W^*_{FF1}$ in \cref{eq:W*_FF1} when $kp\geqslant\ln(2)$. When $kp<\ln(2)$, $\tilde{\delta}^*_2$ and $\tilde{\delta}^*_4$ can be inserted into $W_{FF}(\delta)$ which gives the following expression for $W_{FF}^*$:
\begin{align}\label{eq:W*_FF}
	\begin{array}{lll}
		W_{FF2_0}^* &= \frac{\tau_S}{\sqrt{k}} \left[Ck(1-p) +\sqrt{e^{kp}-1}\right]
		\qquad &  C^2\leqslant \tilde{C}^2_{1}\\[0ex]
		W_{FF2}^* &= \frac{\tau_S}{\sqrt{k}} \left[Ck(1-p) +C(1-\ln(2)) +C\ln \left(1+(e^{kp}-1)/C^2 \right)\right]
		\qquad &  \tilde{C}^2_{1}<C^2< \tilde{C}^2_{2}\\[0ex]
		W_{FF4}^* &= \frac{\tau_S}{\sqrt{k}} \sqrt{C^2(e^{k(1-p)}-1) + (e^k - e^{k(1-p)})}
		\qquad &  \tilde{C}^2_{2}\leqslant C^2 \;,
	\end{array}
\end{align}
where $\tilde{C}^2_{1}\equiv e^{kp}-1$ and $\tilde{C}^2_{2}\equiv (e^{kp} - 1)/(2e^{-k(1-p)}-1)$.

\section{Proofs}\label{app:proofs}
There are 13 different combinations of FF and SF subfunctions which we treat in the subsections below in order to show that $W^*_{FF}-W^*_{SF}>0$ (for $k<\ln(2)$ there are 3 combinations, for
$k>\ln(2)\cap kp>\ln(2)$ there are 2 combinations and for
$k>\ln(2)\cap kp<\ln(2)$ there are 8 combinations).

\subsection{Proof that $W^*_{FF}-W^*_{SF}>0$ when $k<\ln(2)$}
Assume $0<k<\ln(2)$ and $p,C\in(0,1)$. The $(p,C)$-subdomains for the subfunctions of $W^*_{SF}$ and $W^*_{FF}$ are illustrated in the diagrams in the leftmost column in Fig.~\ref{fig:SF_FF_activesubfunctions}. From \cref{eq:W*_SF} $W_{SF}^*=W_{SF4}^*=\frac{\tau_S}{\sqrt{k}} \sqrt{(e^{kp}-1)+C^2(e^k-e^{kp})}$, and $W_{FF}^*$ is given by \cref{eq:W*_FF} since $W^*_{FF1}$ cannot be a solution when $k<\ln(2)$. 
We now show that $W_{FF}^*-W_{SF}^*>0$ for all three cases.

\underline{FF2$_0$ vs. SF4.}  
Set $x\equiv k(1-p) \in (0,\ln(2))$. From \cref{eq:W*_FF}, $C^2\leqslant e^{kp}-1$. Then
\begin{align*}
	\frac{k}{\tau_S^2}(W_{FF2_0}^{*2} - W_{SF4}^{*2})
	&= C^2k^2(1-p)^2 + 2kC(1-p)\sqrt{e^{kp}-1} - C^2(e^k-e^{kp})\\[0ex]
	&\geqslant C^2 \left[k^2(1-p)^2 + 2k(1-p) - e^k(1-e^{-k(1-p)})\right]\\[0ex]
	&= C^2 \left[x^2 + 2x - e^k(1-e^{-x})\right]\\[0ex]
	&> C^2 \left[x^2 + 2x - 2(1-e^{-x})\right] \equiv C^2 g(x) \;,
\end{align*}
where we use that $C^2\leqslant e^{kp}-1$ and $k<\ln(2)$ in the first and second inequality, respectively.
$g(0)=0$, and since $x \in (0,\ln(2))$, $g'(x)=2x+2-2e^{-x}> 2x > 0$. Hence $g(x)> 0$ for $x>0$. QED.

\underline{FF2 vs. SF4.}  
Set $x\equiv (e^{kp}-1)/C^2$, and $z\equiv 2e^{-k(1-p)}-1$. Since $k<\ln(2)$,  $z\in(0,1)$. In \cref{eq:W*_FF} the lower bound $C^2 > e^{kp}-1$ gives $x\in(0,1)$. The upper bound $C^2< (e^{kp} - 1)/(2e^{-k(1-p)}-1)$ gives $x>z$. Then, since $k<\ln(2)$
\begin{align*}
	\frac{\sqrt{k}}{C \tau_S}(W_{FF2}^{*} - W_{SF4}^{*})
	&= k(1-p) +1 -\ln(2) +\ln \left(1+(e^{kp}-1)/C^2 \right)\\[0ex]
	&\qquad\qquad\qquad - \sqrt{(e^{kp}-1)/C^2 + e^k(1-e^{-k(1-p)})}\\[0ex]
	&= \ln(2)-\ln(z+1) +1 -\ln(2) + \ln(1+x) -\sqrt{x+e^k(1-z)/2}  \\[0ex]
	&> 1+\ln(1+x) - \ln(1+z) -\sqrt{1+x-z} \equiv g(x,z) \;.
\end{align*}
For $x=z$, $g(x,z)=0$. For any $z\in(0,1)$, $g(x,z)$ is increasing in $x$ since
\begin{align*}
	\frac{\partial g}{\partial x} = \frac{1}{1+x} - \frac{1}{2\sqrt{1+x-z}} 
	\quad>\quad \frac{1}{1+x} - \frac{1}{2} 
	\quad>\quad 0 \;.
\end{align*}
The first inequality is from $x>z$, and the second from $x<1$. This means that $g(x,z)> 0$ on the triangle domain $z\in(0,1)$ and $x\in(z,1)$. QED.

\underline{FF4 vs. SF4.}
It follows straightforwardly that
\begin{align*}
	\frac{k}{\tau_S^2}(W_{FF4}^{*2} - W_{SF4}^{*2})
	&= (1-C^2)(e^{kp}-1)(e^{k(1-p)}-1)>0 \;.
\end{align*}

\subsection{Proof that $W^*_{FF}-W^*_{SF}>0$ when $k>\ln(2)$ and $kp>\ln(2)$}\label{ass:proof_kpgtln1}
Assume $kp>\ln(2)$ and $p,C\in(0,1)$. Then $W_{SF}^*-W_{FF}^*$ is given by the two combinations in \cref{eq:FF1-SF1}.

\underline{FF1 vs. SF1.}  
Set $z\equiv \frac{\tau_S}{\tau_S+\tau_F} \in (1/2,1)$. We can rewrite \cref{eq:FF1-SF1}:
\begin{align}
	\frac{\sqrt{k}}{\tau_S+\tau_F}(W^*_{FF1} - W^*_{SF1})
	&= \frac{\tau_S}{\tau_S+\tau_F} \ln \left( \frac{\tau_S}{\tau_S+\tau_F} \right) \nonumber
	- \frac{\tau_F}{\tau_S+\tau_F} \ln \left( \frac{\tau_F}{\tau_S+\tau_F} \right)\\[0ex]
	&= z \ln z - (1-z) \ln (1-z) \equiv f(z) \;.
	\label{eq:fzx}
\end{align}
The function $f(z)$ vanishes at the ends of the interval, i.~e., $f(1/2)=\lim_{z \to 1^-} f(z) = 0$.
The extremum points are calculated from $f'(z)=0$ or 
\begin{equation}
	z(1-z) = e^{-2} \;.
	\label{eq:rootsx}
\end{equation}
Hence, there is a single extremum point on the interval, located at $z=z^*=\frac{1}{2} + \sqrt{\frac{1}{4}-e^{-2}}$.
Moreover, from \cref{eq:fzx,eq:rootsx} we obtain $f''(z^*) = e^2 (1-2z^*)<0$, so that this extremum is a maximum. According to these properties, $f(z)>0$ holds within $1/2 < z < 1$.

\underline{FF1 vs. SF2.}  
Set $x\equiv k(1-p) \in (0,\infty)$. The condition $C\leqslant C_1\equiv (e^{k(1-p)}-1)^{-1}$ gives that $x\leqslant\ln(1+C^{-1})\equiv x_u$. This means that $x\in (0,x_u]$. Then \cref{eq:FF1-SF2} can be written as
\begin{align*}
	\frac{\sqrt{k}}{\tau_S}(W_{FF1}^{*} - W_{SF2}^{*})
	&= Cx - \ln \left(1+C^2(e^{x}-1) \right) \equiv g(x,C) \;.
\end{align*}
It follows that $g(x,C)$ is positive on the interval $x\in(0,x_u]$, since for any fixed $C\in (0,1)$, $g(0,C)=0$, and $g(x,C)$ is monotonically increasing in $x$. The latter is due to 
\begin{align*}
	\frac{\partial g}{\partial x} &= C-\frac{C^2 e^x}{C^2(e^x-1)+1} 
	\quad=\quad C^2(1-C)\frac{(1+C^{-1})-e^x}{C^2(e^x-1)+1} \;.
\end{align*}
being positive for $x\in(0,x_u]$, except from in the upper end when $\frac{\partial g}{\partial x}(x_u,C)=0$.

\subsection{Proof that $W^*_{FF}-W^*_{SF}>0$ when $k>\ln(2)$ and $kp<\ln(2)$}
This is when FF2 and FF20 in Fig.~\ref{fig:SF_FF_activesubfunctions} overlaps with the corresponding parts of the SF-diagrams in the two rightmost columns of Fig.~\ref{fig:SF_FF_activesubfunctions}. This gives eight different combinations.

We will now show that the difference $W^*_{FF}-W^*_{SF}>0$ within each of these regions, mainly by considering the partial derivatives with respect to $p$. We use that the difference is smooth on the boundaries between the subdomains. This can be demonstrated quite straightforwardly by differentiation. Moreover, the difference $W^*_{FF}-W^*_{SF}=0$ on the boundaries of the parameter region $p\times C \in [0,1]\times [0,1]$.

The outline of the proof is as follows: Firstly, the difference is positive in the SF3 region, including on the SF3-SF1 and SF3-SF4 boundaries as seen in Fig.~\ref{fig:SF_FF_activesubfunctions}. Then we show that the difference increases in $p$ on the whole SF1-domain, starting out positively on the SF1-SF3 boundary. 
Now, the difference is positive at the SF3-SF4 border and SF1-SF2 border, and it is also positive when $p$ is at its maximum at $p=\ln(2)/k$ (at the FF1 border, see Sec.~\ref{ass:proof_kpgtln1}). We show that the difference has negative curvature in $p$ in both the SF2- and the SF4-regions. This gives that the difference must be positive for all values of $p$ on $p$-intervals starting at the SF3-SF4 or SF1-SF2 borders and ending at $p=\ln(2)/k$ (using that the difference is smooth between regions).

We first show that the difference is increasing in $p$ in the SF1-region. Then the negative curvature in the SF2- and SF4-regions is shown. To show that the difference is positive in the SF3-domain requires a number of subtleties and is left for the last subsection.

\subsubsection{SF1 subdomain}
We show that the differences are increasing when $p$ increases for all values of $k,p,C$ on the SF1-domain. 

\underline{FF2$_0$ vs. SF1.}  
From \cref{eq:W*_FF} and \cref{eq:W*_SF} we have that for all values of $k,p,C$
\begin{align*}
	\frac{\sqrt{k}}{\tau_S}(W_{FF2_0}^{*} - W_{SF1}^{*})
	&=\sqrt{e^{kp}-1}    -  kp - 1 - C\ln\left(\frac{C}{1+C}\right) + \ln\left(\frac{2}{1+C}\right)\\
	\frac{\sqrt{k}}{\tau_S}\frac{\partial (W_{FF2_0}^{*} - W_{SF1}^{*})}{\partial p}
	&=\frac{k(\sqrt{e^{kp}-1}-1)^2}{2\sqrt{e^{kp}-1}}  > 0.
\end{align*}

\underline{FF2 vs. SF1.}
For all values of $k,p,C$ satisfying the FF2-condition in \cref{eq:W*_FF}, $C^2>e^{kp}-1$, 
\begin{align*}
	\frac{\sqrt{k}}{\tau_S}(W_{FF2}^{*} - W_{SF1}^{*})
	&=C(1-\ln(2)) +C\ln \left(1+(e^{kp}-1)/C^2 \right) -  kp - 1 \\
	&\qquad\qquad\qquad- C\ln\left(\frac{C}{1+C}\right) +\ln\left(\frac{2}{1+C}\right)\\
	\frac{\sqrt{k}}{\tau_S}\frac{\partial (W_{FF2}^{*} - W_{SF1}^{*})}{\partial p}
	&= \frac{Cke^{kp}}{C^2 + e^{kp}-1} - k 
	= k(1-C)\frac{C - (e^{kp}-1)}{C^2 + (e^{kp}-1)}\\
	&\geqslant k(1-C)\frac{C^2 - (e^{kp}-1)}{C^2 + (e^{kp}-1)}>0.	
\end{align*}
In the last line $C\geqslant C^2$ and the FF2-condition is used in the first and last inequality, respectively.

\subsubsection{SF2 subdomain}
As for SF4 below we show that the curvature of $W^*_{FF}-W^*_{SF}$ is negative in $p$ for all values of $k,p,C$ on the SF2-domain. 

\underline{FF2$_0$ vs. SF2.}
For all values of $k,p,C$ satisfying the FF2/FF20-condition $kp<\ln(2)$, 
\begin{align*}
	\frac{\sqrt{k}}{\tau_S}(W_{FF20}^{*} - W_{SF2}^{*})
	&=Ck(1-p) +\sqrt{e^{kp}-1}  - kp -(1-\ln(2)) - \ln\left({1+C^2(e^{k(1-p)}-1)}\right)\\
	\frac{\sqrt{k}}{\tau_S}\frac{\partial^2 (W_{FF20}^{*} - W_{SF2}^{*})}{\partial p^2}
	&= -\frac{k^2 e^{kp}(2-e^{kp})}{4(e^{kp}-1)^{\frac{3}{2}}} 
	- \frac{k^2 C^2 (1-C^2)e^{-k(1-p)}}{\left(C^2 + e^{-k(1-p)}(1-C^2)\right)^2} <0,
\end{align*}
where the $kp<\ln(2)$ condition is used in the last inequality.

\underline{FF2 vs. SF2.}  
For all values of $k,p,C$
\begin{align*}
	\frac{\sqrt{k}}{\tau_S}(W_{FF2}^{*} - W_{SF2}^{*})
	&=Ck -(1+C)kp - (1-C)(1-\ln(2))\\
	&\qquad\qquad+ C\ln\left((C^2+e^{kp}-1)/C^2\right) - \ln\left({1+C^2(e^{k(1-p)}-1)}\right)
	\\
	\frac{\sqrt{k}}{\tau_S}\frac{\partial^2 (W_{FF2}^{*} - W_{SF2}^{*})}{\partial p^2}
	&=-k^2 C(1-C^2) \left[ \frac{e^{-kp}}{(1-e^{-kp}(1-C^2))^2}    
	+\frac{C e^{-k(1-p)}}{(C^2+e^{-k(1-p)}(1-C^2))^2}\right]\\
	&<0.
\end{align*}

\subsubsection{SF4 subdomain}
As for SF2 above we show that the curvature of the difference is negative for increasing $p$ on the SF4-domain. 

\underline{FF2$_0$ vs. SF4.}
For all values of $k,p,C$ satisfying the FF2/FF20-condition $kp<\ln(2)$,
\begin{align*}
	\frac{\sqrt{k}}{\tau_S}(W_{FF20}^{*} - W_{SF4}^{*})
	&= Ck(1-p) + \sqrt{e^{kp}-1} - \sqrt{(e^{kp}-1) + C^2(e^k - e^{kp})}\\
	\frac{\sqrt{k}}{\tau_S}\frac{\partial^2 (W_{FF20}^{*} - W_{SF4}^{*})}{\partial p^2}
	&=-\frac{k^2 e^{kp}(2-e^{kp})}{4(e^{kp}-1)^\frac{3}{2}}
	+ \frac{k^2 e^{kp}(1-C^2)}{4} \cdot 
	\frac{(2-e^{kp})-C^2(2e^k-e^{kp})}{(e^{kp}-1 + C^2(e^k-e^{kp}))^\frac{3}{2}}\\
	&<-\frac{k^2 e^{kp}(1-C^2)}{4} \cdot
	\frac{C^2 (2e^k-e^{kp})}{(e^{kp}-1 + C^2(e^k-e^{kp}))^\frac{3}{2}}<0,
\end{align*}
where the $kp<\ln(2)$ condition is used in the first inequality in the last line.

\underline{FF2 vs. SF4.}  
For all values of $k,p,C$ satisfying the FF2/FF20-condition $kp<\ln(2)$,
\begin{align*}
	\frac{\sqrt{k}}{\tau_S}(W_{FF2}^{*} - W_{SF4}^{*})
	&= Ck(1-p) +C(1-\ln(2)) +C\ln \left(1+(e^{kp}-1)/C^2 \right)\\
	&\qquad\qquad\qquad -\sqrt{(e^{kp}-1) + C^2(e^k - e^{kp})}\\
	\frac{\sqrt{k}}{\tau_S}\frac{\partial^2 (W_{FF2}^{*} - W_{SF4}^{*})}{\partial p^2}
	&=-\frac{k^2e^{kp}(1-C^2)}{C^3}\left[
	\frac{1}{(\frac{e^{kp}-1}{C^2} + 1)^2}
	- \frac{(1-C^2)\left(2-e^{kp}\right) - 2C^2(e^{k}-1)}{4((1-C^2)\frac{e^{kp}-1}{C^2} + (e^k-1))^{\frac{3}{2}}}\right]\\
	&\leqslant
	-\frac{k^2e^{kp}(1-C^2)}{C^3}\left[
	\frac{1}{4}
	- \frac{(1-C^2)\left(2-e^{kp}\right)}{4(e^k-1)}\right]\\
	&=
	-\frac{k^2e^{kp}(1-C^2)}{4C^3}\left[
	\frac{(e^k-1)-(1-C^2)\left(1-(e^{kp}-1)\right)}{e^k-1}\right]<0.
\end{align*}
In the first inequality we use the FF2-condition $C^2>e^{kp}-1$ in the first fraction. In the second fraction, we first remove the negative term in the numerator, then the exponent in the denominator is set to 1 (the denominator is larger than 1 since $k>\ln(2)$), and finally the first positive term in the denominator is removed. 
In the last inequality, we use that $k>\ln(2)$ and $(1-C^2)(1-(e^{kp}-1))<1$.

\subsubsection{SF3 subdomain}
Set $y\equiv e^{kp}-1\in (0,1)$, since $kp<\ln(2)$. 
Set $r\equiv C/\sqrt{(e^{kp}-1)(1-C^2)}=C/\sqrt{y(1-C^2)}>0$. 
Set $d\equiv \sqrt{\delta_3^*}$ in \cref{eq:delta_SF3}, such that
\begin{align}\label{eq:SF3d}
	d &=\frac{\sqrt{(e^{kp}-1)(1-C^2)}  - C(e^{kp}-1)}{\sqrt{(e^{kp}-1)(1-C^2)}  + C}
	= \frac{1-ry}{1+r}
	<\frac{1}{1+r},
\end{align}
where $d\in(0,1)$ since $\delta_3^*\in(0,1)$. The inequality is due to $ry>0$ and will be used later in the proof.

\underline{FF2$_0$ vs. SF3.} 
The SF3-condition (vs SF1) $C<({2e^{-kp}-1})$ converts into $y<1/(2r+1)$, while the FF20-condition $C\leqslant \sqrt{e^{kp}-1}$ in \cref{eq:W*_FF} becomes $y\geqslant(r^2-1)/r^2$. This gives that $r\in(0,r^*)$, where $r^*\approx 1.19$ is the solution of $2r^3 - 2r - 1=0$. From \cref{eq:W*_FF} and \cref{eq:W*_SF},
\begin{align*}
	\frac{\sqrt{k}}{\tau_S}(W_{FF20}^{*} - W_{SF3}^{*})
	&=\sqrt{e^{kp}-1} - \sqrt{\frac{(e^{kp}-1)(e^{kp}-1 +d^2)}{e^{kp}-1 + 1}} 
	-C\left[1-\ln(2) +\ln\left({1-d}\right)+d\right]\\
	&> \sqrt{y}\left(1 - \sqrt{\frac{y + d^2}{y + 1}} \right)  
	-C[1-\ln(2)]\\
	&> \sqrt{y}\left(1 - 
	\sqrt{\frac{y + \frac{1}{(1+r)^2}}{y + 1}} \right)  -\sqrt{y}\sqrt{1-C^2}\frac{C}{\sqrt{y(1-C^2)}}[1-\ln(2)]\\		
	&> \sqrt{y}\left[\left(1 - 
	\sqrt{\frac{y + \frac{1}{(1+r)^2}}{y + 1}} \right)  
	- r[1-\ln(2)]\right]
	\equiv \sqrt{y}\cdot H(r,y).		
\end{align*}
In the first inequality we use that $(\ln(1-d)+d)< 0$ for $d\in(0,1)$ (easily shown by differentiation). In the second inequality, we use that $d<1/(1+r)$ in \cref{eq:SF3d}. In the third inequality, we use that $\sqrt{1-C^2}<1$ in the last term.

We now show that for fixed $r$, $H(r,y)$ is decreasing when $y$ increases from $y=0$ towards the SF3-SF1 border where $y=1/(2r+1)$. We then show that $H(r,y)>0$ on the SF3-SF1 border, which means that $H(r,y)>0$ on the whole SF3/FF20-domain. 
\begin{align*}
	\frac{\partial H}{\partial y}
	= -\frac{1}{2}\sqrt{\frac{y+1}{y + a^2}}\cdot \frac{(1-a^2)}{(y + 1)^2}<0,
\end{align*}
where $a\equiv 1/(1+r)<1$. At the SF3-SF1 border,
\begin{align*}
	H(r,y=1/(2r+1))
	= 1- \sqrt{\frac{2+4r+r^2}{2(1+r)^3}}  -r(1-\ln(2)) \equiv h(r),
\end{align*}
and
\begin{align*}
	h'(r) = \frac{1}{2}\sqrt{\frac{1+r}{1+2r+0.5r^2}} \cdot \frac{1+3r+0.5r^2}{1+3r + 3r^2 + r^3}  -(1-\ln(2)).
\end{align*}
Both denominators are increasing faster in $r$ than their respective numerators, which gives that $h''(r)<0$. Moreover, $h(0)=0$ and $h(1.20)=0.0097>0$. This means that $h(r)>0$ on the whole range of $r\in (0,r^*)$.

\underline{FF2 vs. SF3.}  
Set $R\equiv\sqrt{y}/C = \sqrt{e^{kp}-1}/C>0$, such that $r=1/\sqrt{R^2-y}$. The FF2-condition $C> \sqrt{e^{kp}-1}$ in \cref{eq:W*_FF} converts into $R < 1$, while the SF3-condition $C<({2e^{-kp}-1})$ becomes $R>\sqrt{y}(1+y)/(1-y)$.
This gives that $y\in(0,y^*)$, where $y^*\approx 0.30$ is the solution of $y^3 +y^2 +3y - 1=0$.

From \cref{eq:W*_FF} and \cref{eq:W*_SF}, 
\begin{align*}
	\frac{\sqrt{k}}{\tau_S}(W_{FF2}^{*} - W_{SF3}^{*})
	&= C\ln \left(1+\frac{e^{kp}-1}{C^2}\right)-\sqrt{\frac{(e^{kp}-1)(e^{kp}-1 +d^2)}{e^{kp}-1 + 1}} 
	-  C\left[d + \ln(1-d)\right] \\  
	&> C\ln \left(1+\frac{y}{C^2}\right)-\sqrt{y}\sqrt{\frac{y+d^2}{y+1}}\\  
	&> C\ln \left(1+R^2\right)-C\frac{\sqrt{y}}{C}\sqrt{\frac{y+\frac{1}{(1+r)^2}}{y+1}}\\  
	&= C\left( \ln \left(1+R^2\right)-R\sqrt{\frac{y+\frac{1}{\left(1+({R^2-y})^{-0.5}\right)^2}}{y+1}}\right) 
	\equiv C\cdot G(R,y).   
\end{align*}
In the first inequality we use that $(\ln(1-d)+d)< 0$ for $d\in(0,1)$. In the second inequality, we use that $d<1/(1+r)$ in \cref{eq:SF3d}.

We now show that for fixed $R$, $G(R,y)$ is decreasing when $y$ increases from $y=0$ towards the SF3-SF1 border where $R=\sqrt{y}(1+y)/(1-y)$. We then show that $G(R,y)>0$ on the SF3-SF1 border, which means that $G(R,y)>0$ on the whole SF3/FF2-domain.

Set $K(R,y)\equiv \sqrt{R^2-y}$. Then the expression inside the square root in $G(R,y)$ can be written
\begin{align*}
	F(y,K)&\equiv \frac{y+{\left(1+\frac{1}{K}\right)^{-2}}}{y+1}\\
	\frac{d F}{d y}
	&=\frac{\partial F}{\partial y} + \frac{\partial F}{\partial K}\frac{\partial K}{\partial y} =\frac{(K+1)(2K+1)-\sqrt{y}(y+1)}{(y+1)^2(K+1)^3} > 0.
\end{align*}
The inequality is due to $(K+1)>1>\sqrt{y}$ and $(2K+1)>(y+1)$ in the nominator. The latter stems from the SF3-condition:
\begin{align*}
	K^2 &=  R^2-y  > \frac{y(1+y)^2}{(1-y)^2} - y > y(1+y) - y = y^2.
\end{align*}
Since $\frac{d F}{d y}>0$,
\begin{align*}
	\frac{\partial G}{\partial y}
	&=-R\frac{\frac{d F}{d y}}{2\sqrt{F(R,y)}}<0.
\end{align*}

It remains to show that $G(R,y)>0$ on the SF3-SF1 border:
\begin{align*}
	G(R=\frac{\sqrt{y}(1+y)}{(1-y)},y)
	&= \ln \left(1+R^2\right) - \frac{y}{1-y}\sqrt{1+y+\frac{4y}{y+1}} \equiv g(y),
\end{align*}
where $y \in [0,y^*]$ and $y^* \approx 0.3$.
The condition $g(y) \ge 0$ is equivalent to $f(y) \equiv \exp(g(y)) - 1 \ge 0$. 
We use $e^{-x} \ge 1 - x + \frac{x^2}{2} - \frac{x^3}{6}$, which yields
\begin{align}
	f(y) &\ge R^2 + \left( 1 + R^2 \right) \left( - \frac{y \sqrt{Q(y)}}{1-y}  
	+ \frac{1}{2} \frac{y^2 Q(y)}{(1-y)^2} - \frac{1}{6} \frac{y^3 Q(y)^{3/2}}{(1-y)^3}\right) \nonumber\\
	&\ge R^2 + \left( 1 + R^2 \right) \left( - \frac{y \sqrt{Q(y)}}{1-y}  
	+ \frac{1}{2} \frac{y^2 Q(y)}{(1-y)^2} (1- \mathcal{A} y) \right) \;,
	\label{eq:fy2}
\end{align}
where $Q(y)\equiv 1+y+ \frac{4y}{1+y}$ and 
$\mathcal{A} \equiv \frac{\sqrt{Q(y^*)}}{3(1-y^*)} = 
\mathrm{max} \left( \frac{\sqrt{Q(y)}}{3(1-y)} \right) \approx 0.71$.
Inserting $R^2 = y \left( \frac{1+y}{1-y} \right)^2$ into \cref{eq:fy2} and  multiplying
by $(1-y)^4 (1+y)/y$, we obtain the condition
\begin{eqnarray*}
	&& (1+y)^3 (1-y)^2 + (1-y+3y^2+y^3) \left[ (y-1) \sqrt{(1+y)^3 + 4y(1+y)} \right. \nonumber \\
	&+& \left. \frac{y}{2} \left( 1+6y + y^2 \right) (1- \mathcal{A} y) \right] \ge 0
\end{eqnarray*}
for $f(y) \ge 0$. We move the term with square root to the right hand side and
then take square of the expressions on both sides of the inequality. Since
$1- \mathcal{A}y>0$ and $1-y+3y^2+y^3 = (1+R^2)(1-y)^2 >0$, the resulting inequality is
the sufficient condition for $f(y) \ge 0$. It can be written as
$\phi(y) \ge 0$, where
\begin{align}
	&\hspace*{-6ex} \phi(y) \equiv \left[ (1+y)^3 (1-y)^2 + \frac{y}{2} \left( 1-y+3y^2+y^3 \right) 
	\left( 1 + 6y + y^2 \right) (1- \mathcal{A} y) \right]^2 \nonumber \\
	&- (1-y)^2 \left( 1-y+3y^2+y^3 \right)^2 \left( 1+7y+7y^2+y^3 \right) \nonumber\\
	&= \left[ 1+ \frac{3y}{2} +  \frac{1 -\mathcal{A}}{2} \, y^2 
	- \left(3+ \frac{5 \mathcal{A}}{2} \right) y^3 + (10+ \mathcal{A}) y^4 
	+ \left( \frac{11}{2} - 9 \mathcal{A} \right) y^5 \right. \nonumber \\
	&+ \left. \frac{1 - 9 \mathcal{A}}{2} \, y^6 - \frac{\mathcal{A} y^7}{2} \right]^2
	- \left(1 -4y + 12 y^2 - 20 y^3 + 22 y^4 -12 y^5 \right. \nonumber \\ 
	&- \left. 4 y^6 + 4 y^7 + y^8 \right) \left( 1+7y+7y^2+y^3 \right) \label{uuu} \\
	&\ge \left[ 1+ \frac{3y}{2} + a y^2 - b y^3 + c y^4 \right]^2 \nonumber \\
	&- \left( 1 -4y + 12 y^2 - 20 y^3 + 22 y^4 \right) \left( 1+7y+7y^2+y^3 \right) \;,
	\label{aaa}
\end{align}
where $a\equiv\frac{1}{2} (1 -\mathcal{A})$, $b\equiv 3+ \frac{5 \mathcal{A}}{2}$
and $c\equiv 10 + \mathcal{A} + \left( \frac{11}{2} - 9 \mathcal{A} \right) y^*
+ \frac{1}{2} (1- 9 \mathcal{A}) {y^*}^2 - \frac{1}{2} \mathcal{A} {y^*}^3$.
The inequality in \cref{aaa} is obtained replacing $y^{4+n}$ by ${y^*}^n y^4$ for $n=1,2,3$ 
in square brackets of \cref{uuu} (the corresponding coefficients of these terms are negative,
and the resulting expression in the brackets becomes smaller but still positive,
since $a>0$, $b>0$, $c>0$ and $b y^3 <1$),
as well as omitting the terms $-12 y^5 + 4y^7 = 4 y^5 (-3+y^2) <0$ and
$-4 y^6 + y^8 = y^6 (-4+y^2) <0$ in the following brackets.

Further exact calculation of \cref{aaa} yields
\begin{eqnarray}
	\phi(y) &\ge& y^2 \left[ A_0 - A_1 z + A_2 z^2 + A_3 z^3 - A_4 z^4
	- A_5 z^5 + c^2 y^6 \right] \label{eq:ineq1} \nonumber\\
	&\ge& y^2 \left[ A_0 - A_1 z + B z^2  \right] = y^2 \, \varphi(z)\;,
	\label{ineq}
\end{eqnarray}
where $z=y/y^*$,  $A_0 \equiv \frac{45}{4} + 2a$, $A_1\equiv (37+2b-3a) y^*$,
$A_2 \equiv (a^2-3b+2c+38) {y^*}^2$, $A_3\equiv (3c-2ab-26) {y^*}^3$,
$A_4\equiv (134-b^2-2ac) {y^*}^4$, $A_5=(22+2bc) {y^*}^5$
and $B\equiv A_2-A_4-A_5$. 
The inequality~(\ref{ineq}) holds for $y^* = 0.3$ because the
coefficients $A_n$ are positive in this case and $z \le 1$.
The values of coefficients at $y^*=0.3$ are
$a \approx 0.145$, $b \approx 4.775$, $c \approx 10.1909$,
$A_0  \approx 11.54$, $A_1 \approx 13.8345$, $A_2 \approx 3.967$,
$A_3 \approx 0.0861$, $A_4 \approx 0.8768$, $A_5 \approx 0.29$
and $B \approx 2.8003$. 

The function $\varphi(z)$
has a single extremum (minimum) at $z= \frac{A_1}{2B} \approx 2.47 \not\in [0,1]$,
so that it changes monotonously from $\varphi(0) = A_0 \approx 11.54$ to $\varphi(1)=A_0-A_1+B \approx 0.5058$ 
within $z \in [0,1]$, implying that $\varphi(z) >0$ holds within this interval.
Hence, $g(y) \ge 0$ holds for $y \in [0,y^*]$.

\subsection{Proof of maximum relative distance between Fast-First and Slow-First}
We propose that the maximum relative distance $(W^*_{FF}-W^*_{SF})/W^*_{SF}=W^*_{FF}/W^*_{SF}-1$ is in the intersection of the SF4- and the FF20-region.
We therefore seek the maximum of:
\begin{align*}
\frac{W^*_{FF2_0}}{W^*_{SF4}}
&=\frac{\sqrt{e^{kp}-1}+Ck(1-p)}{\sqrt{(e^{kp}-1) + C^2(e^k - e^{kp})}}
=\frac{\sqrt{e^{kp}-1} +\sqrt{C^2(e^k - e^{kp})}\frac{k(1-p)}{\sqrt{(e^k - e^{kp})}}}
{\sqrt{(e^{kp}-1) + C^2(e^k - e^{kp})}}\\[2ex]
&=\frac{x+yd}{\sqrt{x^2 + y^2}}
\equiv g(x,y,d),
\end{align*}
where
\begin{align}\label{eq:xyd-defintions}
x &\equiv \sqrt{e^{kp}-1}\nonumber \\
y &\equiv \sqrt{C^2(e^k - e^{kp})}\\
d &\equiv \frac{k(1-p)}{\sqrt{e^k - e^{kp}}}. \nonumber
\end{align}
For fixed $d$, $g(x,y,d)$ is maximized when $y=dx$ (giving $\frac{\partial g}{\partial x}=0=\frac{\partial g}{\partial y}$). This gives $g(x,y=dx,d)=\sqrt{1+d^2}$, which is maximized by maximizing $d$. For fixed $k$, $d=d(k,p)$ is decreasing in $p$ since, when $kp<\ln(2)$,
\begin{align*}
\frac{\partial d}{\partial p} 
&= -\frac{k}{2(e^k-e^{kp})^{\frac{3}{2}}}\left[2(e^k -e^{kp}) - e^{kp}k(1-p)\right]\\
&< -\frac{k}{2(e^k-e^{kp})^{\frac{3}{2}}}\left[2(e^k -k) - 2(e^{kp}-kp)\right] <0. 
\end{align*}
However, $p=0$ implies that $x=0=y$, which is not in the domain of $g$. This means that $p$ should be small, but positive. For fixed $p$, $d$ is maximized by solving
\begin{align}\label{eq:partial_dk}
\frac{\partial d}{\partial k}  
&= \frac{(1-p)}{2(e^k-e^{kp})^{\frac{3}{2}}}\left[(2-k)e^k - (2-kp)e^{kp} \right]=0.
\end{align}
This equation has only one solution, $k=k(p)$, when we restrict to $k>0$.  
For $p=0$, $k(0)=k^*\approx 1.5936$. Moreover, $k(p)$ is decreasing in $p$ towards $k=1$ when $p\rightarrow 1$.  For small $p$, the solution of \cref{eq:partial_dk} can be approximated by $k(p)\approx k^* - pk^*/(e^{k^*}-2)\approx k^* -0.55p$. 
In the limit, when $p\rightarrow 0$, the maximum value of $g$ is given by:
\begin{align*}
\frac{W^*_{FF2_0}}{W^*_{SF4}}
&=\sqrt{1+d^2(k^*,p)} 
=\sqrt{1+\frac{k^{*2}(1-p)^2}{{e^{k^*} - e^{k^*p}}}}
\quad \overset{p\rightarrow 0}{\longrightarrow} \quad 
\sqrt{1+\frac{k^{*2}}{e^{k^*}-1}}
\approx 1.28359.
\end{align*}

The value of $C$ giving this maximum is found by setting $y=dx$, and using the expressions in \cref{eq:xyd-defintions},
which gives
\begin{align*}
	C&= \frac{k(1-p)}{e^k - e^{kp}}\sqrt{e^{kp}-1}
	\approx \frac{k^\frac{3}{2}}{e^k - 1}\sqrt{p} \approx 0.513\sqrt{p}.
\end{align*}
The approximation holds for small $p$, and $k=k^*$ in the last expression. The solution $(k,p,C)=(k^*,p,0.513\sqrt{p})$, where $p\approx 0$, is within the domains of FF20 and SF4.

\clearpage
\section{Supplementary figures}\label{app:figures}
\begin{figure}[htb]
	\begin{center}
		\includegraphics[width=7cm]{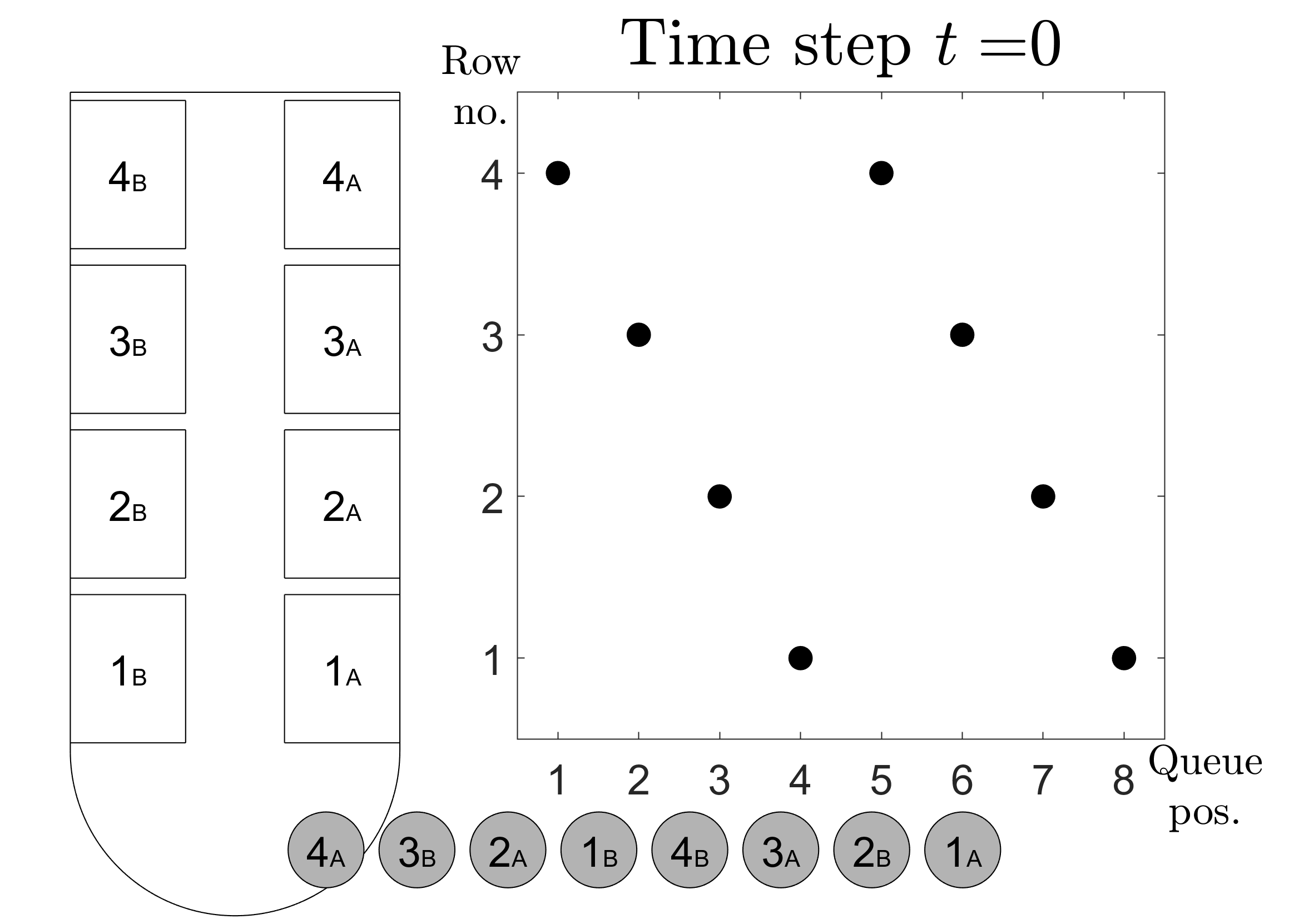}
		\parbox[b][3.6cm][c]{1cm}{$\quad$}
		\includegraphics[width=7cm]{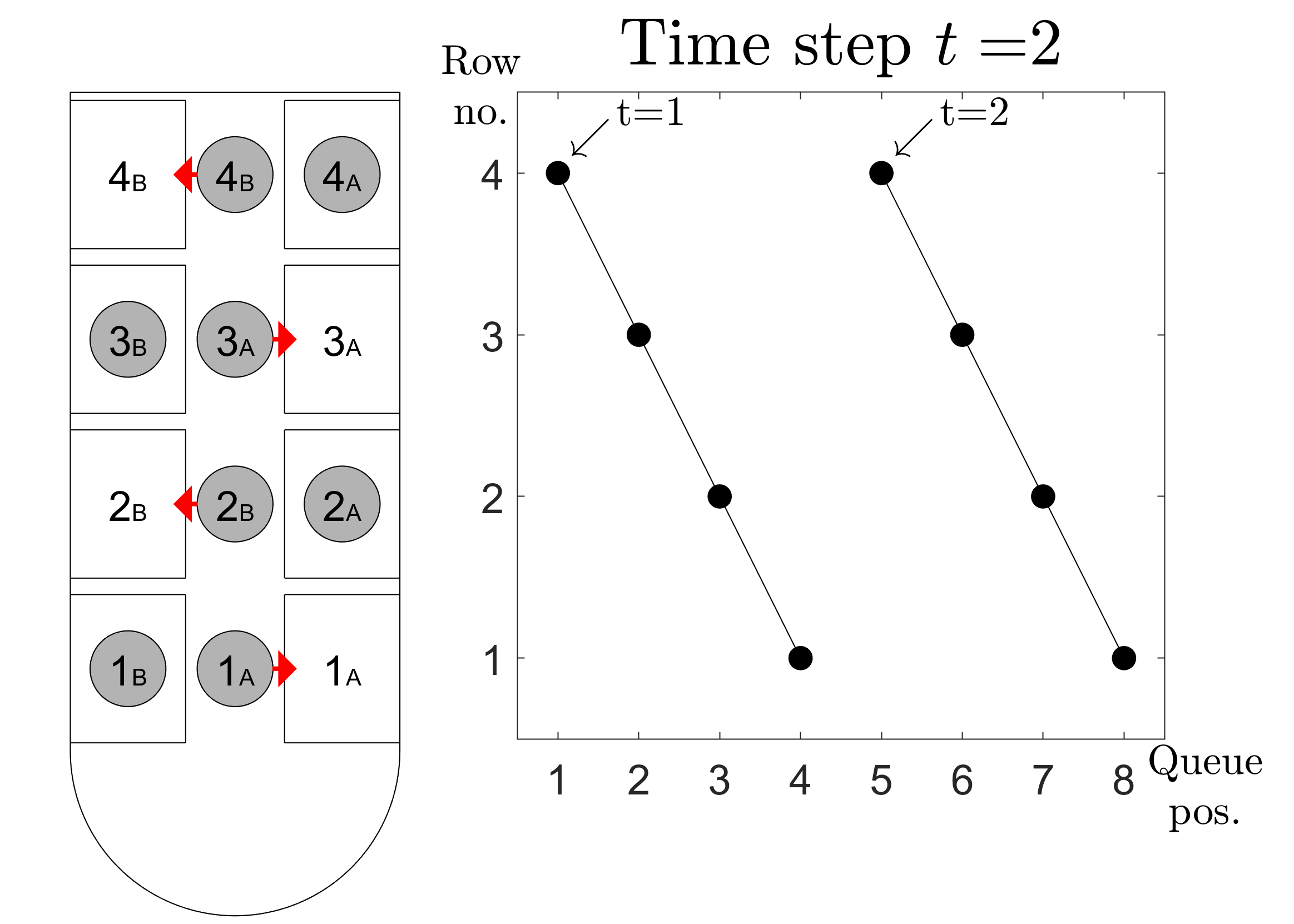}\\
		\mbox{\makebox[7.5cm][c]{(a)}\makebox[7.5cm][c]{(b)}}
	\end{center}
	\caption{\label{fig:boarding_illustration_optimal} (a) Initial position of the optimal queue ordering, with $N=8$ passengers, $4$ rows and $2$ seats per row. (b) The symmetry of the optimal queue ordering enables four passengers to sit down in each wave, which gives a minimal boarding time of $T=2$. Time step $t=1$ is not explicitly shown.}
\end{figure}

\bibliography{ABbib}

\begin{thebibliography}{24}%
\makeatletter
\providecommand \@ifxundefined [1]{%
 \@ifx{#1\undefined}
}%
\providecommand \@ifnum [1]{%
 \ifnum #1\expandafter \@firstoftwo
 \else \expandafter \@secondoftwo
 \fi
}%
\providecommand \@ifx [1]{%
 \ifx #1\expandafter \@firstoftwo
 \else \expandafter \@secondoftwo
 \fi
}%
\providecommand \natexlab [1]{#1}%
\providecommand \enquote  [1]{``#1''}%
\providecommand \bibnamefont  [1]{#1}%
\providecommand \bibfnamefont [1]{#1}%
\providecommand \citenamefont [1]{#1}%
\providecommand \href@noop [0]{\@secondoftwo}%
\providecommand \href [0]{\begingroup \@sanitize@url \@href}%
\providecommand \@href[1]{\@@startlink{#1}\@@href}%
\providecommand \@@href[1]{\endgroup#1\@@endlink}%
\providecommand \@sanitize@url [0]{\catcode `\\12\catcode `\$12\catcode
  `\&12\catcode `\#12\catcode `\^12\catcode `\_12\catcode `\%12\relax}%
\providecommand \@@startlink[1]{}%
\providecommand \@@endlink[0]{}%
\providecommand \url  [0]{\begingroup\@sanitize@url \@url }%
\providecommand \@url [1]{\endgroup\@href {#1}{\urlprefix }}%
\providecommand \urlprefix  [0]{URL }%
\providecommand \Eprint [0]{\href }%
\providecommand \doibase [0]{http://dx.doi.org/}%
\providecommand \selectlanguage [0]{\@gobble}%
\providecommand \bibinfo  [0]{\@secondoftwo}%
\providecommand \bibfield  [0]{\@secondoftwo}%
\providecommand \translation [1]{[#1]}%
\providecommand \BibitemOpen [0]{}%
\providecommand \bibitemStop [0]{}%
\providecommand \bibitemNoStop [0]{.\EOS\space}%
\providecommand \EOS [0]{\spacefactor3000\relax}%
\providecommand \BibitemShut  [1]{\csname bibitem#1\endcsname}%
\let\auto@bib@innerbib\@empty
\bibitem [{\citenamefont {Bachmat}\ \emph {et~al.}(2006)\citenamefont
  {Bachmat}, \citenamefont {Berend}, \citenamefont {Sapir}, \citenamefont
  {Skiena},\ and\ \citenamefont
  {Stolyarov}}]{Bachmat/Berend/Sapir/Skiena/Stolyarov:2006}%
  \BibitemOpen
  \bibfield  {author} {\bibinfo {author} {\bibfnamefont {E.}~\bibnamefont
  {Bachmat}}, \bibinfo {author} {\bibfnamefont {D.}~\bibnamefont {Berend}},
  \bibinfo {author} {\bibfnamefont {L.}~\bibnamefont {Sapir}}, \bibinfo
  {author} {\bibfnamefont {S.}~\bibnamefont {Skiena}}, \ and\ \bibinfo {author}
  {\bibfnamefont {N.}~\bibnamefont {Stolyarov}},\ }\href {\doibase
  10.1088/0305-4470/39/29/L01} {\bibfield  {journal} {\bibinfo  {journal} {J.
  Phys. A: Math. Gen.}\ }\textbf {\bibinfo {volume} {39}},\ \bibinfo {pages}
  {L453} (\bibinfo {year} {2006})}\BibitemShut {NoStop}%
\bibitem [{\citenamefont {Frette}\ and\ \citenamefont
  {Hemmer}(2012)}]{Frette/Hemmer:2012}%
  \BibitemOpen
  \bibfield  {author} {\bibinfo {author} {\bibfnamefont {V.}~\bibnamefont
  {Frette}}\ and\ \bibinfo {author} {\bibfnamefont {P.~C.}\ \bibnamefont
  {Hemmer}},\ }\href {\doibase 10.1103/PhysRevE.85.011130} {\bibfield
  {journal} {\bibinfo  {journal} {Phys. Rev. E}\ }\textbf {\bibinfo {volume}
  {85}},\ \bibinfo {pages} {011130} (\bibinfo {year} {2012})}\BibitemShut
  {NoStop}%
\bibitem [{\citenamefont {Bernstein}(2012)}]{Bernstein:2012}%
  \BibitemOpen
  \bibfield  {author} {\bibinfo {author} {\bibfnamefont {N.}~\bibnamefont
  {Bernstein}},\ }\href {\doibase 10.1103/PhysRevE.86.023101} {\bibfield
  {journal} {\bibinfo  {journal} {Phys. Rev. E}\ }\textbf {\bibinfo {volume}
  {86}},\ \bibinfo {pages} {023101} (\bibinfo {year} {2012})}\BibitemShut
  {NoStop}%
\bibitem [{\citenamefont {Bachmat}\ \emph {et~al.}(2013)\citenamefont
  {Bachmat}, \citenamefont {Khachaturov},\ and\ \citenamefont
  {Kuperman}}]{Bachmat/Khachaturov/Kuperman:2013}%
  \BibitemOpen
  \bibfield  {author} {\bibinfo {author} {\bibfnamefont {E.}~\bibnamefont
  {Bachmat}}, \bibinfo {author} {\bibfnamefont {V.}~\bibnamefont
  {Khachaturov}}, \ and\ \bibinfo {author} {\bibfnamefont {R.}~\bibnamefont
  {Kuperman}},\ }\href {\doibase 10.1103/PhysRevE.87.062805} {\bibfield
  {journal} {\bibinfo  {journal} {Phys. Rev. E}\ }\textbf {\bibinfo {volume}
  {87}},\ \bibinfo {pages} {062805} (\bibinfo {year} {2013})}\BibitemShut
  {NoStop}%
\bibitem [{\citenamefont {Baek}\ \emph {et~al.}(2013)\citenamefont {Baek},
  \citenamefont {Ha},\ and\ \citenamefont {Jeong}}]{Baek/Ha/Jeong:2013}%
  \BibitemOpen
  \bibfield  {author} {\bibinfo {author} {\bibfnamefont {Y.}~\bibnamefont
  {Baek}}, \bibinfo {author} {\bibfnamefont {M.}~\bibnamefont {Ha}}, \ and\
  \bibinfo {author} {\bibfnamefont {H.}~\bibnamefont {Jeong}},\ }\href
  {\doibase 10.1103/PhysRevE.87.052803} {\bibfield  {journal} {\bibinfo
  {journal} {Phys. Rev. E}\ }\textbf {\bibinfo {volume} {87}},\ \bibinfo
  {pages} {052803} (\bibinfo {year} {2013})}\BibitemShut {NoStop}%
\bibitem [{\citenamefont {Brics}\ \emph {et~al.}(2013)\citenamefont {Brics},
  \citenamefont {Kaupu\ifmmode~\check{z}\else \v{z}\fi{}s},\ and\ \citenamefont
  {Mahnke}}]{Martins/Kaupuzs/Mahnke:2013}%
  \BibitemOpen
  \bibfield  {author} {\bibinfo {author} {\bibfnamefont {M.}~\bibnamefont
  {Brics}}, \bibinfo {author} {\bibfnamefont {J.}~\bibnamefont
  {Kaupu\ifmmode~\check{z}\else \v{z}\fi{}s}}, \ and\ \bibinfo {author}
  {\bibfnamefont {R.}~\bibnamefont {Mahnke}},\ }\href {\doibase
  10.1103/PhysRevE.87.042117} {\bibfield  {journal} {\bibinfo  {journal} {Phys.
  Rev. E}\ }\textbf {\bibinfo {volume} {87}},\ \bibinfo {pages} {042117}
  (\bibinfo {year} {2013})}\BibitemShut {NoStop}%
\bibitem [{\citenamefont {Bachmat}(2014)}]{Bachmat:2014}%
  \BibitemOpen
  \bibfield  {author} {\bibinfo {author} {\bibfnamefont {E.}~\bibnamefont
  {Bachmat}},\ }\href {\doibase 10.1007/978-3-319-09513-4} {\emph {\bibinfo
  {title} {Mathematical Adventures in Performance Analysis: From Storage
  Systems, Through Airplane Boarding, to Express Line Queues}}}\ (\bibinfo
  {publisher} {Springer International Publishing},\ \bibinfo {address} {Cham},\
  \bibinfo {year} {2014})\ Chap.~\bibinfo {chapter} {3}\BibitemShut {NoStop}%
\bibitem [{\citenamefont {Mahnke}\ \emph {et~al.}(2015)\citenamefont {Mahnke},
  \citenamefont {Kaupu{\v{z}}s},\ and\ \citenamefont
  {Brics}}]{Mahnke/Kaupuzs/Brics:2015}%
  \BibitemOpen
  \bibfield  {author} {\bibinfo {author} {\bibfnamefont {R.}~\bibnamefont
  {Mahnke}}, \bibinfo {author} {\bibfnamefont {J.}~\bibnamefont
  {Kaupu{\v{z}}s}}, \ and\ \bibinfo {author} {\bibfnamefont {M.}~\bibnamefont
  {Brics}},\ }in\ \href {\doibase 10.1007/978-3-319-10629-8_37} {\emph
  {\bibinfo {booktitle} {Traffic and Granular Flow '13}}},\ \bibinfo {editor}
  {edited by\ \bibinfo {editor} {\bibfnamefont {M.}~\bibnamefont {Chraibi}},
  \bibinfo {editor} {\bibfnamefont {M.}~\bibnamefont {Boltes}}, \bibinfo
  {editor} {\bibfnamefont {A.}~\bibnamefont {Schadschneider}}, \ and\ \bibinfo
  {editor} {\bibfnamefont {A.}~\bibnamefont {Seyfried}}}\ (\bibinfo
  {publisher} {Springer International Publishing},\ \bibinfo {address} {Cham},\
  \bibinfo {year} {2015})\ pp.\ \bibinfo {pages} {305--314}\BibitemShut
  {NoStop}%
\bibitem [{\citenamefont {Bombelli}\ \emph {et~al.}(1987)\citenamefont
  {Bombelli}, \citenamefont {Lee}, \citenamefont {Meyer},\ and\ \citenamefont
  {Sorkin}}]{Bombelli/Lee/Meyer/Sorkin:1987}%
  \BibitemOpen
  \bibfield  {author} {\bibinfo {author} {\bibfnamefont {L.}~\bibnamefont
  {Bombelli}}, \bibinfo {author} {\bibfnamefont {J.}~\bibnamefont {Lee}},
  \bibinfo {author} {\bibfnamefont {D.}~\bibnamefont {Meyer}}, \ and\ \bibinfo
  {author} {\bibfnamefont {R.~D.}\ \bibnamefont {Sorkin}},\ }\href {\doibase
  10.1103/PhysRevLett.59.521} {\bibfield  {journal} {\bibinfo  {journal} {Phys.
  Rev. Lett.}\ }\textbf {\bibinfo {volume} {59}},\ \bibinfo {pages} {521}
  (\bibinfo {year} {1987})}\BibitemShut {NoStop}%
\bibitem [{\citenamefont {Myrheim}(1978)}]{myrheim:1978}%
  \BibitemOpen
  \bibfield  {author} {\bibinfo {author} {\bibfnamefont {J.}~\bibnamefont
  {Myrheim}},\ }\href {http://cdsweb.cern.ch/record/293594/files/197808143.pdf}
  {\enquote {\bibinfo {title} {Statistical geometry},}\ } (\bibinfo {year}
  {1978}),\ \bibinfo {note} {{CERN} preprint TH. 25 38-CERN}\BibitemShut
  {NoStop}%
\bibitem [{\citenamefont {t'Hooft}(1979)}]{tHooft:1979}%
  \BibitemOpen
  \bibfield  {author} {\bibinfo {author} {\bibfnamefont {G.}~\bibnamefont
  {t'Hooft}},\ }\enquote {\bibinfo {title} {Quantum gravity: A fundamental
  problem and some radical ideas},}\ in\ \href {\doibase
  10.1007/978-1-4613-2955-8_8} {\emph {\bibinfo {booktitle} {Recent
  Developments in Gravitation}}},\ \bibinfo {series} {NATO Advanced Study
  Institutes Series (Series B: Physics)}, Vol.~\bibinfo {volume} {44},\
  \bibinfo {editor} {edited by\ \bibinfo {editor} {\bibfnamefont
  {M.}~\bibnamefont {Lévy}}\ and\ \bibinfo {editor} {\bibfnamefont
  {S.}~\bibnamefont {Deser}}}\ (\bibinfo  {publisher} {Springer},\ \bibinfo
  {year} {1979})\ pp.\ \bibinfo {pages} {323--345}\BibitemShut {NoStop}%
\bibitem [{\citenamefont {Brightwell}\ and\ \citenamefont
  {Gregory}(1991)}]{Brightwell/Gregory:1991}%
  \BibitemOpen
  \bibfield  {author} {\bibinfo {author} {\bibfnamefont {G.}~\bibnamefont
  {Brightwell}}\ and\ \bibinfo {author} {\bibfnamefont {R.}~\bibnamefont
  {Gregory}},\ }\href {\doibase 10.1103/PhysRevLett.66.260} {\bibfield
  {journal} {\bibinfo  {journal} {Phys. Rev. Lett.}\ }\textbf {\bibinfo
  {volume} {66}},\ \bibinfo {pages} {260} (\bibinfo {year} {1991})}\BibitemShut
  {NoStop}%
\bibitem [{\citenamefont {Jaehn}\ and\ \citenamefont
  {Neumann}(2015)}]{Jaehn/Neumann:2015}%
  \BibitemOpen
  \bibfield  {author} {\bibinfo {author} {\bibfnamefont {F.}~\bibnamefont
  {Jaehn}}\ and\ \bibinfo {author} {\bibfnamefont {S.}~\bibnamefont
  {Neumann}},\ }\href {\doibase 10.1016/j.ejor.2014.12.008} {\bibfield
  {journal} {\bibinfo  {journal} {Eur. J. Oper. Res.}\ }\textbf {\bibinfo
  {volume} {244}},\ \bibinfo {pages} {339 } (\bibinfo {year}
  {2015})}\BibitemShut {NoStop}%
\bibitem [{\citenamefont {Neumann}(2019)}]{neumann:2019}%
  \BibitemOpen
  \bibfield  {author} {\bibinfo {author} {\bibfnamefont {S.}~\bibnamefont
  {Neumann}},\ }\href {\doibase 10.1016/j.ejor.2019.02.001} {\bibfield
  {journal} {\bibinfo  {journal} {Eur. J. Oper. Res.}\ }\textbf {\bibinfo
  {volume} {277}},\ \bibinfo {pages} {128 } (\bibinfo {year}
  {2019})}\BibitemShut {NoStop}%
\bibitem [{\citenamefont {Reed}(2013)}]{Reed:2013}%
  \BibitemOpen
  \bibfield  {author} {\bibinfo {author} {\bibfnamefont {T.}~\bibnamefont
  {Reed}},\ }\href
  {https://www.forbes.com/sites/tedreed/2013/05/18/will-americans-new-boarding-process-work-it-failed-at-virgin-america/#376878b8480c}
  {\enquote {\bibinfo {title} {Will {A}merican's new boarding process work?
  {I}t failed at {V}irgin {A}merica},}\ } (\bibinfo {year} {2013}),\ \bibinfo
  {note} {[Online; posted 15-May-2013]}\BibitemShut {NoStop}%
\bibitem [{\citenamefont {Audenaert}\ \emph {et~al.}(2009)\citenamefont
  {Audenaert}, \citenamefont {Verbeeck},\ and\ \citenamefont
  {Berghe}}]{Audenaert/Verbeeck/Berghe:2009}%
  \BibitemOpen
  \bibfield  {author} {\bibinfo {author} {\bibfnamefont {J.}~\bibnamefont
  {Audenaert}}, \bibinfo {author} {\bibfnamefont {K.}~\bibnamefont {Verbeeck}},
  \ and\ \bibinfo {author} {\bibfnamefont {G.}~\bibnamefont {Berghe}},\ }in\
  \href {http://wwwis.win.tue.nl/bnaic2009/papers/bnaic2009_paper_38.pdf}
  {\emph {\bibinfo {booktitle} {The 21st Benelux Conference on Artificial
  Intelligence}}}\ (\bibinfo {year} {2009})\ pp.\ \bibinfo {pages}
  {3--10}\BibitemShut {NoStop}%
\bibitem [{\citenamefont {Milne}\ and\ \citenamefont
  {Kelly}(2014)}]{Milne/Kelly:2014}%
  \BibitemOpen
  \bibfield  {author} {\bibinfo {author} {\bibfnamefont {R.~J.}\ \bibnamefont
  {Milne}}\ and\ \bibinfo {author} {\bibfnamefont {A.~R.}\ \bibnamefont
  {Kelly}},\ }\href {\doibase 10.1016/j.jairtraman.2013.08.006} {\bibfield
  {journal} {\bibinfo  {journal} {J. Air Transp. Manag.}\ }\textbf {\bibinfo
  {volume} {34}},\ \bibinfo {pages} {93 } (\bibinfo {year} {2014})}\BibitemShut
  {NoStop}%
\bibitem [{\citenamefont {Qiang}\ \emph {et~al.}(2014)\citenamefont {Qiang},
  \citenamefont {Jia}, \citenamefont {Xie},\ and\ \citenamefont
  {Gao}}]{Qiang/Jia/Xie/Gao:2014}%
  \BibitemOpen
  \bibfield  {author} {\bibinfo {author} {\bibfnamefont {S.-J.}\ \bibnamefont
  {Qiang}}, \bibinfo {author} {\bibfnamefont {B.}~\bibnamefont {Jia}}, \bibinfo
  {author} {\bibfnamefont {D.-F.}\ \bibnamefont {Xie}}, \ and\ \bibinfo
  {author} {\bibfnamefont {Z.-Y.}\ \bibnamefont {Gao}},\ }\href {\doibase
  10.1016/j.jairtraman.2014.05.007} {\bibfield  {journal} {\bibinfo  {journal}
  {J. Air Transp. Manag.}\ }\textbf {\bibinfo {volume} {40}},\ \bibinfo {pages}
  {42 } (\bibinfo {year} {2014})}\BibitemShut {NoStop}%
\bibitem [{\citenamefont {Notomista}\ \emph {et~al.}(2016)\citenamefont
  {Notomista}, \citenamefont {Selvaggio}, \citenamefont {Sbrizzi},
  \citenamefont {Maio}, \citenamefont {Grazioso},\ and\ \citenamefont
  {Botsch}}]{Notomistaetal:2016}%
  \BibitemOpen
  \bibfield  {author} {\bibinfo {author} {\bibfnamefont {G.}~\bibnamefont
  {Notomista}}, \bibinfo {author} {\bibfnamefont {M.}~\bibnamefont
  {Selvaggio}}, \bibinfo {author} {\bibfnamefont {F.}~\bibnamefont {Sbrizzi}},
  \bibinfo {author} {\bibfnamefont {G.~D.}\ \bibnamefont {Maio}}, \bibinfo
  {author} {\bibfnamefont {S.}~\bibnamefont {Grazioso}}, \ and\ \bibinfo
  {author} {\bibfnamefont {M.}~\bibnamefont {Botsch}},\ }\href {\doibase
  10.1016/j.jairtraman.2016.02.012} {\bibfield  {journal} {\bibinfo  {journal}
  {J. Air Transp. Manag.}\ }\textbf {\bibinfo {volume} {53}},\ \bibinfo {pages}
  {140 } (\bibinfo {year} {2016})}\BibitemShut {NoStop}%
\bibitem [{\citenamefont {Bachmat}(2019)}]{Bachmat:2019}%
  \BibitemOpen
  \bibfield  {author} {\bibinfo {author} {\bibfnamefont {E.}~\bibnamefont
  {Bachmat}},\ }\href {\doibase 10.1016/j.ejor.2018.12.017} {\bibfield
  {journal} {\bibinfo  {journal} {Eur. J. Oper. Res.}\ }\textbf {\bibinfo
  {volume} {275}},\ \bibinfo {pages} {1165 } (\bibinfo {year}
  {2019})}\BibitemShut {NoStop}%
\bibitem [{\citenamefont {Bachmat}\ \emph {et~al.}(2009)\citenamefont
  {Bachmat}, \citenamefont {Berend}, \citenamefont {Sapir}, \citenamefont
  {Skiena},\ and\ \citenamefont
  {Stolyarov}}]{Bachmat/Berend/Sapir/Skiena/Stolyarov:2009}%
  \BibitemOpen
  \bibfield  {author} {\bibinfo {author} {\bibfnamefont {E.}~\bibnamefont
  {Bachmat}}, \bibinfo {author} {\bibfnamefont {D.}~\bibnamefont {Berend}},
  \bibinfo {author} {\bibfnamefont {L.}~\bibnamefont {Sapir}}, \bibinfo
  {author} {\bibfnamefont {S.}~\bibnamefont {Skiena}}, \ and\ \bibinfo {author}
  {\bibfnamefont {N.}~\bibnamefont {Stolyarov}},\ }\href {\doibase
  10.1287/opre.1080.0630} {\bibfield  {journal} {\bibinfo  {journal} {Oper.
  Res.}\ }\textbf {\bibinfo {volume} {57}},\ \bibinfo {pages} {499} (\bibinfo
  {year} {2009})}\BibitemShut {NoStop}%
\bibitem [{\citenamefont {Vershik}\ and\ \citenamefont
  {Kerov}(1977)}]{Vershik/Kerov:1977}%
  \BibitemOpen
  \bibfield  {author} {\bibinfo {author} {\bibfnamefont {A.~M.}\ \bibnamefont
  {Vershik}}\ and\ \bibinfo {author} {\bibfnamefont {S.~V.}\ \bibnamefont
  {Kerov}},\ }\href@noop {} {\bibfield  {journal} {\bibinfo  {journal} {Sov.
  Math. Dokl.}\ }\textbf {\bibinfo {volume} {18}},\ \bibinfo {pages} {527}
  (\bibinfo {year} {1977})}\BibitemShut {NoStop}%
\bibitem [{\citenamefont {Logan}\ and\ \citenamefont
  {Shepp}(1977)}]{Logan/Shepp:1977}%
  \BibitemOpen
  \bibfield  {author} {\bibinfo {author} {\bibfnamefont {B.}~\bibnamefont
  {Logan}}\ and\ \bibinfo {author} {\bibfnamefont {L.}~\bibnamefont {Shepp}},\
  }\href {\doibase 10.1016/0001-8708(77)90030-5} {\bibfield  {journal}
  {\bibinfo  {journal} {Adv. Math.}\ }\textbf {\bibinfo {volume} {26}},\
  \bibinfo {pages} {206 } (\bibinfo {year} {1977})}\BibitemShut {NoStop}%
\bibitem [{\citenamefont {Deuschel}\ and\ \citenamefont
  {Zeitouni}(1995)}]{Deuschel/Zeitouni:1995}%
  \BibitemOpen
  \bibfield  {author} {\bibinfo {author} {\bibfnamefont {J.}~\bibnamefont
  {Deuschel}}\ and\ \bibinfo {author} {\bibfnamefont {O.}~\bibnamefont
  {Zeitouni}},\ }\href {\doibase 10.1214/aop/1176988293} {\bibfield  {journal}
  {\bibinfo  {journal} {Ann. Probab.}\ }\textbf {\bibinfo {volume} {23}},\
  \bibinfo {pages} {852} (\bibinfo {year} {1995})}\BibitemShut {NoStop}%
\end{thebibliography}%
\end{document}